\documentclass[twocolumn,twoside,noshowpacs,amsmath,amssymb,byrevtex,altaffilletter,citeautoscript,aps]{revtex4}
\pdfoutput=1
\usepackage{graphicx}
\usepackage{dcolumn}
\usepackage{bm}
\usepackage{units}
\usepackage{color}      
\usepackage{helvet}  
\usepackage{longtable}%
\usepackage{ifthen}
\usepackage{hyperref}

\newcommand{\capfont}{}
\newcommand{\Fsquared}{F$^2$ }
\newcommand{\degreesC}{$^\circ$C}

\newcommand{\setspacing}{\renewcommand{\baselinestretch}{1.8}\large\normalsize}
\newcommand{\onespacing}{\renewcommand{\baselinestretch}{1.15}\large\normalsize}
\newcommand{\atend}{0}
\newcommand{\forSubmission}{0}

\newcommand{\ThisPaper}{Burr et. al., ``Phase change memory technology''}

\newcommand{\SingleFigure}[4]{
\ifthenelse{\forSubmission = 1}{\ifthenelse{\atend=1}{
\begin{figure}[t!]
\includegraphics[width=85mm]{#1}
\vglue 1in \onespacing \caption{\label{#2}
\small \hfill \ThisPaper
\newline \newline #3 }
\setspacing
\end{figure}}{}}{
\begin{figure}
\vglue -0.05in
\includegraphics[width=#4]{#1}
\vglue -0.1in
\capfont \caption{\scriptsize \label{#2} \capfont #3}
\normalsize\vglue -0.1in\hrulefill\vglue -0.1in
\end{figure}}}

\newcommand{\TallSingleFigure}[4]{
\ifthenelse{\forSubmission = 1}{\ifthenelse{\atend=1}{
\begin{figure}[t!]
\includegraphics[width=85mm]{#1}
\vglue 0.25in \onespacing \caption{\label{#2}
\small \hfill \ThisPaper
\newline \newline #3 }
\setspacing
\end{figure}}{}}{
\begin{figure}
\vglue -0.05in
\includegraphics[width=#4]{#1}
\vglue -0.1in
\capfont \caption{\scriptsize \label{#2} \capfont #3}
\normalsize\vglue -0.1in\hrulefill\vglue -0.1in
\end{figure}}}

\newcommand{\DoubleFigure}[5]{
\ifthenelse{\forSubmission = 1}{\ifthenelse{\atend=1}{
\begin{figure}[t!]
\centerline{\epsfig{file=#1, width = 6.25in}}
\onespacing \vglue 0.75in  \caption{\label{#3}(a)
\small \hfill \ThisPaper
\newline \newline #4 \normalsize}
\setspacing
\end{figure}
\pagebreak
\addtocounter{figure}{-1}
\begin{figure}[t!]
\centerline{\epsfig{file=#2, width = 6.25in}}
\vglue 0.75in  \caption{(b)
\small \hfill \ThisPaper}
\end{figure}
}{}}{
\begin{figure}
\centerline{\epsfig{file=#1, width = #5}}
\vglue 0.1in
\centerline{\epsfig{file=#2, width = #5}}
\capfont \caption{\scriptsize \label{#3} \capfont #4}
\normalsize\vglue -0.1in\hrulefill\vglue -0.1in
\end{figure}}}

\newcommand{\FigureCaptionA}{
Programming of a PCM device involves application of electrical power through applied voltage, leading to internal temperature changes that
either melt and then rapidly quench a volume of amorphous material (RESET), or hold this volume at a slightly lower temperature for sufficient time for recrystallization (SET).  A low voltage is used to sense the device resistance (READ), so that the device state is not perturbed.
}
\newcommand{\FigureA}{
\SingleFigure{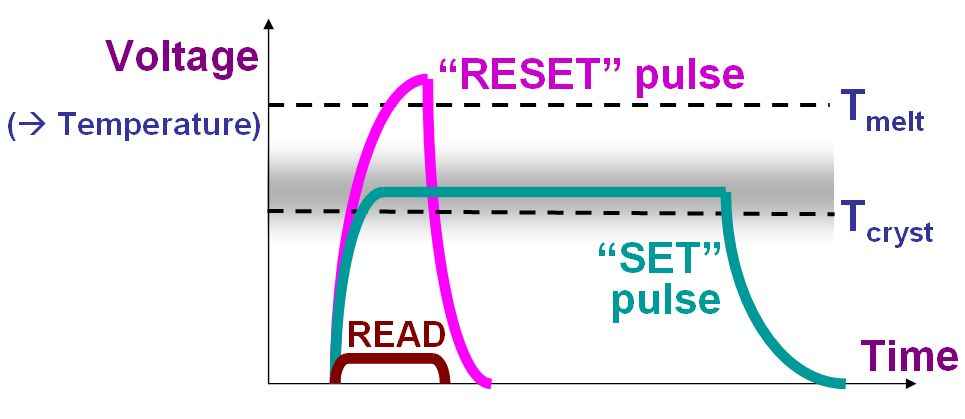}{fig:WhatIsPCM:SET,RESET,READ}{\FigureCaptionA}{\columnwidth}}

\newcommand{\FigureCaptionB}{
The memory hierarchy in computers spans orders of magnitude in read-write performance, ranging from small amounts of
expensive yet high-performance memory sitting near the Central Processing Unit (CPU) to vast amounts of low-cost yet very slow off-line storage.
}
\newcommand{\FigureB}{
\SingleFigure{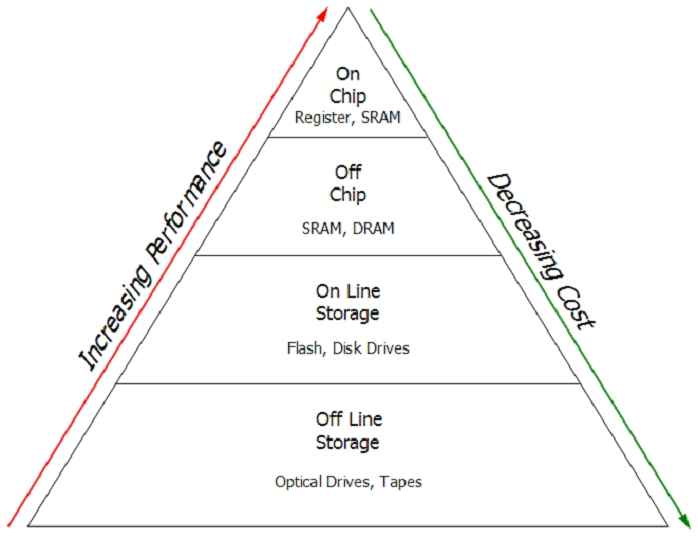}{fig:deviceApplications:MemoryHierarchy}{\FigureCaptionB}{\columnwidth}}

\newcommand{\FigureCaptionC}{
Qualitative representation of the cost and performance of various memory and storage technologies, ranging from extremely dense yet slow Hard Disk Drives (HDD) to ultra-fast but expensive SRAM. $F$ is the size of the smallest lithographic feature, and a smaller device footprint leads to higher density and thus lower cost.
}
\newcommand{\FigureC}{
\SingleFigure{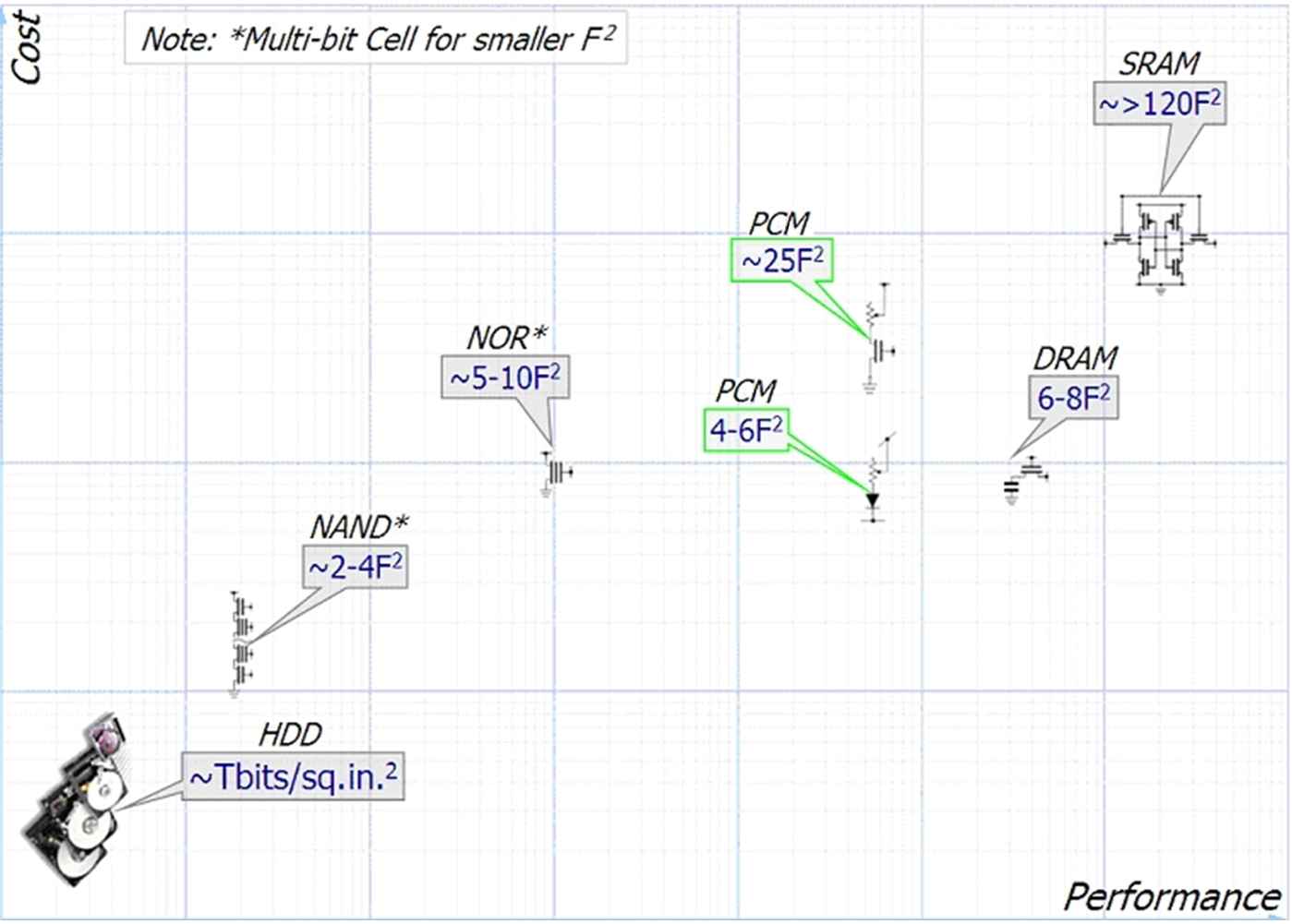}{fig:deviceApplications:CostPerformance}{\FigureCaptionC}{\columnwidth}}

\newcommand{\FigureCaptionD}{
Access times for various storage and memory technologies, both in nanoseconds and in terms of human perspective.  For the latter,
all times are scaled by 10$^9$ so that the fundamental unit of a single CPU operation is analogous to a human making a one-second decision.
In this context, writing data to Flash memory can require more than ``1 week'' and obtaining data from an offline tape cartridge takes ``1000 years.''\cite{WinfriedInfo, FAST:2009}.
}
\newcommand{\FigureD}{
\SingleFigure{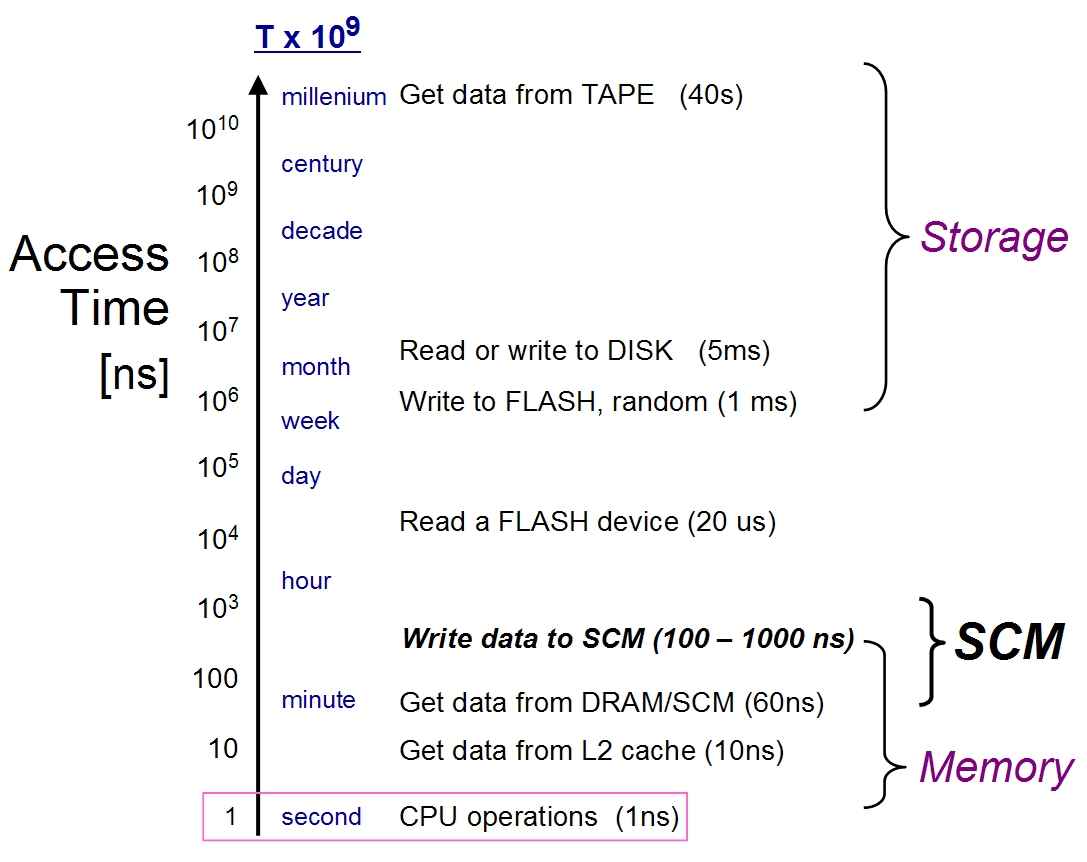}{fig:deviceApplications:HumanPerspectiveAccessTimes}{\FigureCaptionD}{\columnwidth}}

\newcommand{\FigureCaptionE}{
A semiconductor device technology node is commonly described by the minimum feature size $F$ that is available via lithographic patterning.
Thus the smallest device area that can be envisioned which is still accessible by lithographically-defined wiring is 4\Fsquared.
To increase \emph{effective bit density} beyond this, either sub-lithographic wiring, multiple bits per device (analogous to Multi-Level Cell (MLC) Flash technology), or multiple layers of stacked memory arrays are required, as described in Section~\ref{subsec:ultraHighDensity}.
}
\newcommand{\FigureE}{
\SingleFigure{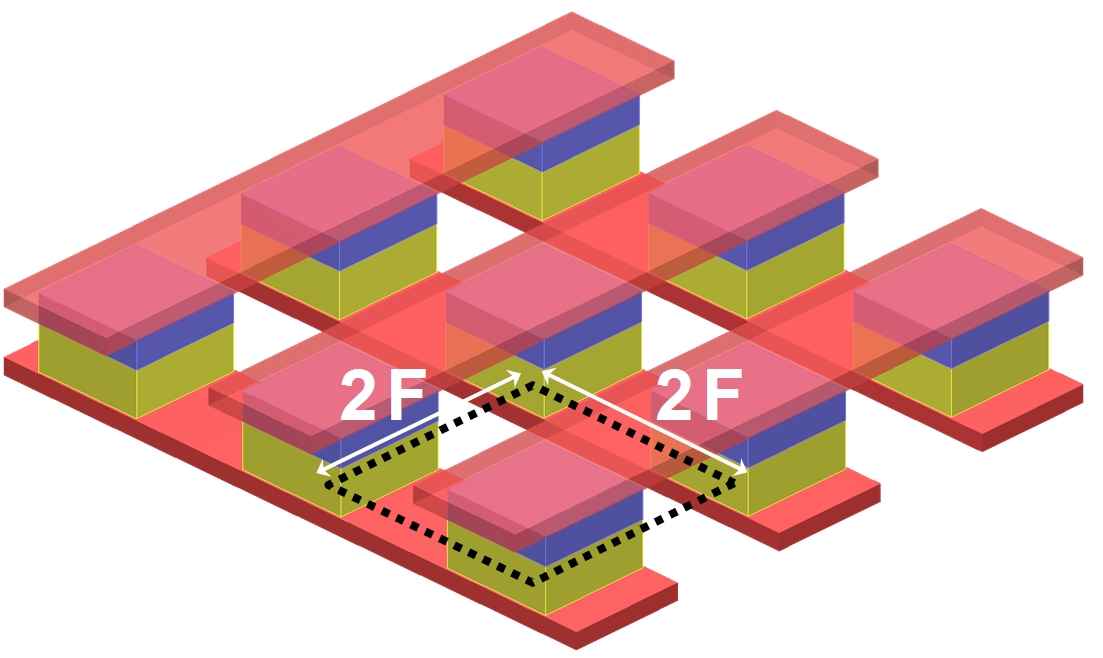}{fig:deviceApplications:FourFsquared}{\FigureCaptionE}{\columnwidth}}

\newcommand{\FigureCaptionF}{
(a) Phase change nanoparticles of Ge-Sb with 15 at. \% Ge, fabricated by electron-beam lithography, diameter about 40 nm. Reprinted with permission from Reference~\cite{Raoux:2007a} ($\copyright$ 2007 American Institute of Physics). (b) GeTe nanoparticles synthesized by solution-based chemistry, diameter about 30 nm\cite{Caldwell:2007}. (c) Nanoparticles of Ge-Sb with 15 at. \% Ge, fabricated by self-assembly based lithography and sputter deposition, diameter about 15 nm. Reprinted with permission from Reference~\cite{Zhang:2007h} ($\copyright$ 2007 American Institute of Physics). (d) Nanoparticles of Ge-Sb-Se, fabricated by self-assembly based lithography and spin-on deposition, diameter about 30 nm. Reprinted with permission from Reference~\cite{Milliron:2007} ($\copyright$ 2007
Macmillan Publishers Ltd).
}
\newcommand{\FigureF}{
\TallSingleFigure{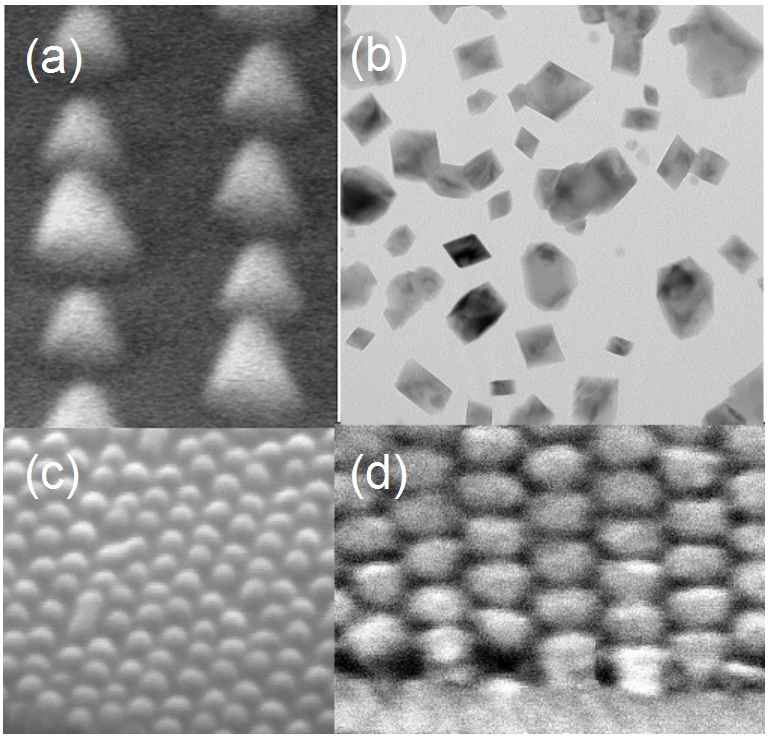}{fig:materScaling:Nanoparticles}{\FigureCaptionF}{\columnwidth}}

\newcommand{\FigureCaptionG}{
(a) Relative change in reflectivity $\Delta$R/R in \% of a crystalline Ge-Sb thin film with 15 at. \% Ge as a function of laser power and duration. The film was first crystallized by heating it in a furnace for 5 min at 300\degreesC. A first pulse of fixed time and power (100 ns, 50 mW) was  applied to create melt-quenched spots in the crystalline film, and then a second laser spot of variable power and duration at the same location was used to recrystallize the amorphous spots. (b) Normalized change in reflectivity (in \%) integrated over a power range between 24 and 25 mW from (a) as a function of laser pulse length. The dots are experimental data, the line is a fit to $1 \;-\; \exp -\left(\frac{t}{\tau}\right)^a$, with $t$ being the time, $\tau$ = 7 ns, and $a$ = 3.
(c) Relative change in reflectivity $\Delta$R/R in \% of an amorphous Ge-Sb thin film with 15 at. \% Ge as a function
of laser power and duration.  Note that much longer pulses are required.
Reprinted from Reference~\cite{Krebs:2009} ($\copyright$ 2009 American Institute of Physics).
}
\newcommand{\FigureG}{
\TallSingleFigure{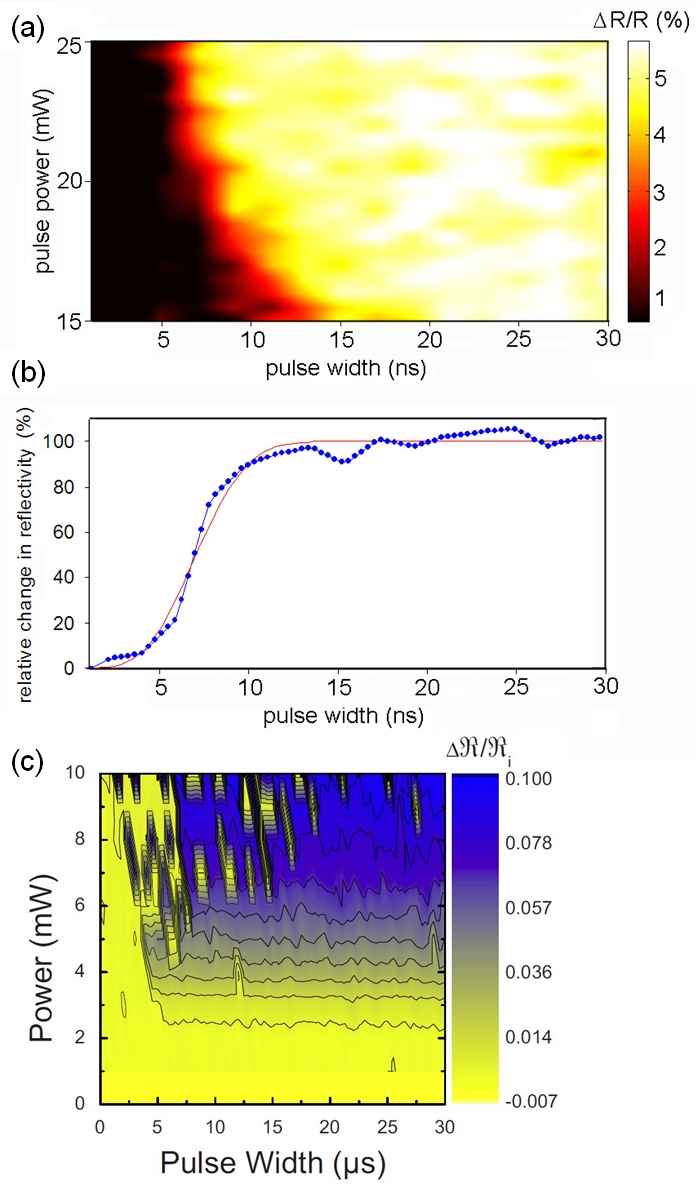}{fig:speed:Reflectivity}{\FigureCaptionG}{\columnwidth}}

\newcommand{\FigureCaptionH}{
Crystal growth velocity (red line) for GeSb as inferred by matching between simulation and empirical measurements.  Low temperature crystal growth speed was  measured by monitoring the slow growth of crystalline nuclei for growth-dominated (AIST) material\cite{Kalb:2004}; high-temperature crystal growth speeds represent the best match between the
measured optically-induced recrystallization of amorphous marks on thin-film GeSb and simulations of this process\cite{Kalb:2004, Burr:2008b}.
}
\newcommand{\FigureH}{
\SingleFigure{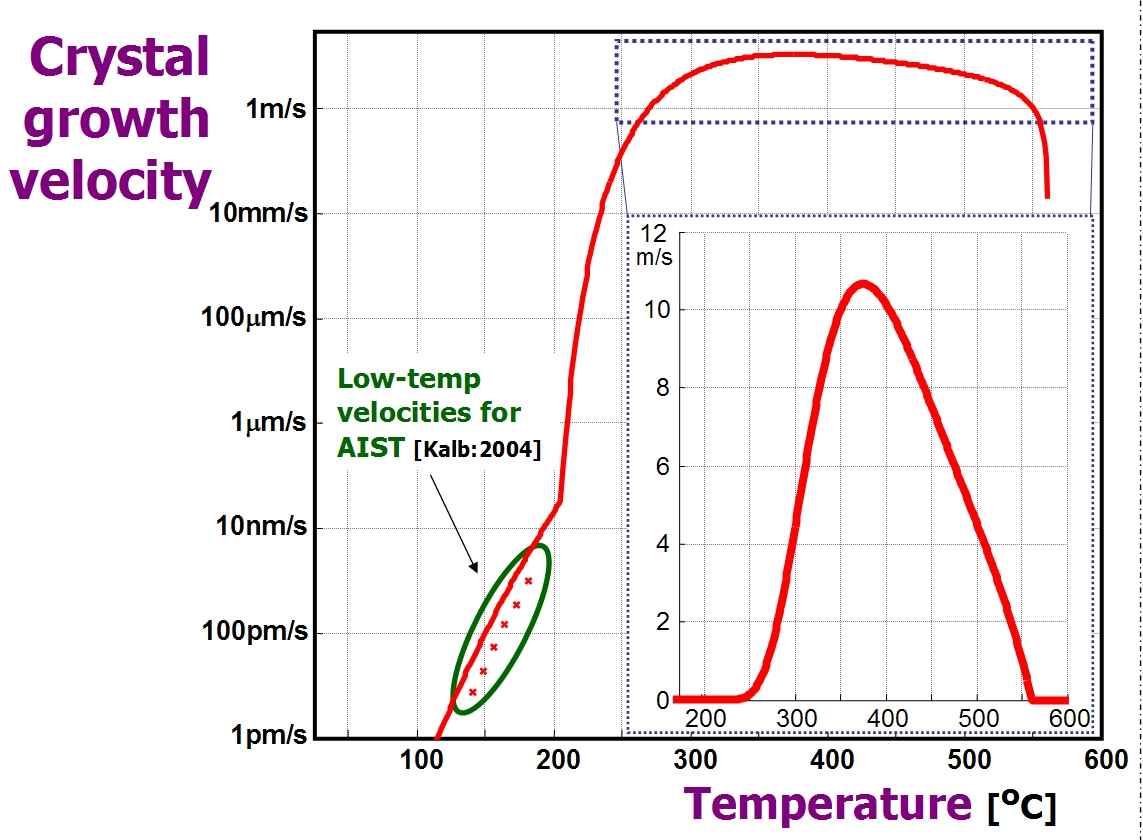}{fig:speed:GrowthSpeed}{\FigureCaptionH}{\columnwidth}}

\newcommand{\FigureCaptionI}{
(a) Scanning electron microscope image of a phase change bridge and its TiN electrodes. (b) Cross-sectional transmission electron microscope image of a 3nm-thick GeSb bridge. Reprinted with permission from Reference~\cite{Chen:2006t} ($\copyright$ 2006 IEEE).
}
\newcommand{\FigureI}{
\SingleFigure{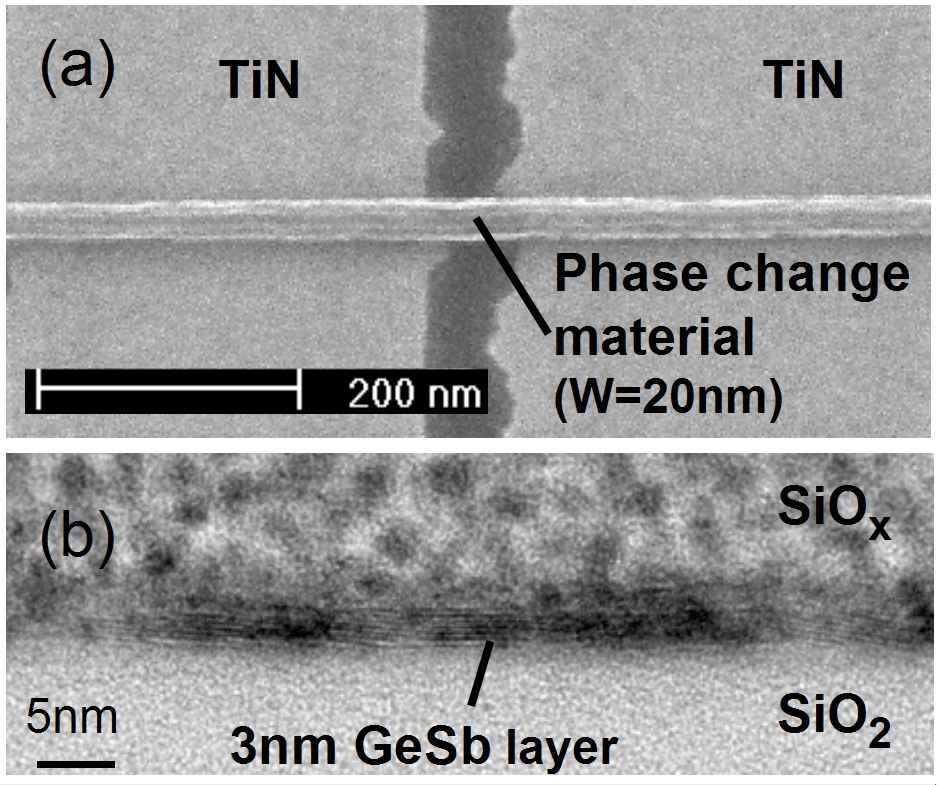}{fig:devScaling:PCBTEM}{\FigureCaptionI}{\columnwidth}}

\newcommand{\FigureCaptionJ}{
RESET current of doped-GeSb phase change bridge devices vs. cross-sectional area defined by the lithographic bridge width $W$ and the ultra-thin film thickness $H$.  Reprinted with permission from Reference~\cite{Chen:2006t} ($\copyright$ 2006 IEEE).
}
\newcommand{\FigureJ}{
\SingleFigure{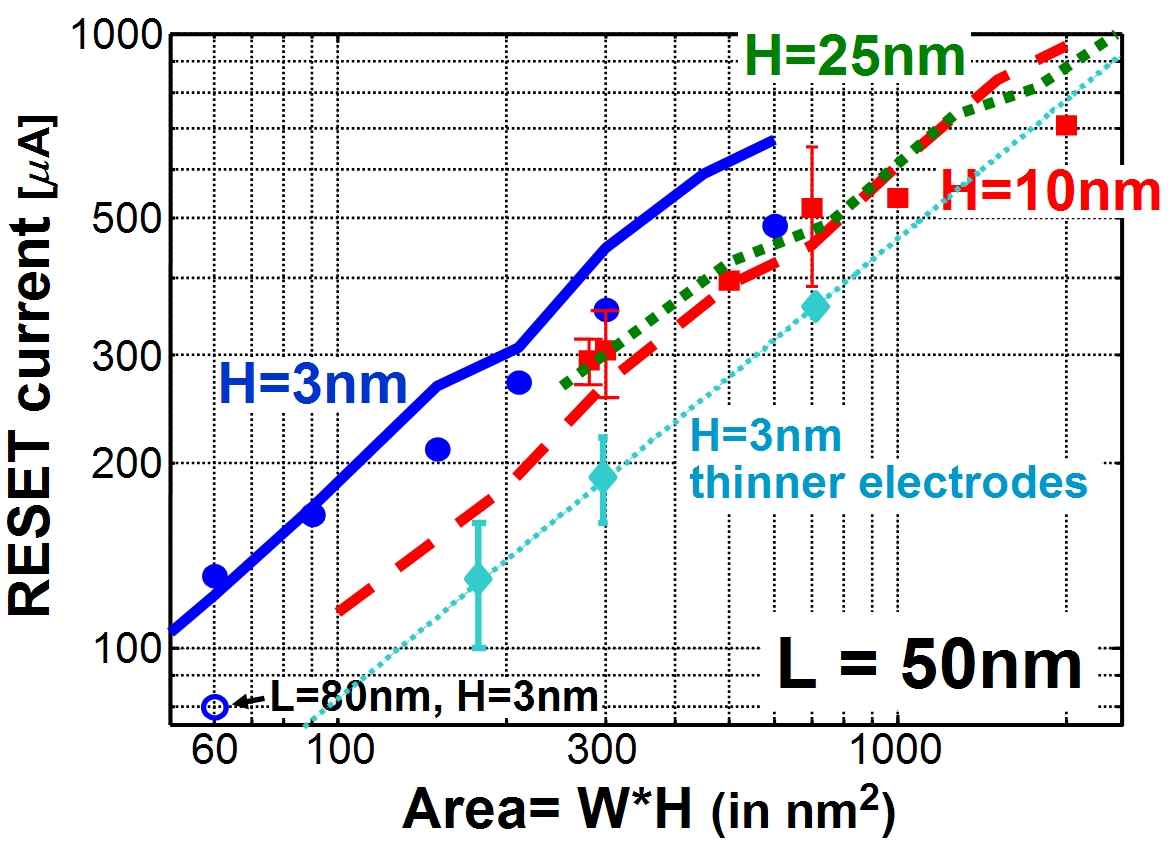}{fig:materScaling:PCBRESETcurrentScaling}{\FigureCaptionJ}{\columnwidth}}

\newcommand{\FigureCaptionK}{
Back-of-the-envelope estimate for the expected pulse width and associated
current density as function of the pitch $2F$ of the active volume of the phase change material.
Also shown are the empirical current densities for a phase-change bridge device (300$\mu$A for a $H=10$nm, $W=30$nm bridge\cite{Chen:2006t}, with the equivalent pitch for lithographic definition estimated to be $2\sqrt{10*30} \sim 35$nm pitch), and 160$\mu$A for a 7.5nm $\times$ 65nm dash-type cell (plotted for an equivalent 45nm pitch).
}
\newcommand{\FigureK}{
\SingleFigure{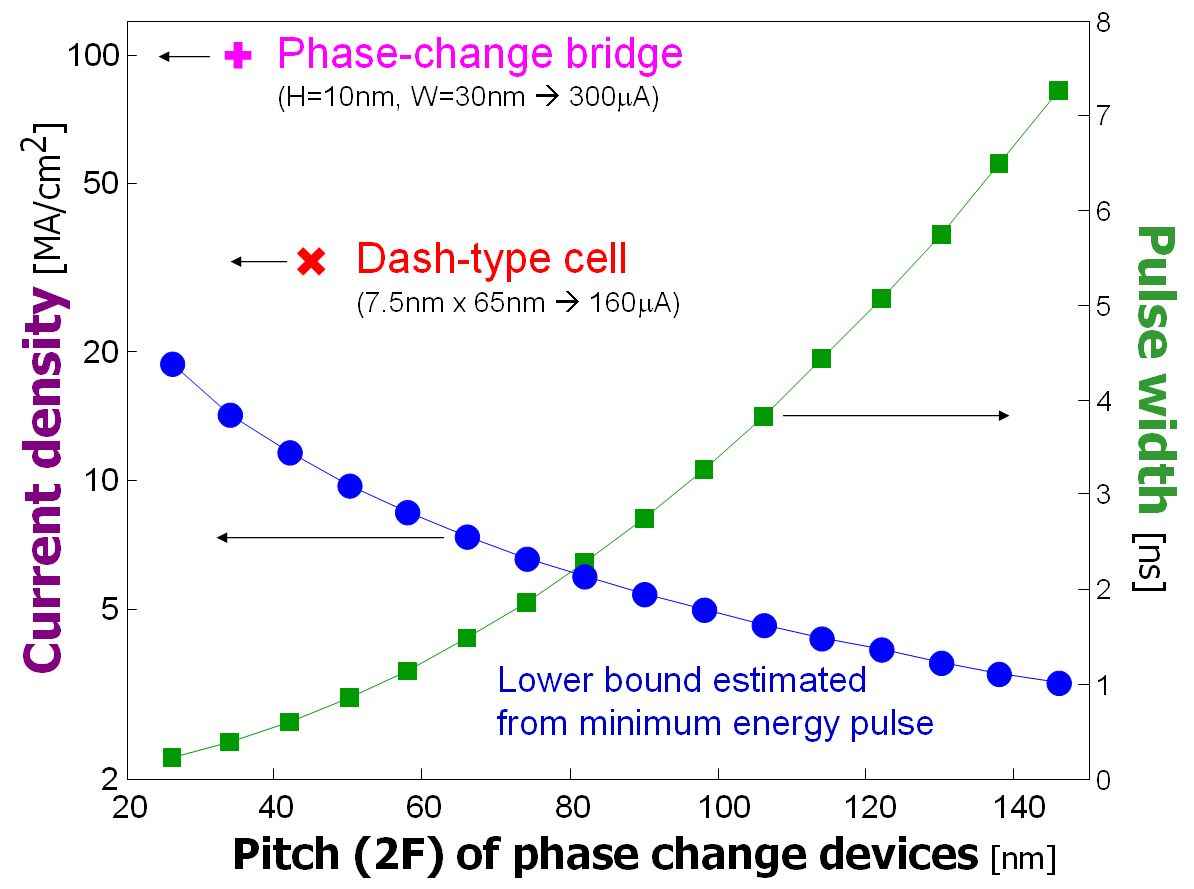}{fig:cellStructures:ExpectedCurrentVsPitch}{\FigureCaptionK}{\columnwidth}}

\newcommand{\FigureCaptionL}{
Phase-change device archetypes: (a) A typical contact-minimized cell, the mushroom
cell, forces current to pass through a small aperture formed by the intersection of one electrode and the phase change material. (b) A typical
volume-minimized cell, the ``pore'' cell, confines the volume of the phase change material in order to create a small cross-section within the
PCM device. Reprinted with permission from Reference~\cite{Raoux:2008a} ($\copyright$ 2008 IBM).
}
\newcommand{\FigureL}{
\SingleFigure{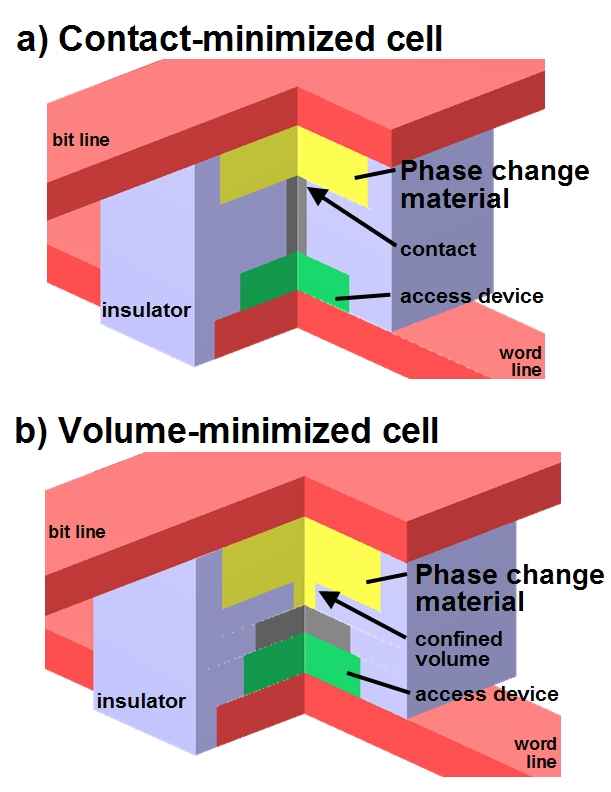}{fig:cellStructures:twoTypes}{\FigureCaptionL}{\columnwidth}}

\newcommand{\FigureCaptionM}{
TEM cross-sections of a mushroom cell PCM element in the (a) SET state and (b) RESET state. In the SET state, the phase change material is polycrystalline throughout.  In the RESET state, a ``mushroom cap'' of amorphous phase change material restricts the current flow through the bottom electrode. Reprinted with permission from Reference~\cite{Breitwisch:2009} ($\copyright$ 2009 Springer).
}
\newcommand{\FigureM}{
\SingleFigure{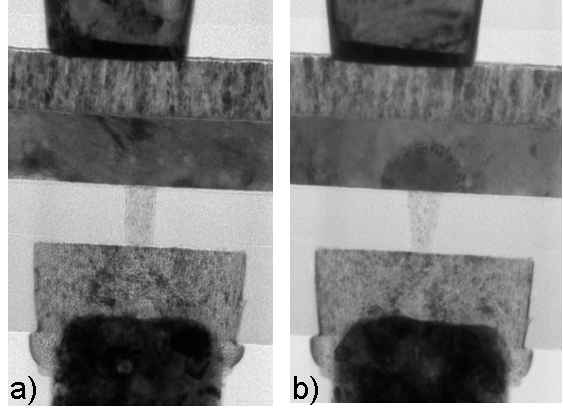}{fig:cellStructures:TEM_mushroom}{\FigureCaptionM}{\columnwidth}}

\newcommand{\FigureCaptionN}{
(a) TEM cross-section of a pillar cell with an FET access device.  (b) Close-up TEM cross-section of GST/TiN pillar.
The simulated RESET current dependence of this device is shown in Figure~\ref{fig:variability:RESETcurrentScaling}(b).
Reprinted with permission from Reference~\cite{Happ:2006} ($\copyright$ 2006 IEEE).
}
\newcommand{\FigureN}{
\SingleFigure{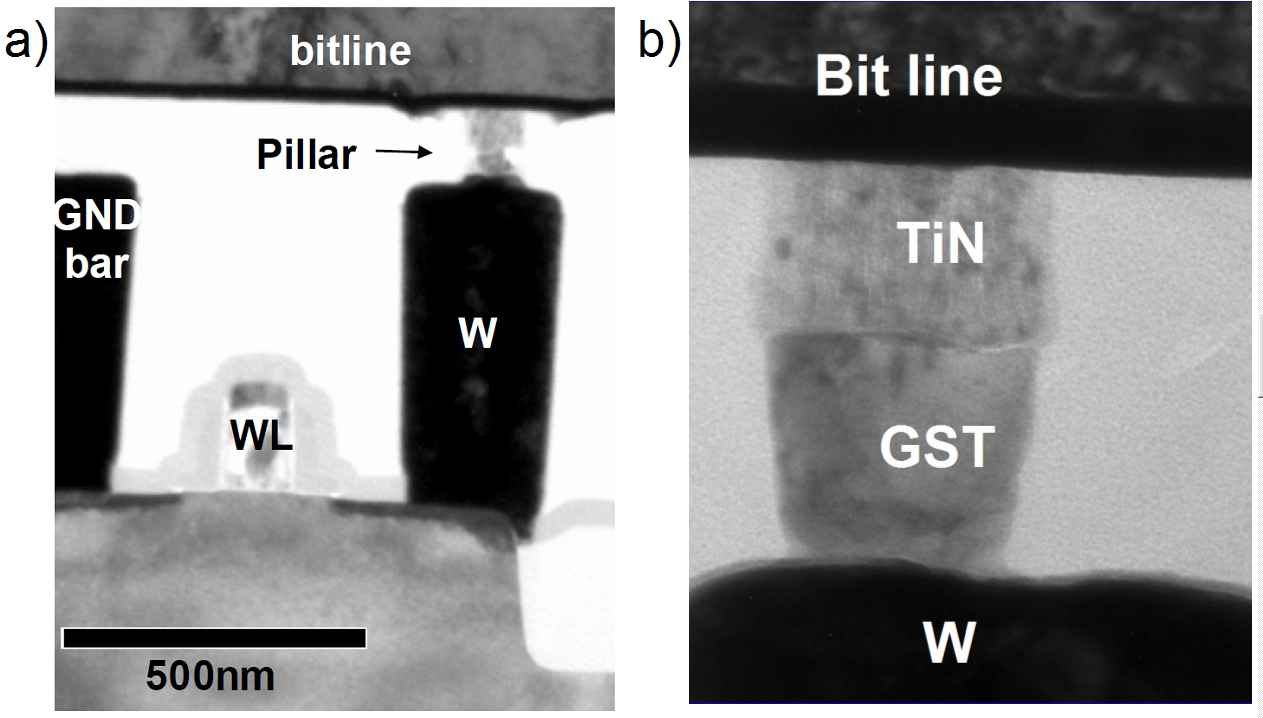}{fig:cellStructures:FETaccessDevice}{\FigureCaptionN}{\columnwidth}}

\newcommand{\FigureCaptionO}{
TEM cross-section of a 45 nm bottom CD low aspect ratio pore cell filled with a PVD GST process. The simulated RESET current dependence of this device is shown in Figure~\ref{fig:variability:RESETcurrentScaling}(c).
Reprinted with permission from Reference~\cite{Breitwisch:2007}, ($\copyright$ 2007 IEEE).
}
\newcommand{\FigureO}{
\SingleFigure{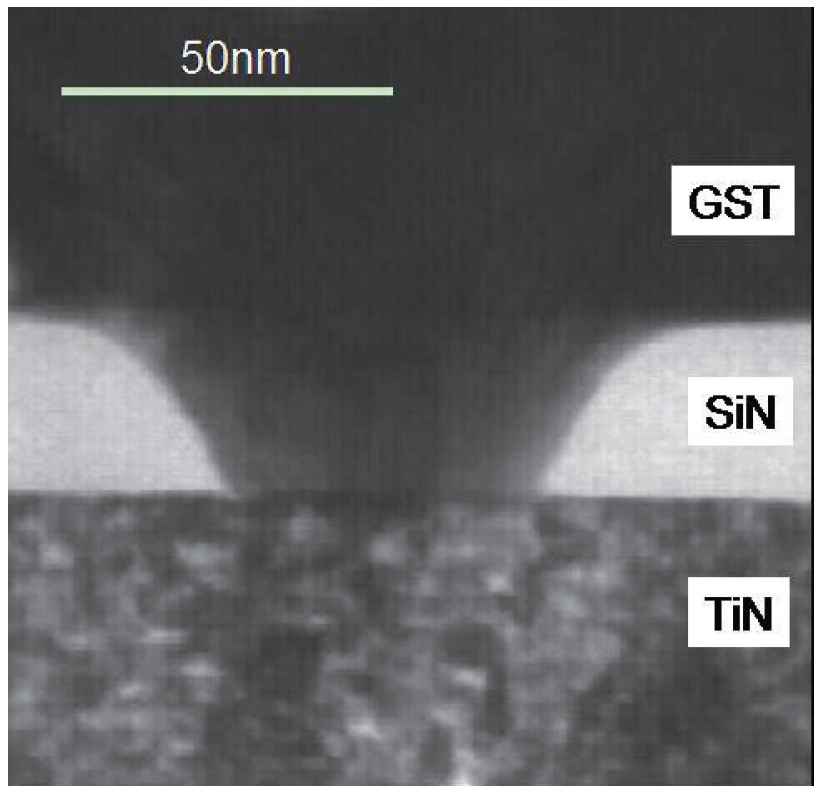}{fig:cellStructures:TEM_subLithoPore}{\FigureCaptionO}{\columnwidth}}

\newcommand{\FigureCaptionP}{
Illustration of the $\mu$-trench cell, showing two neighboring devices with a common top electrode
(e.g., along the bitline).  Current passes through an aperture which is limited in one dimension by the thickness of the underlying sidewall-deposited metal ``heater,'' and in the other dimension by the width of the narrow trench in which phase change material (here, GST) is deposited.
Reprinted with permission from Reference~\cite{Pellizzer:2004} ($\copyright$ 2004 IEEE).
}
\newcommand{\FigureP}{
\SingleFigure{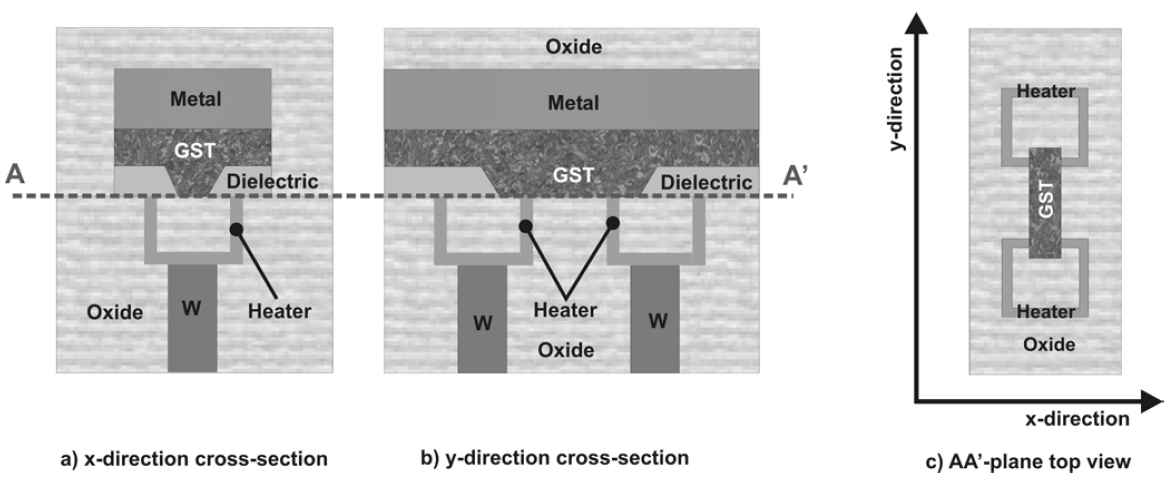}{fig:cellStructures:MicroTrench}{\FigureCaptionP}{\columnwidth}}

\newcommand{\FigureCaptionQ}{
TEM cross-section of the dash-confined cell,
showing devices fabricated by a spacer process that are only 7.5nm wide, and 65nm deep in the orthogonal direction.
 Reprinted with permission from Reference~\cite{Im:2008} ($\copyright$ 2008 IEEE).
}
\newcommand{\FigureQ}{
\SingleFigure{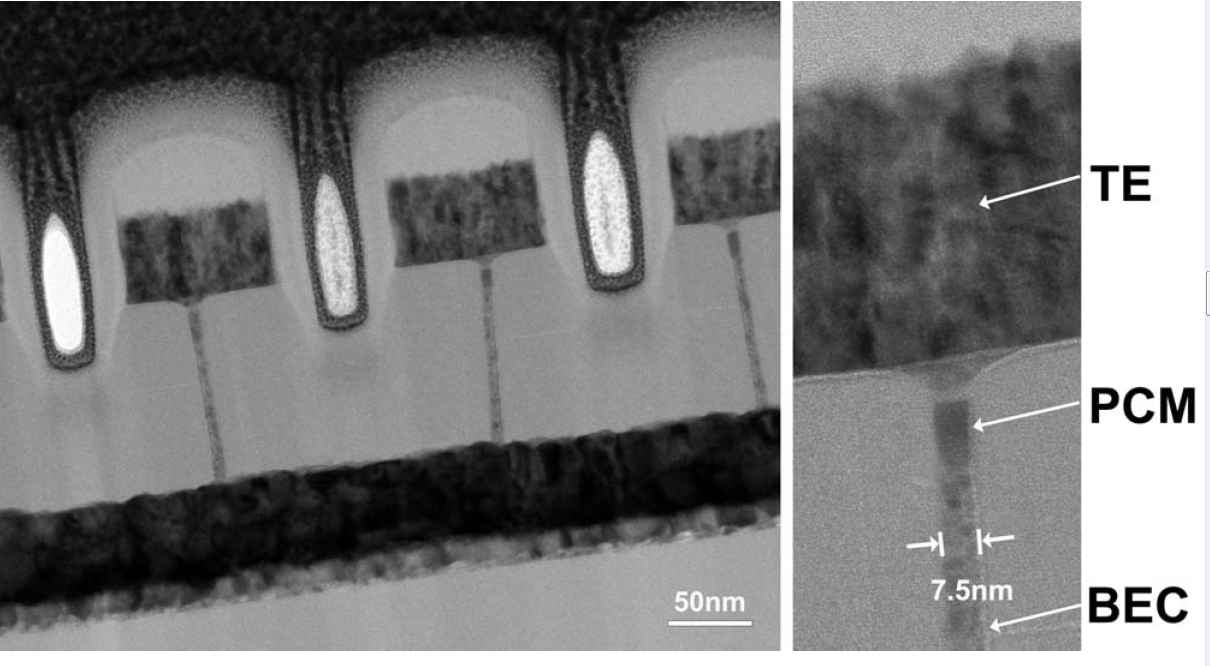}{fig:cellStructures:DashConfinedCell}{\FigureCaptionQ}{\columnwidth}}

\newcommand{\FigureCaptionR}{
a) RESET current for the Pillar cell and the Mushroom cell both show a strong dependence on the critical aperture size. Reprinted with permission from Reference~\cite{Happ:2006} ($\copyright$ 2006 IEEE).  b) The Pore cell RESET current is both strongly dependent on the aperture size and shape (pore slope). Reprinted with permission from Reference~\cite{Breitwisch:2007} ($\copyright$ 2007 IEEE).  (Figure~\ref{fig:materScaling:PCBRESETcurrentScaling} shows how the RESET current of the Bridge cell scales directly with the cross-sectional area of the phase change material.)
}
\newcommand{\FigureR}{
\TallSingleFigure{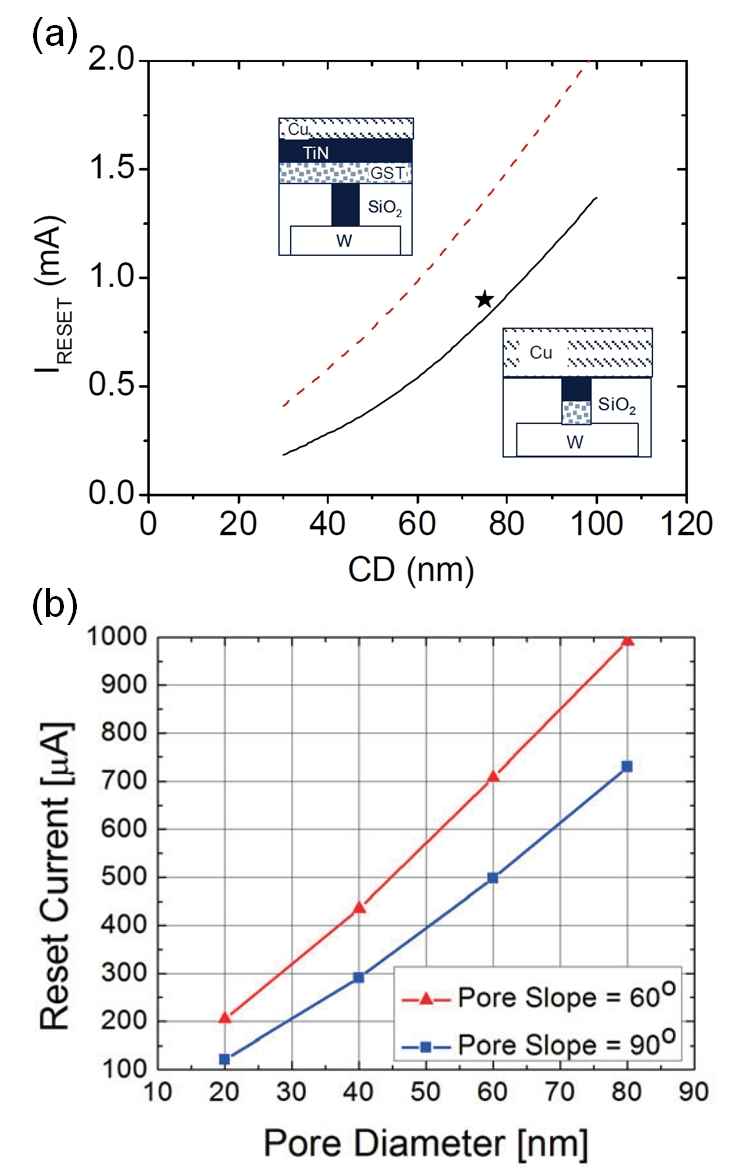}{fig:variability:RESETcurrentScaling}{\FigureCaptionR}{\columnwidth}}

\newcommand{\FigureCaptionS}{
(a) A collar process is used to create a sub-lithographically sized TiN bottom electrode.  First, a lithographically defined hole of diameter $D$ is etched into an SiON/SiN stack.  A first collar is formed by depositing a conformal SiON layer followed by a collar RIE step.  A second collar is formed in the same manner.  Next, the CVD TiN is deposited to fill the hole.  Finally, a series of CMP (Chemical-Mechanical Polishing) and oxide etchback processes are performed, resulting in a cylindrical TiN bottom electrode.    (b) A TiN ring electrode is constructed in a similar manner except that only a thin layer of CVD TiN is deposited into the hole, and then the center of the hole is filled with oxide.  Reprinted with permission from Reference~\cite{Breitwisch:2009} ($\copyright$ Springer 2009),
original figure from Reference~\cite{Ahn:2005a} ($\copyright$ IEEE 2005).)
}
\newcommand{\FigureS}{
\SingleFigure{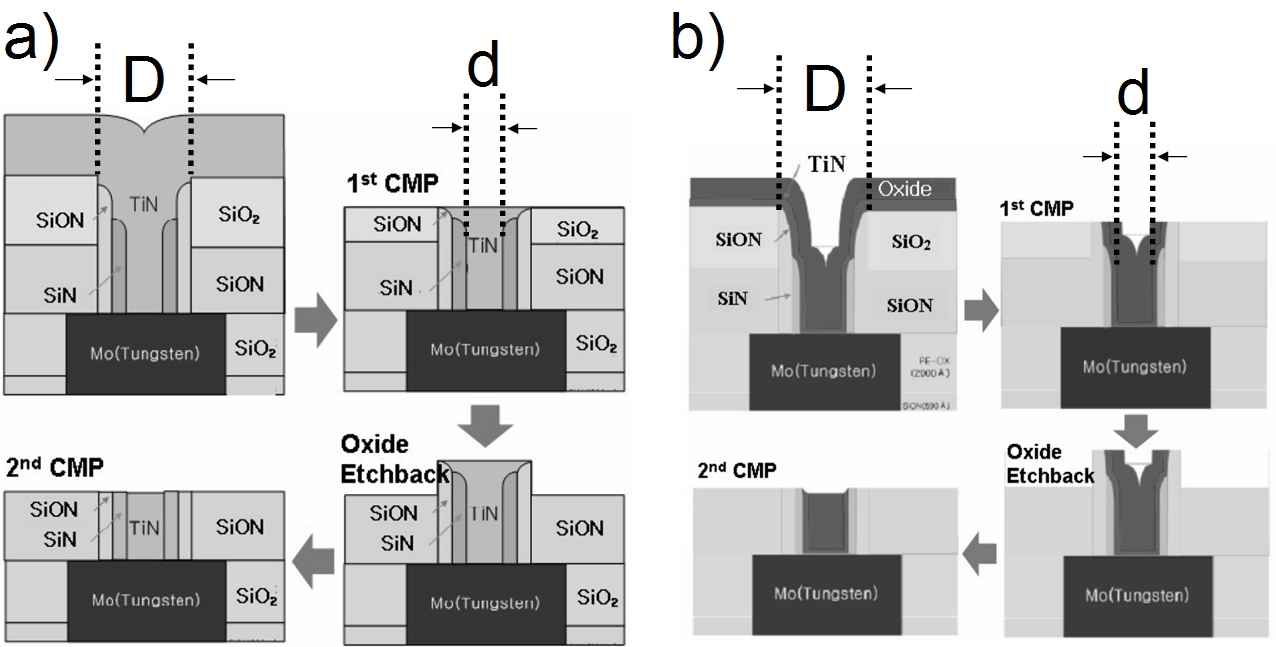}{fig:variability:CollarProcesses}{\FigureCaptionS}{\columnwidth}}

\newcommand{\FigureCaptionT}{
(a) A sub-lithographic and lithography-independent feature is fabricated using the keyhole-transfer process: 1) A lithographically-defined hole is etched, and 2) the middle SiO$_2$ layer is recessed.  3) A highly conformal poly-Si film is deposited, producing a sub-lithographic keyhole whose diameter is equal to the recess of the SiO$_2$ layer. 4) The keyhole is transferred into the underlying SiN layer to define a pore, followed by 5) removal of the SiO$_2$ and poly-Si.  6) The phase change and top electrode (TiN) materials are then deposited and the cell is patterned for isolation.  (b) An SEM cross-section corresponding to step 3), showing keyholes for two different sized lithographically-defined holes.  Since the keyhole size does not depend on lithography, the phase change CD can be successfully decoupled from any lithographic variability. Reprinted with permission from Reference~\cite{Breitwisch:2009} ($\copyright$ Springer 2009).
}
\newcommand{\FigureT}{
\SingleFigure{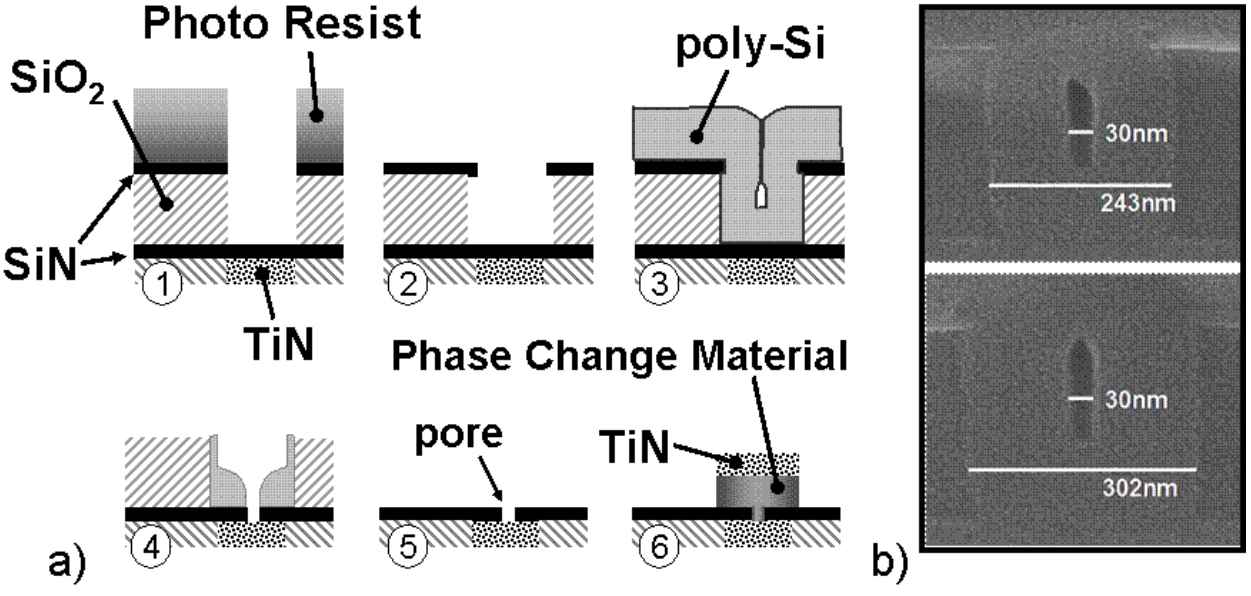}{fig:variability:LithoPoreIdea}{\FigureCaptionT}{\columnwidth}}

\newcommand{\FigureCaptionU}{
SET resistance and RESET resistance distributions as a function of the programming pulse width.  In this example, while 100ns is sufficient to SET most of the cells to below 2 k$\Omega$, many cells still have a resistance greater than 10k$\Omega$.  However, extending the 50ns RESET pulse to 100ns has no noticeable effect on increasing the resistance of the RESET tail.  GST refers to the phase change material used in this experiment, Ge$_2$Sb$_2$Te$_5$. Reprinted with permission from Reference~~\cite{Kang:2007a} ($\copyright$ 2007 IEEE).
}
\newcommand{\FigureU}{
\SingleFigure{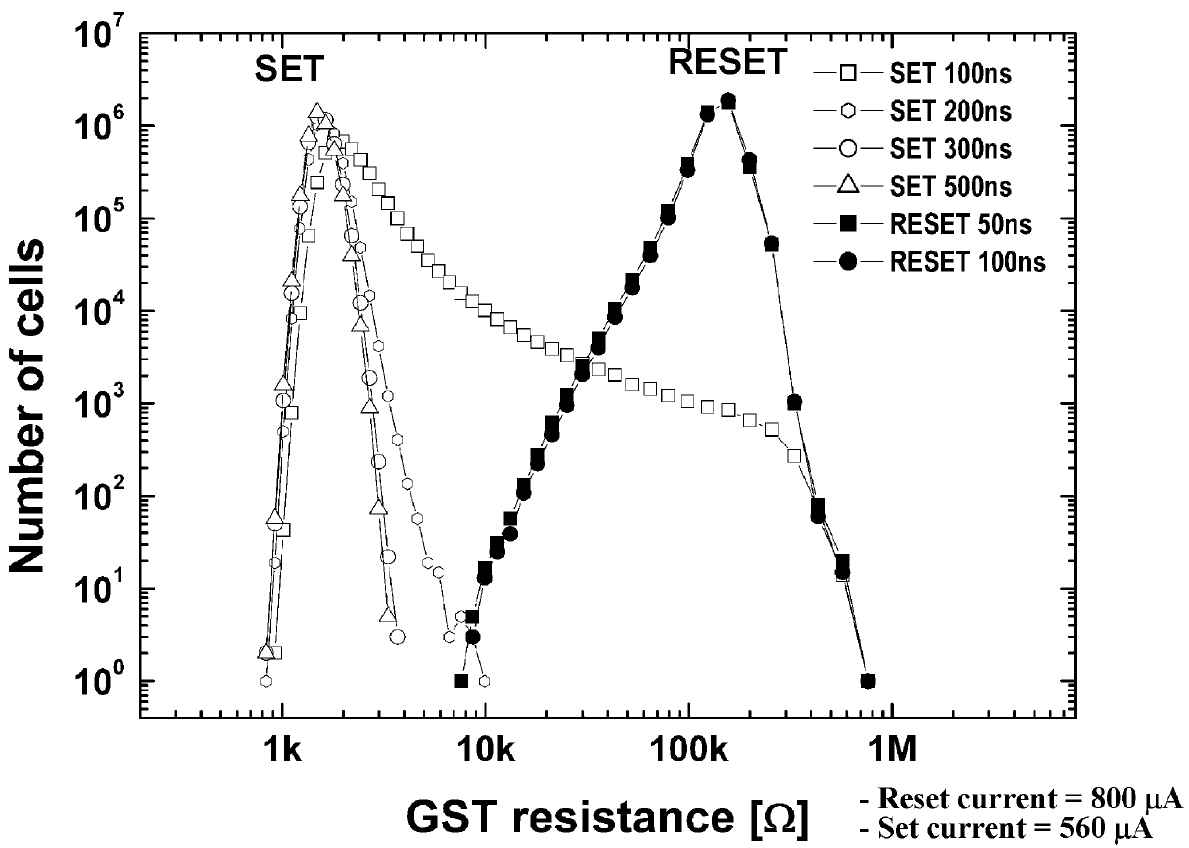}{fig:variability:Rdistributions}{\FigureCaptionU}{\columnwidth}}

\newcommand{\FigureCaptionV}{
RESET tail modulation by the quenching time $t_Q$. A long quench time results in a partial SET of a small fraction of devices within a large array. Reprinted from Reference~\cite{Mantegazza:2008} ($\copyright$ 2008 Elsevier).
}
\newcommand{\FigureV}{
\SingleFigure{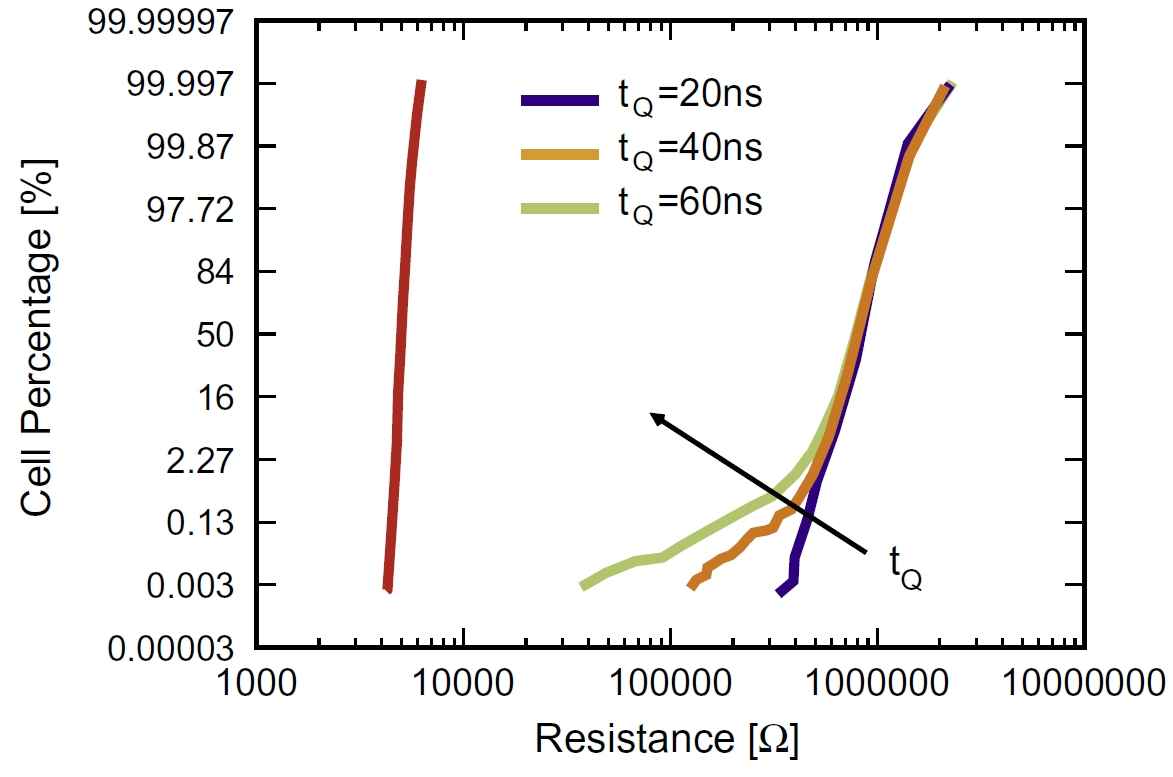}{fig:variability:RESETtailModulation}{\FigureCaptionV}{\columnwidth}}

\newcommand{\FigureCaptionW}{
Cycling performance of the SET and RESET state of a single PCM cell. Reprinted with permission from Reference~\cite{Lai:2001} ($\copyright$ 2001 IEEE).
}
\newcommand{\FigureW}{
\SingleFigure{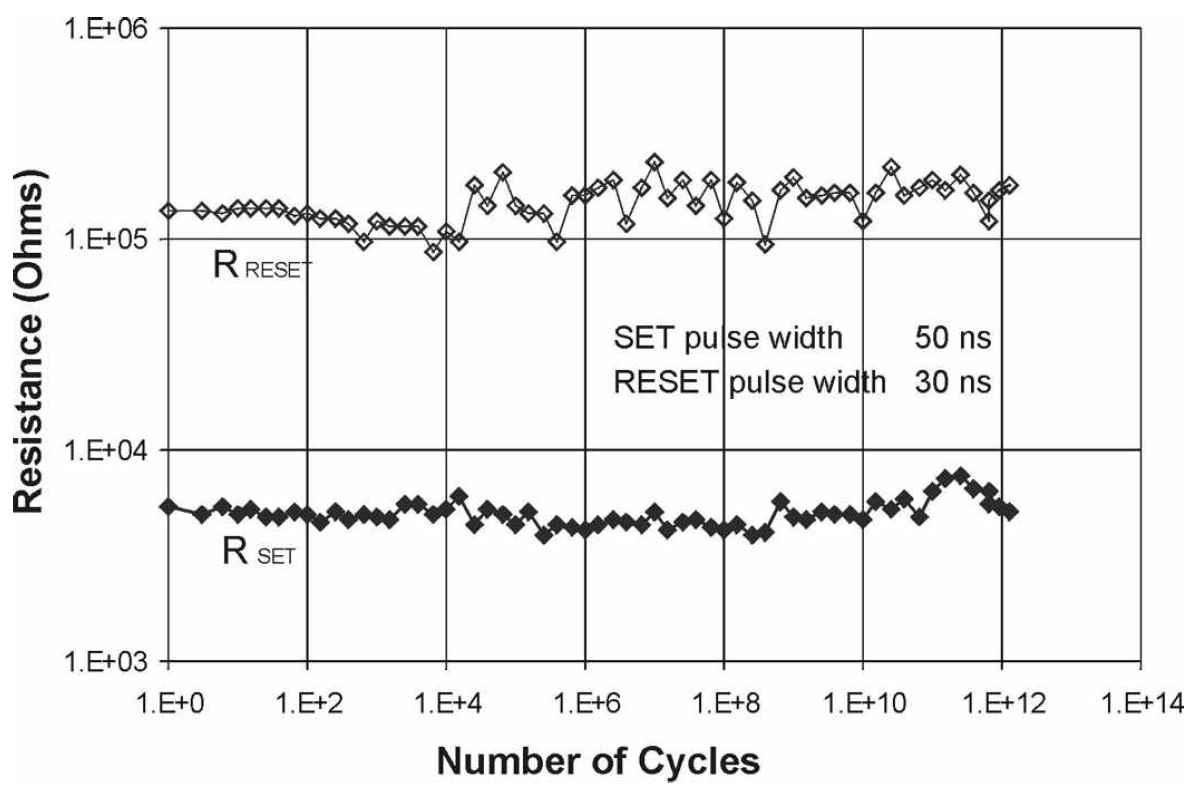}{fig:variability:CyclingPerformance}{\FigureCaptionW}{\columnwidth}}

\newcommand{\FigureCaptionX}{
Resistance distribution of a four-level cell using single pulse programming.  Process-induced variations cause distributions to overlap because the same applied voltage pulse leads to different temperatures in different cells.  Reprinted with permission from Reference~\cite{Nirschl:2007} ($\copyright$ 2007 IEEE).
}
\newcommand{\FigureX}{
\SingleFigure{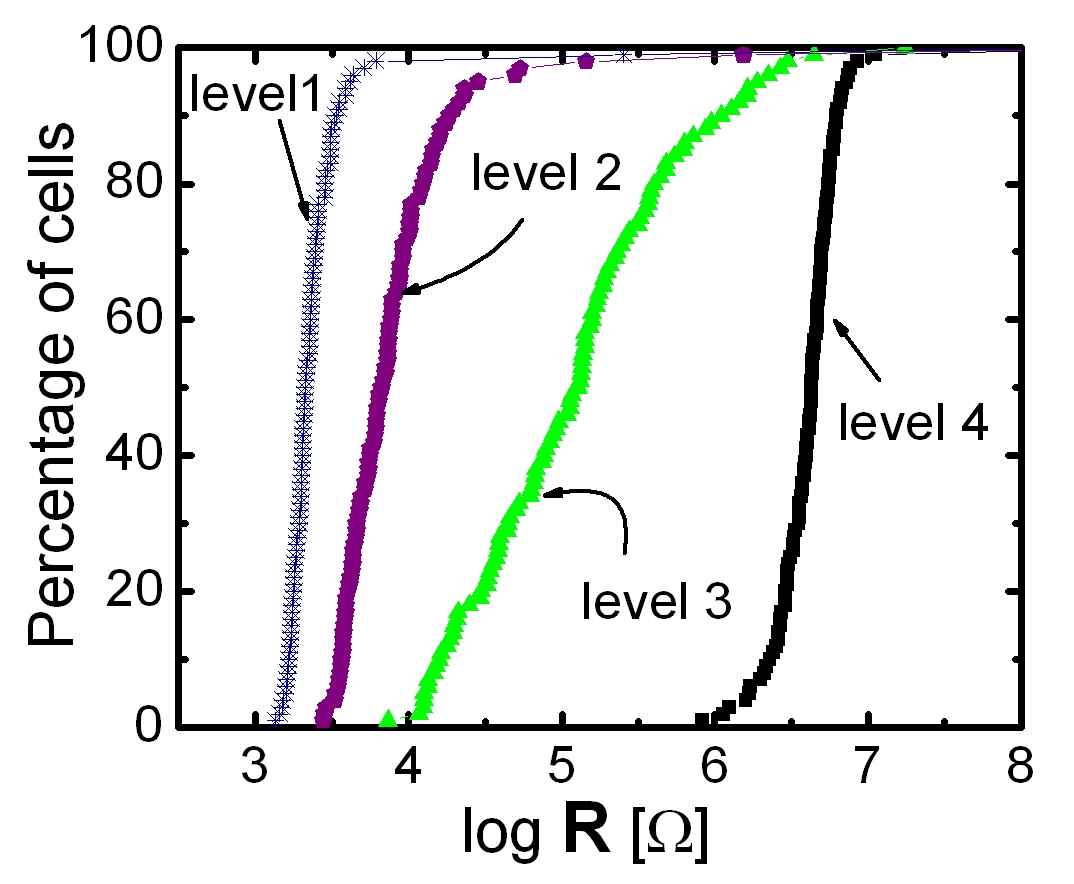}{fig:variability:BeforeMLCdistributions}{\FigureCaptionX}{\columnwidth}}

\newcommand{\FigureCaptionY}{
10$\times$10 array (100 devices) test structure programmed into 16 levels. Tight, well-controlled distributions allow 4 bits/cell. Iterative adjustment of pulse slopes depending on the programmed resistances is one method for achieving narrow distributions.  Reprinted with permission from Reference~\cite{Nirschl:2007} ($\copyright$ 2007 IEEE).
}
\newcommand{\FigureY}{
\SingleFigure{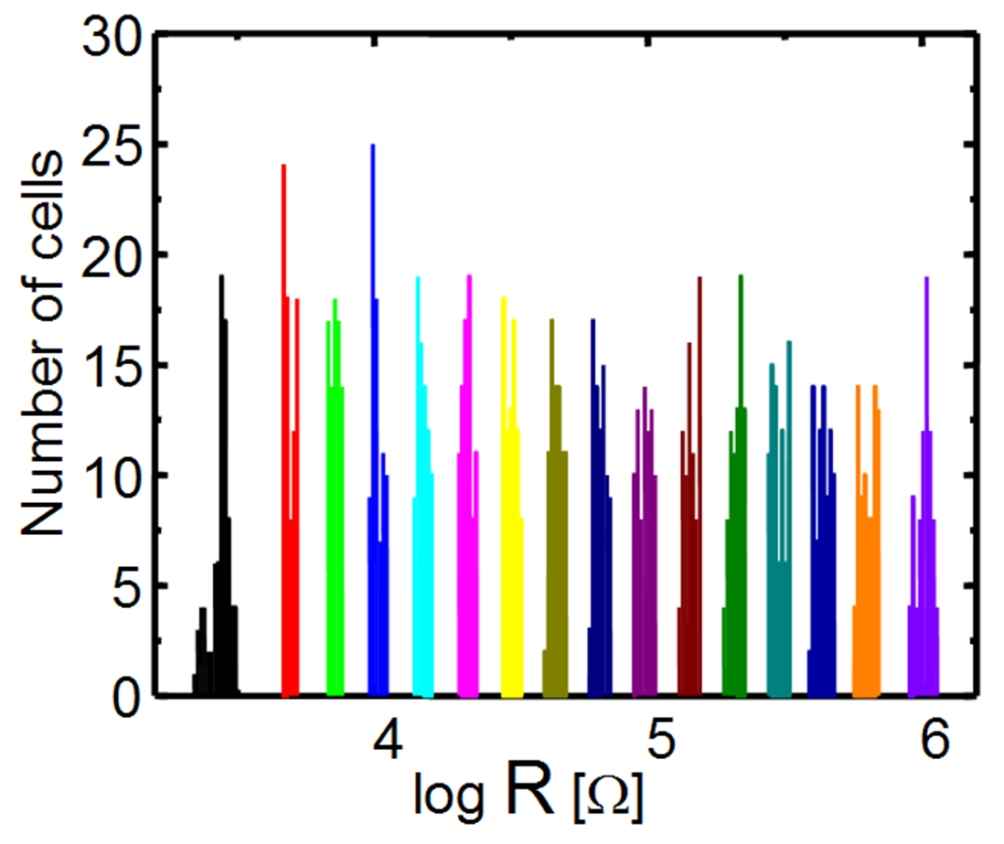}{fig:variability:AfterMLCdistributions}{\FigureCaptionY}{\columnwidth}}

\newcommand{\FigureCaptionZ}{
Accelerated failure of a $\mu$-trench PCM cell, showing decrease in RESET resistance as a function of
time at 210\degreesC. Reprinted with permission from Reference~\cite{Russo:2006} ($\copyright$ 2006 IEEE).
}
\newcommand{\FigureZ}{
\SingleFigure{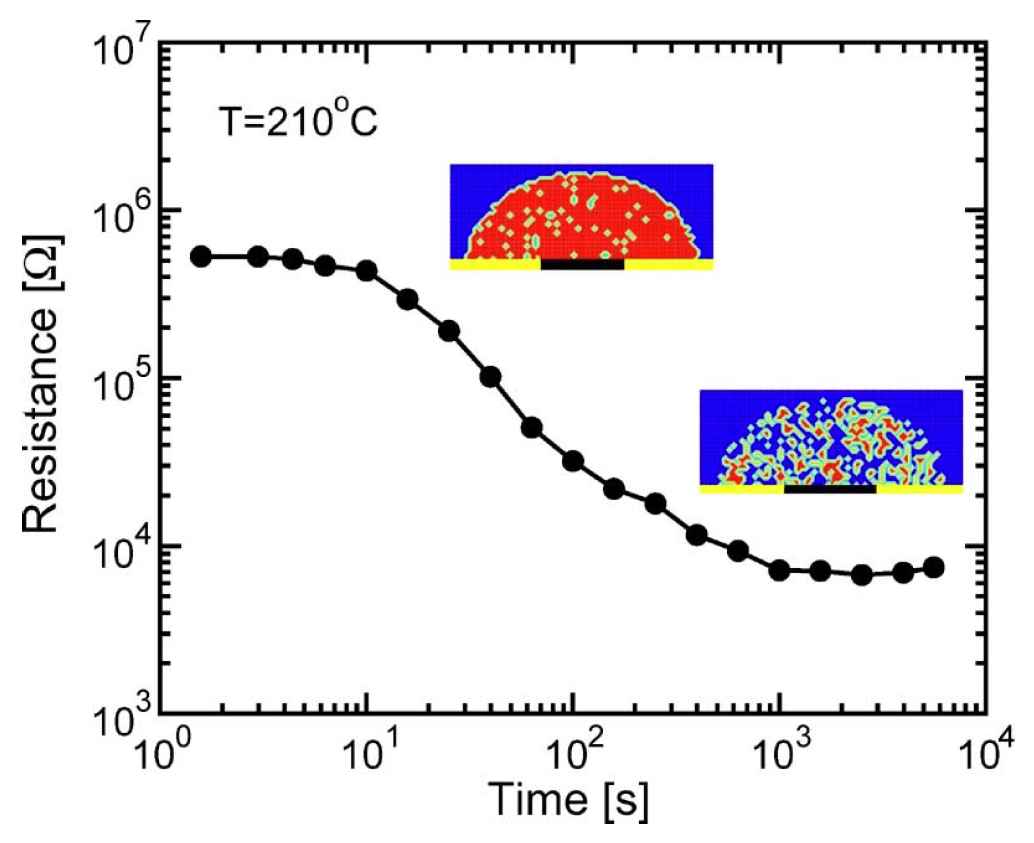}{fig:retention:simpleLoss}{\FigureCaptionZ}{\columnwidth}}

\newcommand{\FigureCaptionAA}{
Simulated temperature profiles for PCM devices (``micro-trench''-type devices) for the
180nm and 65nm technology nodes.  Reprinted with permission from Reference~\cite{Pirovano:2003} ($\copyright$ 2003 IEEE).
Note that while the transient temperatures become close to the steady-state temperature, the expected temperature rise at the neighboring
device remains much lower than 100\degreesC.
}
\newcommand{\FigureAA}{
\TallSingleFigure{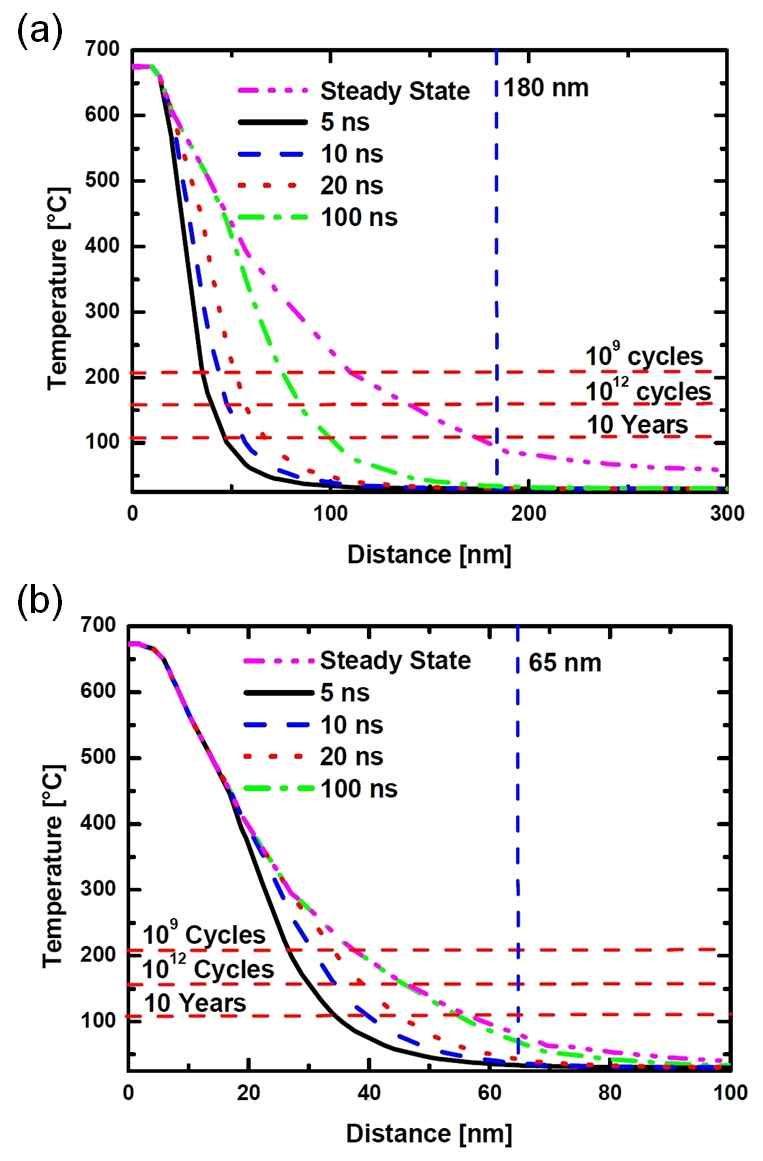}{fig:retention:PirovanoSimulations}{\FigureCaptionAA}{\columnwidth}}

\newcommand{\FigureCaptionBB}{
SET and RESET Resistances during cycling, illustrating the differences between
failure by ``stuck-SET'' and by ``stuck-RESET.'' Reprinted with permission
from Reference~\cite{Gleixner:2007} ($\copyright$ 2007 IEEE).
}
\newcommand{\FigureBB}{
\SingleFigure{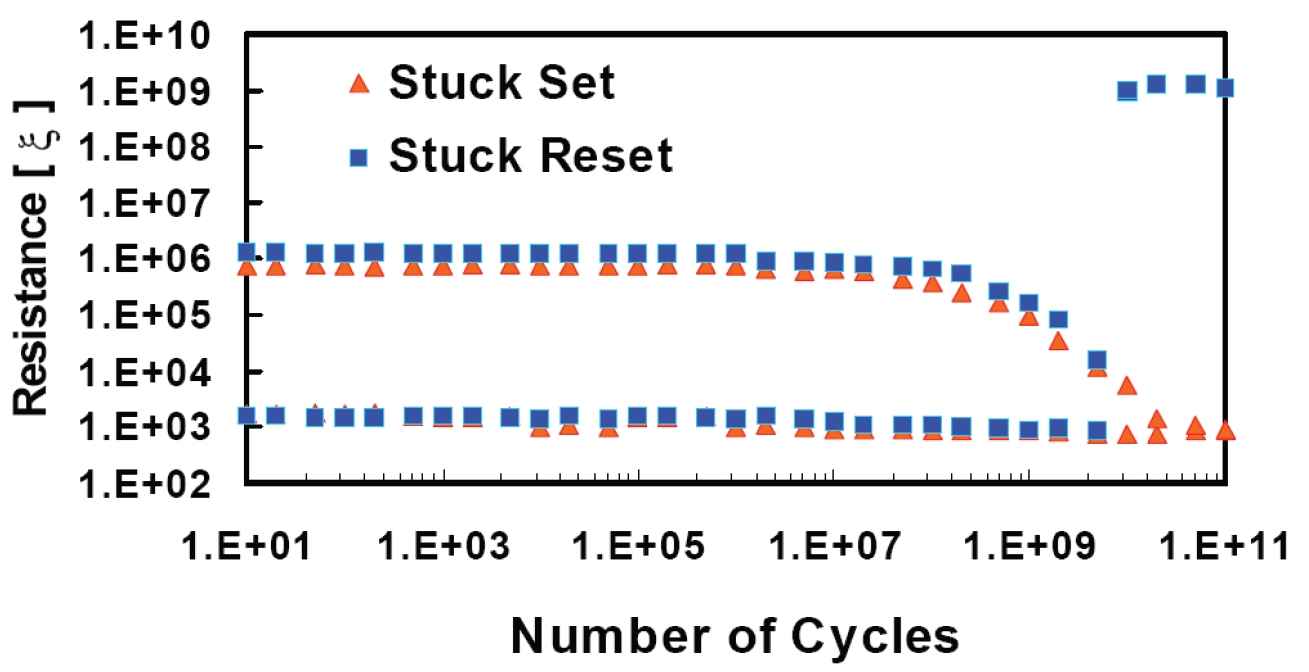}{fig:endurance:cyclingFailures}{\FigureCaptionBB}{\columnwidth}}

\newcommand{\FigureCaptionCC}{
Cycling endurance as a function of pulse energy, showing that device endurance
drops rapidly with prolonged exposure to high temperatures. Reprinted with permission from Reference~\cite{Lai:2003} ($\copyright$ 2003 IEEE).
}
\newcommand{\FigureCC}{
\SingleFigure{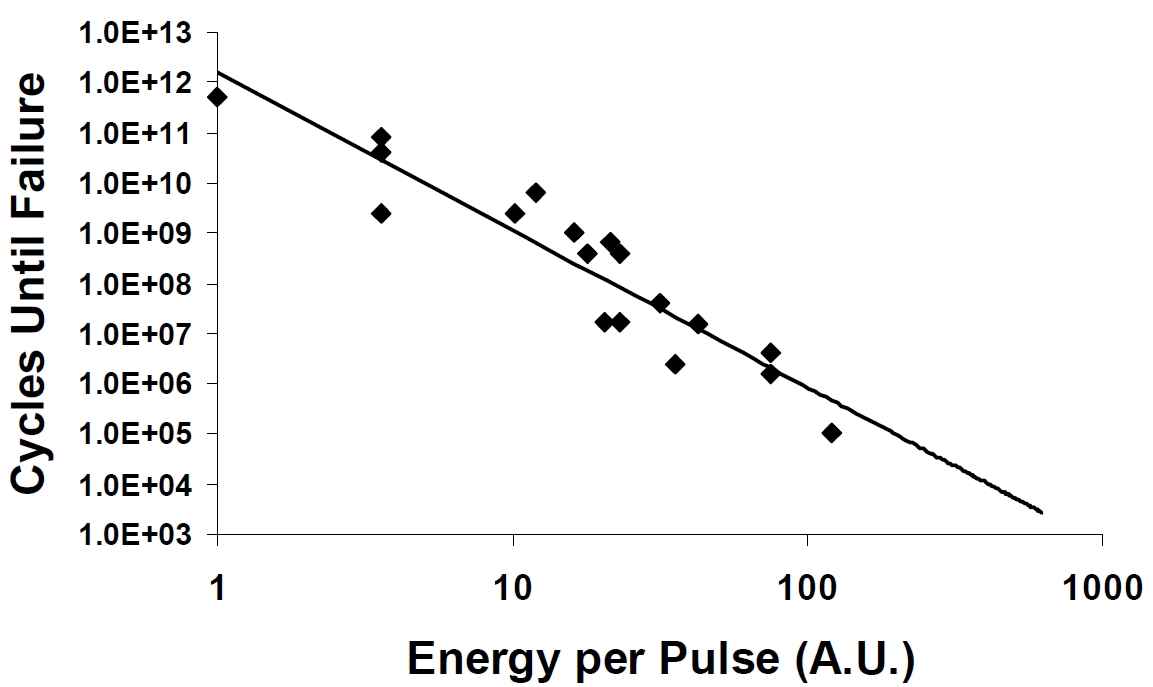}{fig:endurance:vsEnergy}{\FigureCaptionCC}{\columnwidth}}

\newcommand{\FigureCaptionDD}{
Top-down TEM images of large phase change bridge devices ($L$=740nm, $W$=300nm bridges of 20nm thick Ge-doped SbTe material), showing a $\sim$100nm polarity-dependent shift of the amorphous plug towards the anode (+). Reprinted with permission from Reference~\cite{TioCastro:2007} ($\copyright$ 2007 IEEE).
}
\newcommand{\FigureDD}{
\SingleFigure{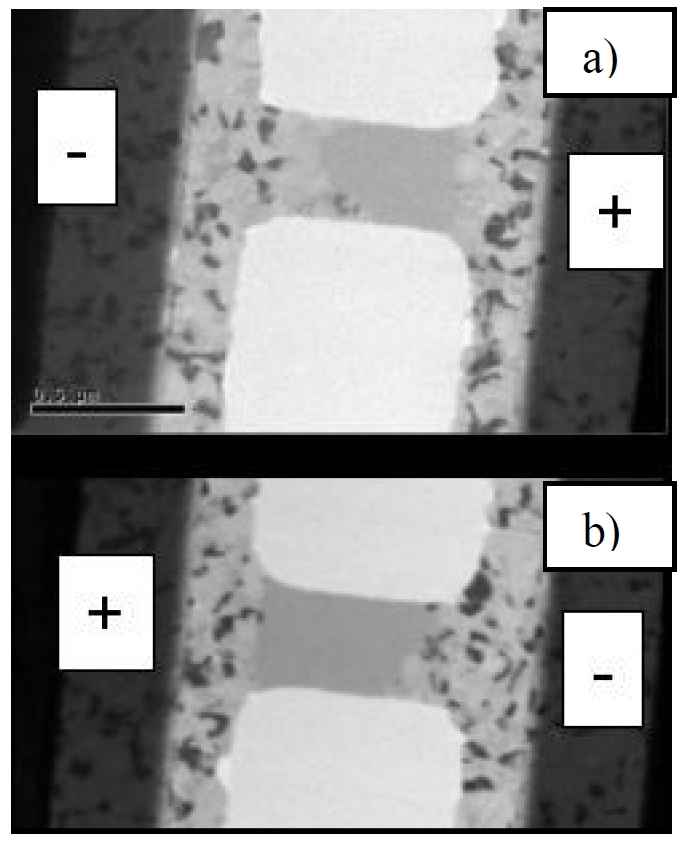}{fig:polarity:TioCastro}{\FigureCaptionDD}{\columnwidth}}

\newcommand{\FigureCaptionEE}{
WDS profiles of elemental concentration (Te, Sb, and Ge) along the length of a 20$\mu$-long Ge$_2$Sb$_2$Te$_5$ bridge at a) 0.17ms and b) 1.27ms after melting was initiated by a voltage-pulse, showing rapid desegregation of elements in the molten state. Reprinted with permission from Reference~\cite{Yang:2009d} ($\copyright$ 2009 American Institute of Physics)
}
\newcommand{\FigureEE}{
\TallSingleFigure{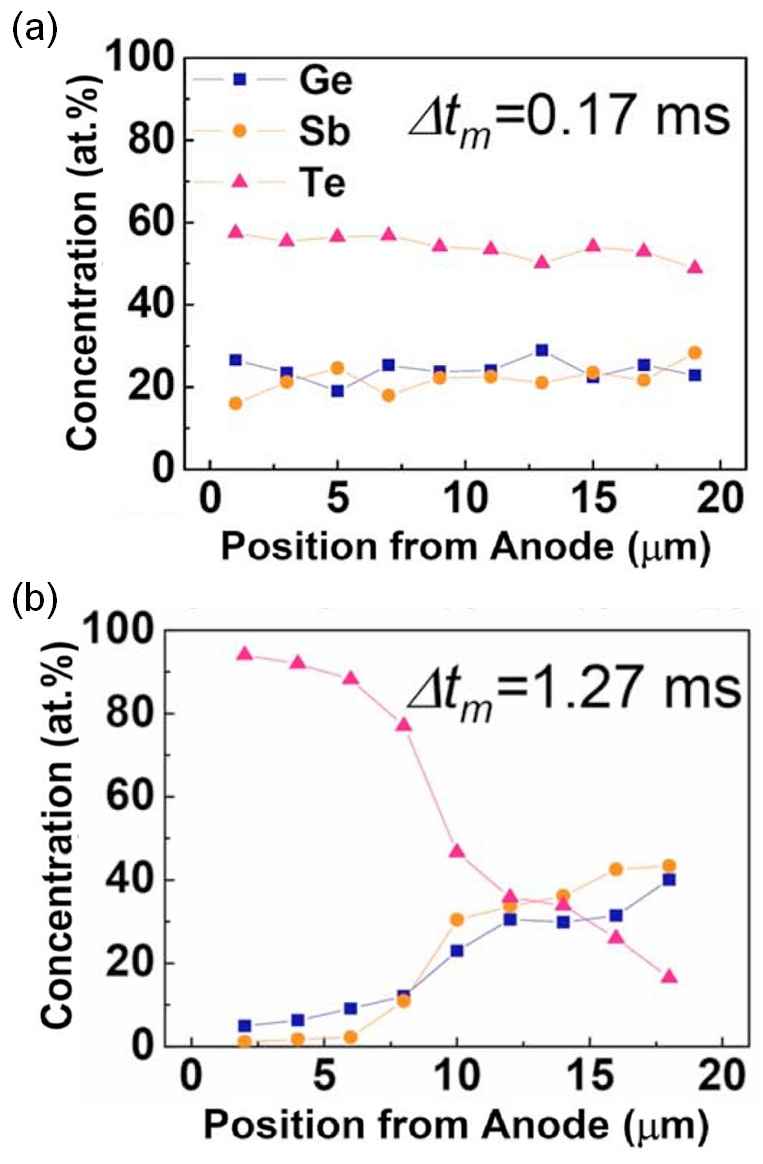}{fig:polarity:rapidDesegregation}{\FigureCaptionEE}{\columnwidth}}

\newcommand{\FigureCaptionFF}{
Cycling of a pore-PCM GST device, showing a stuck-SET failure after $\sim\! 5\! \times\! 10^5$ cycles, followed by ten pulses of reverse polarity (and of slightly-higher magnitude), which proved sufficient to alow cycling to continue for another $10^5$ SET-RESET cycles. Reprinted with permission  from Reference~\cite{Lee:2008k, Lee:2009a} ($\copyright$ 2009 IEEE).
}
\newcommand{\FigureFF}{
\SingleFigure{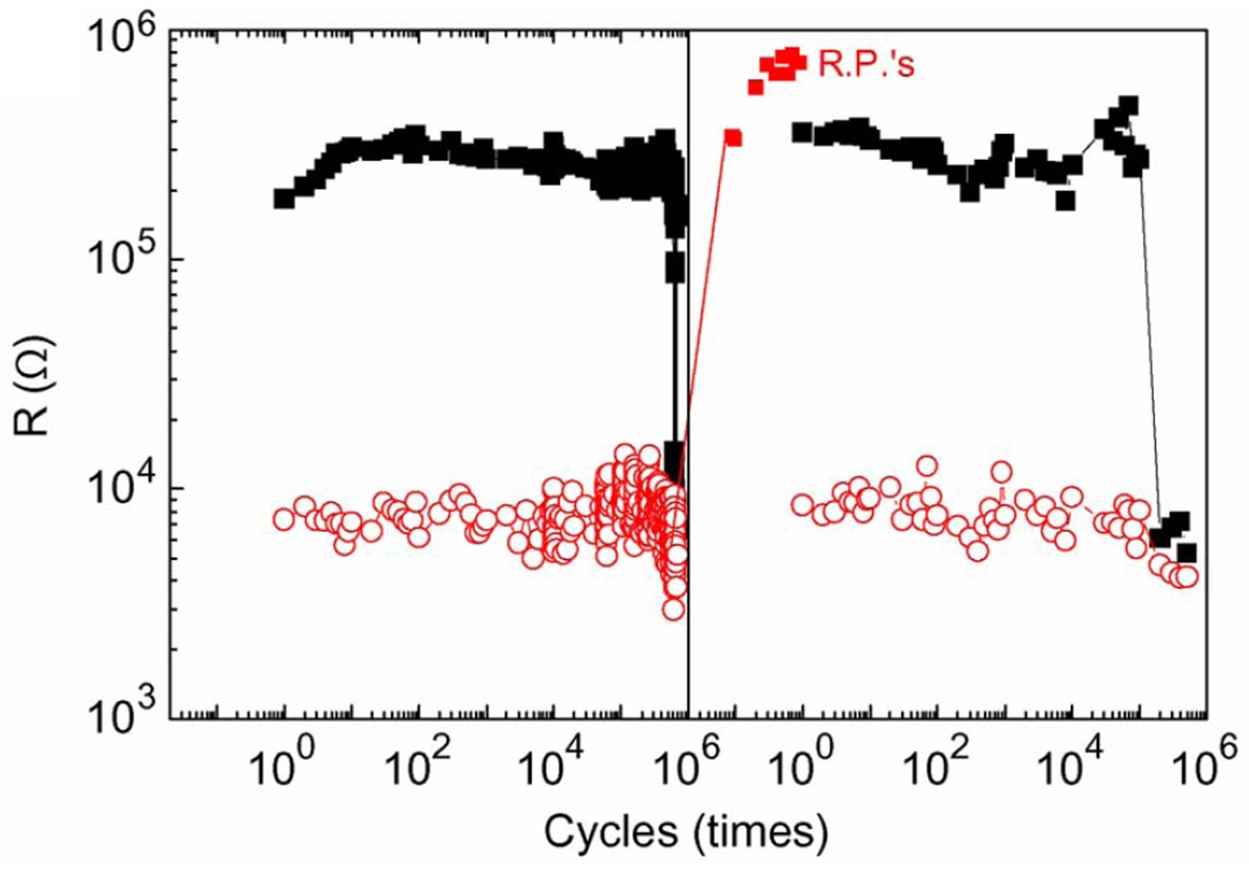}{fig:polarity:rescuePulses}{\FigureCaptionFF}{\columnwidth}}

\newcommand{\FigureCaptionGG}{
Two example families of electrical signals that can be used for MLC programming: (a) rectangular pulses and (b) variable slope pulses.
}
\newcommand{\FigureGG}{
\SingleFigure{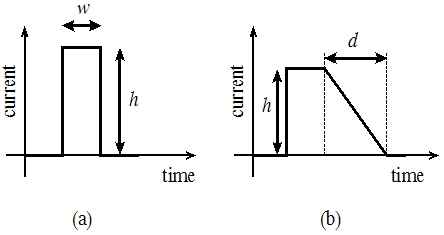}{fig:MLC:pulseFamilies}{\FigureCaptionGG}{\columnwidth}}

\newcommand{\FigureCaptionHH}{
Example distribution of the logarithm of the resistance for each of four possible stored levels, implementing 2-bit MLC.  The distributions shown here would suggest a non-negligible probability of classification error.
}
\newcommand{\FigureHH}{
\SingleFigure{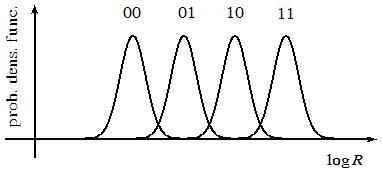}{fig:MLC:histograms}{\FigureCaptionHH}{\columnwidth}}

\newcommand{\FigureCaptionII}{
Distribution tightening by means of a write-and-verify procedure.
}
\newcommand{\FigureII}{
\SingleFigure{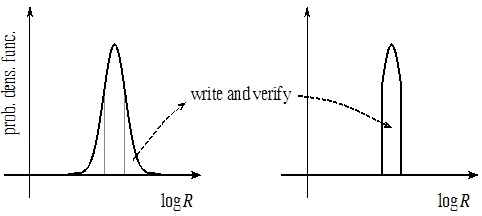}{fig:MLC:writeAndVerify}{\FigureCaptionII}{\columnwidth}}

\newcommand{\FigureCaptionJJ}{
Distributions of four resistance levels immediately after programming, after 400 hours at room temperature, and after an additional thermal annealing at 130\degreesC~for 12 hours. Reprinted with permission from Reference~\cite{Kang:2008} ($\copyright$ 2008 IEEE).
}
\newcommand{\FigureJJ}{
\SingleFigure{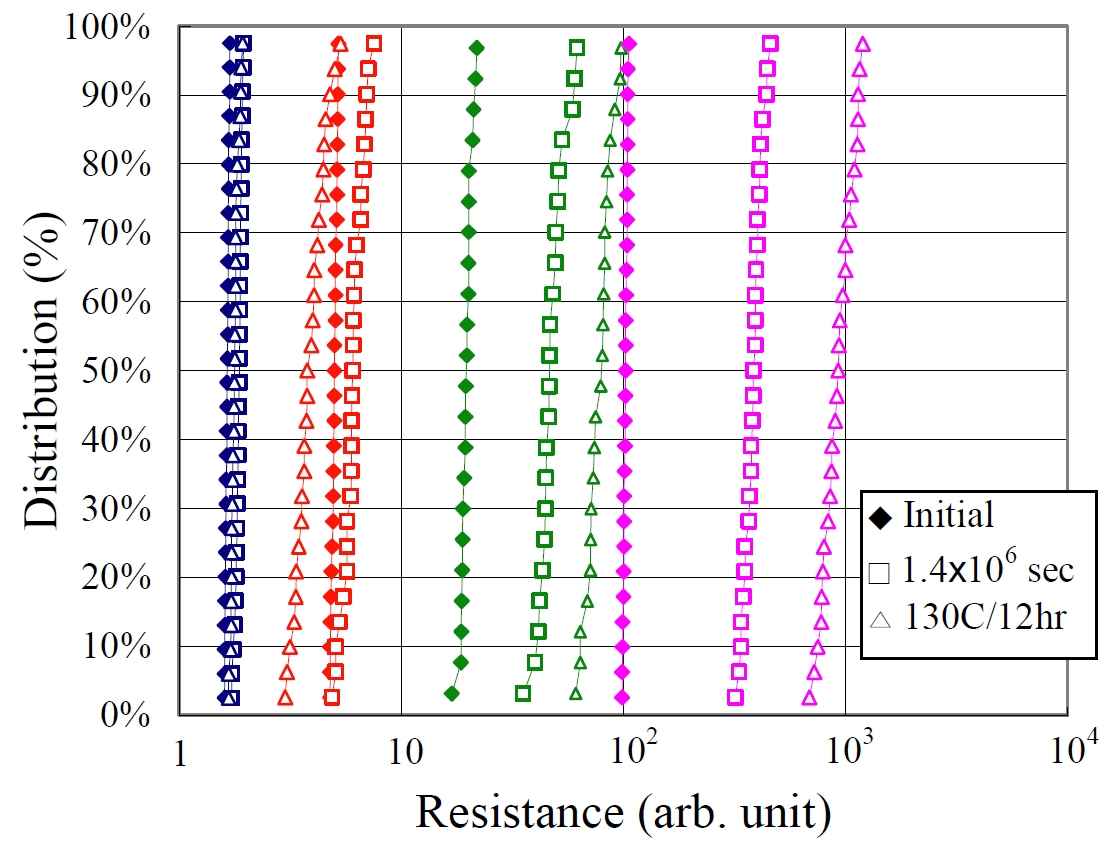}{fig:MLC:KangData}{\FigureCaptionJJ}{\columnwidth}}

\newcommand{\TableA}
{
\ifthenelse{\forSubmission = 1}{\ifthenelse{\atend=1}{
\begin{table*}
\onespacing
\begin{tabular}{rc|cl}
{\bf Phase change material parameter} &~&~& {\bf Influence on PCM device performance}\\[0.5mm]
\hline\\[-2mm]
Crystallization temperature \&
&&& Data retention and archival lifetime\\
thermal stability of the amorphous phase	
&&& SET power\\[1.25mm]
Melting temperature	&&& RESET power \\[1.25mm]
Resistivity in amorphous and crystalline phases	&&& On/off ratio\\
&&& SET and RESET current\\[1.25mm]
Threshold voltage	&&& SET voltage and reading voltage \\[1.25mm]
Thermal conductivity in both phases	&&& SET and RESET power\\[1.25mm]
Crystallization speed &&& SET pulse duration (and thus power) \\
&&& Data rate \\[1.25mm]
Melt-quenching speed &&& RESET pulse duration (and thus power) \\
\end{tabular}
\caption{\label{table:materScaling:MaterialParameters}
Some phase change material parameters and the device performance characteristics they influence.}
\vglue -0.125in
\end{table*}
\setspacing
}{}}{
\begin{table*}
\begin{tabular}{rc|cl}
{\bf Phase change material parameter} &~&~& {\bf Influence on PCM device performance}\\[0.5mm]
\hline\\[-2mm]
Crystallization temperature \&
&&& Data retention and archival lifetime\\
thermal stability of the amorphous phase	
&&& SET power\\[1.25mm]
Melting temperature	&&& RESET power \\[1.25mm]
Resistivity in amorphous and crystalline phases	&&& On/off ratio\\
&&& SET and RESET current\\[1.25mm]
Threshold voltage	&&& SET voltage and reading voltage \\[1.25mm]
Thermal conductivity in both phases	&&& SET and RESET power\\[1.25mm]
Crystallization speed &&& SET pulse duration (and thus power) \\
&&& Data rate \\[1.25mm]
Melt-quenching speed &&& RESET pulse duration (and thus power) \\
\end{tabular}
\caption{\label{table:materScaling:MaterialParameters}
Some phase change material parameters and the device performance characteristics they influence.}
\vglue -0.125in
\end{table*}
}}

\begin{document}

\onecolumngrid
\hglue 0.55in\parbox[t]{5.65in}{
\begin{center}
\small The following {\sc Review Article} appeared in the March/April 2010 issue of the\\ {\em Journal of Vacuum Science and Technology B}, volume 28, issue 2, pages 223--262.\\  The published article can be found at \url{link.aip.org/link/?JVB/28/223}
\end{center}
}
\twocolumngrid
\vglue 0.35in

\title{\Large Phase change memory technology}

\author{Geoffrey W. Burr$^{1,a}$,\footnotetext{e-mail: \emph{burr@almaden.ibm.com}}
Matthew J. Breitwisch$^2$, Michele Franceschini$^2$, Davide Garetto$^1$,
Kailash Gopalakrishnan$^1$, Bryan Jackson$^1$, B\"{u}lent Kurdi$^1$, Chung Lam$^2$,
Luis A. Lastras$^2$, Alvaro Padilla$^1$, Bipin Rajendran$^2$, Simone Raoux$^2$, and
Rohit S. Shenoy$^1$}

\affiliation{%
$^1$ IBM Almaden Research Center, 650 Harry Road, San Jose, California 95120\\
$^2$ IBM T.J. Watson Research Center, Yorktown Heights, NY 10598
}%

\date{\today}

\begin{abstract}
We survey the current state of phase change memory (PCM), a non-volatile solid-state memory technology built around the large electrical contrast between the highly-resistive amorphous and highly-conductive crystalline states in so-called phase change materials. PCM technology has made rapid progress in a short time, having passed older technologies in terms of both sophisticated demonstrations of scaling to small device dimensions, as well as integrated large-array demonstrators with impressive retention, endurance, performance and yield characteristics.

We introduce the physics behind PCM technology, assess how its characteristics match up with various potential applications across the memory-storage hierarchy, and discuss its strengths including scalability and rapid switching speed.  We then address challenges for the technology, including the design of PCM cells for low RESET current, the need to control device-to-device variability, and undesirable changes in the phase change material that can be induced by the fabrication procedure. We then turn to issues related to operation of PCM devices, including retention, device-to-device thermal crosstalk, endurance, and bias-polarity effects. Several factors that can be expected to enhance PCM in the future are addressed, including Multi-Level Cell technology for PCM (which offers higher density through the use of intermediate resistance states), the role of coding, and possible routes to an ultra-high density PCM technology.

\end{abstract}


\maketitle

\tableofcontents

\section{Motivation for PCM}\label{sec:MotivationForPCM}
\subsection{The case for a next-generation memory}

As with many modern technologies, the extent to which non-volatile memory (NVM) has pervaded our day-to-day lives is truly remarkable. From the music on our MP3 players, to the photos on digital cameras, the stored e-mail and text messages on smartphones, the documents we carry on our USB thumb-drives, and the program code that enables everything from our portable electronics to cars, the NVM known as Flash memory is everywhere around us. Both NOR and NAND Flash began humbly enough, as unappreciated side projects of a Toshiba DRAM engineer named Fujio Masuoka\cite{ForbesFlashOrigins}.  But from his basic patents in 1980 and 1987, respectively\cite{ForbesFlashOrigins}, Flash has grown in less than three decades to become a \$20 billion dollar-per-year titan of the semiconductor industry\cite{iSuppliCorp, Lai:2008}.

This market growth has been made possible by tremendous increases in the system functionality (e.g., more GBytes) that can be delivered in the same size package.  These improvements are both a byproduct {\em of} --- and the driving force {\em for} --- the relentless march to smaller device dimensions known as Moore's Law\cite{Moore:1965}.  The history of the solid-state memory industry,  and of the semiconductor industry as a whole, has been dominated by this concept: higher densities at similar cost lead to more functionality, and thus more applications, which then spur investment for the additional research and development needed to implement the ``next size smaller'' device.  Throughout this extensive history, extrapolation from the recent past has proven to be amazingly reliable for predicting near-future developments. Thus the memory products that will be built in the next several years have long been forecast\cite{ITRS:2008}.

Beyond the near-future, however, while the \emph{planned} device sizes may be sketched out, for the first time in many years it is not clear exactly how achievable these goals might be.  This uncertainty is present in many portions of the semiconductor industry, primarily due to the increasing importance of device-to-device variations, and to the common dependence on continued lithographic innovation. New patterning techniques will almost certainly be needed to replace the 193nm immersion and ``double patterning'' techniques now being used to implement the 32nm and even 22nm nodes\cite{Hazelton:2009, Lin:2009}. In addition to such issues common to the larger semiconductor industry, however, the Flash industry faces  additional uncertainties specific to its technology.

Over the past few years, Flash has been wrestling with unpleasant tradeoffs between the scaling of lateral device dimensions, the need to maintain coupling between the control and floating gates, the stress-induced leakage current (SILC) that is incurred by programming with large voltages across ultra-thin oxides, and the cell-to-cell parasitic interference between the stored charge in closely-packed cells\cite{Kim:2007i, Prall:2007, Lai:2008, Lai:2008a}.  Many alternative cell designs were proposed, typically involving replacement of the floating polysilicon gate by some type of charge-trapping layer, such as the Silicon Nitride at the center of the Silicon-Oxide-Nitride-Oxide-Semiconductor (SONOS) cell structure\cite{White:2000}. While early SONOS memory devices used extremely thin tunnel and blocking oxides for acceptable write/erase performance, and thus suffered from data retention issues\cite{Tsai:2001}, recent work seems to have migrated to Tantalum nitride--Alumina--Nitride--Oxide--Semiconductor (TANOS) structures\cite{Lee:2003i, Park:2006g, Sim:2007, Mauri:2008}.  These structures offer improved immunity to both SILC and parasitic interference between cells\cite{Mauri:2008}, while also allowing any defects to gracefully degrade Signal-to-Noise Ratio (SNR) rather than serve as avenues for catastrophic charge leakage\cite{Prall:2007, Mauri:2008}.  TANOS data retention has improved to acceptable levels\cite{Prall:2007}, and the reduced programming efficiency is now understood\cite{Mauri:2008}.

However, TANOS structures cannot help to scale NOR Flash, because the charge injected at one edge of such devices by channel hot-electron injection\cite{ProcIEEE_Flash:2003} must be redistributed throughout the floating gate after programming\cite{Lai:2008a}.  For NAND Flash, the finite and fairly modest number of discrete traps in each TANOS cell has accelerated the onset of new problems, ranging from device-to-device variations in $V_t$\cite{Prall:2007}, stochastic or ``shot-noise'' effects\cite{Prall:2007}, random telegraph noise\cite{Kurata:2006, Ghetti:2008}, and a significant reduction in the number of stored electrons that differentiate one stored analog level from the next\cite{Kim:2006w}. These issues are particularly problematic for Multi-Level Cell (MLC) Flash, where multiple analog levels allow an increase in the effective number of bits per physical device by a factor of 2, 3, or even 4.  Worse yet, such few-electron problems will only increase with further dimensional scaling, leading Flash researchers to explore even more complicated schemes for FinFET Flash devices\cite{Hsu:2007b,Lombardo:2007} or 3-D stacking of Flash memory\cite{Jung:2006d, Lai:2006b, Tanaka:2007, Fukuzumi:2007}.

With these difficulties in scaling to future technology nodes, Flash researchers are already hard-pressed to maintain specifications such as write endurance, retention of heavily cycled cells, and write/erase performance, let alone improve them.  As one indication of these pressures, some authors have pointed out that in cases such as digital photography, larger capacity formats can be expected to be tolerant of even more relaxed endurance specifications\cite{Kim:2006w}. However, at the same time that Flash is struggling to maintain current levels of reliability and performance while increasing density, new applications are opening up for which these specifications are just barely adequate.

The solid-state drive (SSD) market --- long dominated by high-cost, battery-backed DRAM for military and other critical applications --- has grown rapidly since the introduction of Flash-based SSD drives, passing \$400 million in revenues in 2007\cite{IDC_SSDmarket}.  One reason for the time delay between the widespread use of Flash in consumer applications and its appearance in SSD applications was the need to build system controllers that could hide the weaknesses of Flash. Consider that each underlying block of Flash devices takes over a millisecond to erase, and if written to continuously, would start to exhibit significant device failures in mere seconds. Sophisticated algorithms have been developed to avoid unnecessary writes, to perform static or dynamic wear-leveling, to pipeline writes, and to maintain pre-erased blocks in order to finesse or hide the poor write/erase performance\cite{Freitas:2008, FAST:2009}.  Together with simple overprovisioning of extra capacity, these techniques allow impressive system performance.  For instance, the Texas Memory Systems RamSan-500 can write at 2GB/sec with an effective Flash endurance of $>$15 years\cite{TexasMemory_WhitePaper}.  However, it is interesting to note that despite the fact that MLC Flash costs much less than 1-bit/cell Single-Layer Cell (SLC) Flash, for a long time only SLC Flash was used in SSD devices\cite{TexasMemory_WhitePaper}.  This is because MLC Flash tends to have 10$\times$ lower endurance and 2$\times$ lower write speed than SLC Flash\cite{TexasMemory_WhitePaper}, illustrating the importance of these specifications within SSD applications.

Thus there is a need for a new next-generation NVM that might have an easier scaling path than NAND Flash to reach the higher densities offered by future technology nodes.  Simultaneously, there is a need for a memory that could offer better write endurance and I/O performance than Flash, in order to bring down the cost while increasing the performance of NVM-based SSD drives. But the size of the opportunity here is even larger: the emergence of a non-volatile solid-state memory technology that could combine high performance, high density and low-cost could usher in seminal changes in the memory/storage hierarchy throughout all computing platforms, ranging all the way up to high performance computing (HPC). If the cost-per-bit could be driven low enough through ultra-high memory density, ultimately such a Storage-Class Memory (SCM) device could potentially displace magnetic hard disk drives (HDD) in enterprise storage server systems.

Fortunately, new NVM candidate technologies have been under consideration as possible Flash ``replacements'' for more than a decade\cite{Burr:2008a}.  These candidates range from technologies that have reached the marketplace after successful integration in real CMOS fabs (ferroelectric and magnetic RAM), to novel ideas that are barely past the proof-of-principle stage (racetrack memory and organic RAM), to technologies that are somewhere in-between (PCM, resistance RAM, and solid-electrolyte memory)\cite{Burr:2008a}.  Each of these has its strengths and weaknesses. In general, the farther along a technology has progressed towards real integration, the more that is known about it.  And as research gives way to development, it is typically new weaknesses --- previously hidden yet all too quickly considered to be obvious in hindsight --- that tend to be revealed.  In contrast, by avoiding these known pitfalls, fresh new technologies are immediately attractive, at least until their own unique weaknesses are discovered.

In this article, we survey the current state of phase change memory (PCM).  This technology has made rapid progress in a short time, having passed older technologies such as FeRAM and MRAM in terms of sophisticated demonstrations of scaling to small device dimensions.  In addition, integrated large-array demonstrators with impressive retention, endurance, performance and yield characteristics\cite{Burr:2008a} have been built.

The paper is organized into 7 sections, beginning with the current section titled ``Motivation for PCM.'' Section~\ref{sec:MotivationForPCM} also includes a brief overview of PCM technology, and an assessment of how its characteristics match up with various potential applications across the memory-storage hierarchy.  Section~\ref{sec:PhysicsOfPCM} goes into the physics behind PCM in more depth, in terms of the underlying phase change materials and their inherent scalability, and the physical processes affecting the switching speed of PCM devices.  The section concludes with a survey of PCM modeling efforts published to date, and a discussion of scalability as revealed by ultra-small prototype PCM devices.

In Section~\ref{sec:DesignFabricationOfPCM}, we address factors that affect the design and fabrication of PCM devices, including cell design, variability, changes in the phase change material induced by the fabrication procedure, and the design of surrounding access circuitry. We then turn to issues related to operation of PCM devices in Section~\ref{sec:PCMinOperation}, including endurance, retention, and device-to-device crosstalk. Section~\ref{sec:FutureOfPCM} addresses several factors that can be expected to enhance PCM in the future, including Multi-Level Cell technology for PCM, the role of coding, and possible routes to an ultra-high density PCM technology. The Conclusion section (Section~\ref{sec:Conclusions}) is followed by a brief Acknowledgements section (Section~\ref{sec:Acknowledgements}).

\subsection{What is PCM?}\label{subsec:WhatIsPCM}

Phase change memory (PCM) exploits the large resistance contrast between the amorphous and crystalline states in so-called phase change materials\cite{Raoux:2008a}. The amorphous phase tends to have high electrical resistivity, while the crystalline phase exhibits a low resistivity, sometimes 3 or 4 orders of magnitude lower. Due to this large resistance contrast, the change in read current is quite large, opening up the opportunity for the multiple analog levels needed for MLC operations\cite{Raoux:2008a}.

To SET the cell into its low-resistance state, an electrical pulse is applied to heat a significant portion of the cell above the crystallization temperature of the phase change material.  This SET operation tends to dictate the write speed performance of PCM technology, since the required duration of this pulse depends on the crystallization speed of the phase change material (Section~\ref{subsec:SpeedOfPCM}).  SET pulses shorter than 10ns have been demonstrated\cite{Wang:2008b, Krebs:2009a, Krebs:2009,Bruns:2009}. Because the crystallization process is many orders of magnitude slower at low temperatures ($<$ 120\degreesC), PCM is a NVM technology that can offer years of data lifetime.

In the RESET operation, a larger electrical current is applied in order to melt the central portion of the cell.  If this pulse is cut off abruptly enough, the molten material quenches into the amorphous phase, producing a cell in the high-resistance state.  The RESET operation tends to be fairly current-- and power--hungry, and thus care must be taken to choose an access device capable of delivering high current and power without requiring a significantly larger footprint than the PCM element itself. The read operation is performed by measuring the device resistance at low voltage, so that the device state is not perturbed. These operations are summarized in Figure~\ref{fig:WhatIsPCM:SET,RESET,READ}.

\FigureA

Even though the principle of applying phase change materials to electronic memory was demonstrated as long ago as the 1960s\cite{Ovshinsky:1968}, interest in PCM was slow to develop compared to other NVM candidates. However, renewed interest in PCM technology was triggered by the discovery of fast ($<$100 nanosecond) crystallizing materials such as Ge$_2$Sb$_2$Te$_5$ (GST) or Ag-- and In--doped Sb$_2$Te (AIST)\cite{Yamada:1991,Tominaga:1997} by optical storage researchers. Over the past few years, a large number of sophisticated integration efforts have been undertaken in PCM technology, leading to demonstration of high endurance\cite{Lai:2003}, fast speed\cite{Pirovano:2003}, inherent scaling of the phase change process out beyond the 22nm node\cite{Chen:2006t}, and integration at technology nodes down to 90nm\cite{Oh:2006b}. One important remaining unknown for the success of PCM technology is whether the memory access device (diode\cite{Oh:2006b}, transistor\cite{Pellizzer:2006}, etc.) in a dense memory array will be able to supply sufficient current to RESET the PCM cell. Already, in order to try to minimize the RESET current, it is assumed that the dimension of the phase change material will be only 30\% of the lithographic feature size $F$\cite{ITRS:2008}, mandating the use of sub-lithographic techniques for accurate definition of this critical dimension (CD). However, even with this difficult integration task, the success of PCM technology may end up depending on advances in the access device as much as on the PCM cell itself\cite{ITRS:2008}.

Important device characteristics for a PCM cell include widely separated SET and RESET resistance distributions (necessary for sufficient noise margin upon fast readout), the ability to switch between these two states with accessible electrical pulses, the ability to read/sense the resistance states without perturbing them, high endurance (allowing many switching cycles between SET and RESET), long data retention (usually specified as 10 year data lifetime at some elevated temperature), and fast SET speed (the time required to recrystallize the cell from the RESET state).  Data retention usually comes down to the cell's ability to retain the amorphous RESET state by avoiding unintended recrystallization.  An additional aspect that can be of significant importance is the ability to store (and retain over time) more than one bit of data per cell, since this allows one to increase effective density much like MLC Flash without decreasing the feature size.

A critical property of phase change materials is the so-called threshold switching\cite{Pirovano:2004,Adler:1978, Adler:1980, Redaelli:2004}.  Without this effect PCM would simply not be a feasible technology, because in the high resistance state extremely high voltages would be required to deliver enough power to the cell to heat it above the crystallization temperature.  However, when a voltage above a particular threshold $V_t$ is applied to a phase change material in the amorphous phase, the resulting large electrical fields greatly increase the electrical conductivity.  This effect is still not completely understood, but is attributed to a complex interplay between trapped charge, device current, and local electrical fields\cite{Pirovano:2004,Ielmini:2007b}.  With the previously resistive material now suddenly highly conducting, a large current flows --- which can then heat the material.  However, if this current pulse is switched off immediately after the threshold switching, the material returns to the highly resistive amorphous phase after about 30ns\cite{Ielmini:2007}, with both the original threshold voltage $V_t$ and RESET resistance recovering slowly over time\cite{Ielmini:2007, Pirovano:2004a}.  Only when a current sufficient to heat the material above the crystallization temperature, but below the melting point, is sustained for a long enough time does the cell switch to the crystalline state.  The threshold switching effect serves to make this possible with applied voltages of a few volts, despite the high initial resistance of the device in the RESET state.

\subsection{Potential applications of PCM}\label{subsec:potentialApplications}

The ultimate goal of researchers and developers studying emerging memory technologies is to devise a universal memory that could work across multiple layers of the existing memory hierarchy for modern computers.  This memory hierarchy, shown in Figure~\ref{fig:deviceApplications:MemoryHierarchy}, is designed to bridge the performance gap between the fast central processing units and the slower (sometimes {\em much} slower) memory and storage technologies, while keeping overall system costs down.  Figure~\ref{fig:deviceApplications:CostPerformance} shows how PCM is expected to compare to the four major incumbent memory and storage technologies in terms of cost and performance.  The enormous range of cost and performance spanned by these technologies makes a single universal memory --- one capable of replacing all of these well-established memory and storage techniques --- an aggressive goal indeed.

However, Figure~\ref{fig:deviceApplications:HumanPerspectiveAccessTimes} shows that there is currently a gap of more than 3 orders of magnitude between the access time of off-chip DRAM (60ns) and the write-cycle time of Flash (1ms).
To put this into human perspective, this slow write-cycle time is equivalent to a person, who might be making data-based decisions analogous to a single CPU operation every second, having to wait approximately 10 days to record a small block of information.  An interesting region on this chart sits just above off-chip DRAM, where access times of 100--1000ns could potentially be enabled by a ``storage-class memory'' (SCM) made possible by PCM.

\FigureB
\FigureC
\FigureD

In the remainder of this Section, we examine the suitability of PCM for the layers of the memory hierarchy currently served by SRAM (Static), DRAM (Dynamic), NOR and NAND Flash.  We also discuss the emerging area of Storage-Class Memory, for which Flash-based Solid-State Drives are just now becoming available.  While the two principal integration metrics are cost and performance, we also briefly examine critical reliability issues such as data retention and read/write endurance here (leaving more in-depth discussion to Section~\ref{sec:PCMinOperation}).  We do not consider the relative merit of power consumption, assuming instead that all these technologies are roughly comparable within an order of magnitude.  The non-volatility of PCM does compare favorably to volatile memories, both in terms of standby power as well as by enabling easier recovery from system or power failures in critical applications.

\subsubsection{PCM as SRAM}

Much of the SRAM used in computers today is embedded close to the central processor unit (CPU), serving as high performance Level~1 (L1) and Level~2 (L2) cache memories.  Some off-chip Level 3 (L3) cache memories also use SRAM. In consumer electronics, SRAM has been used in combination with NOR Flash in cell phones.  A typical SRAM cell comprises six CMOS transistors, two pMOSFETs and four nMOSFETs, and thus occupies more than 120\Fsquared in chip real estate per bit. (Here $F$ is the size of the smallest lithographic feature, so that this measure of device size is independent of the particular device technology used to fabricate the memory.) Embedded SRAM typically runs at the CPU clock speed, so that access times for these devices must be less than ten nanoseconds.  Commodity SRAM used in cell phones runs at slower clock speeds, allowing access times in the tens of nanoseconds.

While there is no problem for PCM to improve upon the large SRAM cell size, even if a large access device is used for the PCM cell, SRAM performance is hard to match.  The performance limiter for PCM is the SET speed, which in turn depends on the crystallization speed of the phase change material.  As will be described in detail in Section~\ref{subsec:SpeedOfPCM}, while some researchers have demonstrated the use of SET pulses shorter than 10ns\cite{Wang:2008b, Krebs:2009, Krebs:2009a, Bruns:2009}, most of the realistically large array demonstrations tend to use SET pulses that range from roughly 50 to 500ns in length\cite{Kang:2006}.

In any case, the most stringent requirement for any emerging memory technology that seeks to replace SRAM is endurance.  For all practical purposes, the read/write endurance of SRAM is infinite.  While read endurance is not a likely problem for PCM, the required write endurance for SRAM replacement is probably 10$^{18}$ --- out of reach for nearly all NVM technologies.  Storing data semi-permanently with PCM and most other NVM technologies involves some form of ``brute force'' that alters an easily observable material characteristic of the memory device.  For PCM, this ``brute force'' is the melt-quench RESET operation, and at such elevated temperatures, it has been shown that the constituent atoms of a phase change material will tend to migrate over time\cite{Lai:2003, Nam:2009}, as discussed in Sections~\ref{subsec:Polarity} and \ref{subsec:Endurance}.

Since non-volatility is not a requirement for SRAM applications, one might be able to trade some data retention for improved endurance.  Some remote evidence of this trade-off has been demonstrated by showing a strong correlation between the total energy in the RESET pulse and the resulting PCM endurance\cite{Lai:2003}. The best-case endurance, achieved for the lowest-energy RESET pulses, was 10$^{12}$ SET--RESET cycles\cite{Lai:2003}.  Yet this is still 6 orders of magnitude away from the target specification for SRAM.

\subsubsection{PCM as DRAM}

DRAM is used in a more diversified set of applications than SRAM.  Most of the characteristics discussed above for the replacement of SRAM also apply to the replacement of DRAM, although in most cases the specifications are slightly relaxed.  Access times of tens of nanoseconds would be acceptable for most computer and consumer electronics applications of DRAM.  For embedded DRAM used as video RAM and L3 cache\cite{Iyer:2005}, however, an access time of 10 nanoseconds or less is required.  As for write endurance, the requirement can be estimated using the following equation\cite{WinfriedInfo}:
\begin{equation}\label{eq:wearLeveling}
E \;=\; T_\mathrm{life} \, \frac{B}{\alpha C},
\end{equation}
where $E$ is endurance, $T_\mathrm{life}$ is the life expectancy of the system, $B$ is memory bandwidth, $\alpha$ is wear-leveling efficiency, and $C$ is the system memory capacity.  Assuming a typical server with a ten year life expectancy, 1$\,$GB/sec bandwidth, 10\% wear-leveling efficiency and 16$\,$GB capacity, the endurance requirement is approximately 2$\times$10$^8$ --- well within the reach of PCM\cite{Pellizzer:2006,Oh:2006a}.

There is also a power argument to be made when discussing PCM as a potential DRAM replacement.  This might seem to be a difficult case to make for a technology for which every write cycle involves heating to temperatures ranging from 400 to 700 degrees Celsius.  However, DRAM turns out to be a fairly power-hungry technology.  This is not due to its periodic refresh, however, which takes place only infrequently, and is not too strongly related to the underlying physical storage mechanism of charging up a local capacitor.  Instead, power inefficiency in DRAM is due to the simultaneous addressing of multiple banks within the chip.  For every bit that passes into or out of a DRAM chip, 8 or even 16 devices are being internally accessed (read and then re-written), somewhat as if your librarian knocked an entire row of books onto the floor each time you asked for a book.  Low-power DRAM intended for mobile, battery-powered applications tends to have lower performance, although some developments have been made that can combine high performance with low power\cite{IEEESpectrum_RambusLowPowerDRAM}. However, the inherent need to re-write after each read access is unavoidable for a volatile memory like DRAM. Thus simply by being non-volatile, PCM could potentially offer a lower-power alternative to DRAM, despite the inherently power-hungry nature of PCM write operations.

For stand-alone memories, cost is directly proportional to memory cell size.  State of the art DRAM cells occupy 6\Fsquared in chip area.  Thus for PCM to compete in the DRAM arena, PCM cell size would need to be this size or smaller with comparable ``area efficiency (the fraction of the chip area dedicated to memory devices rather than to peripheral circuitry). Fortunately, such small cell sizes have already been demonstrated using a diode select device\cite{Oh:2006a}.  PCM also competes favorably with DRAM in terms of forward scaling into future generations, as DRAM developers are quickly hitting various scaling limits associated with storage interference, device leakage, and challenges in integrating high aspect-ratio capacitors in tight spaces.  Currently, DRAM has fallen behind NAND Flash and standard CMOS logic technologies in terms of scaling to the 45nm technology node and preparation for the 32nm node.  However, DRAM is a proven, reliable technology that has been employed in modern computers since the early 1970s.  It would be a long journey to displace such a stable technology.

\subsubsection{PCM as Flash}

There are two kinds of Flash memories, NOR and NAND.  In (common-source) NOR memory architectures, each cell in a two-dimensional (2D) array is directly connected to its word-line and bit-line input lines (with the source electrode of each cell sharing a common ground), whereas in NAND memory architectures, small blocks of cells are connected in series between a high input signal and ground.  Thus, while NAND flash can inherently be packed more densely (due to its smaller unit cell size) than NOR flash, NOR flash offers significantly faster random access (since each cell in the array is directly connected to the input lines).  However, since NOR memory requires large programming currents (to place charge on the floating gate via channel hot electrons), its programming throughput (measured in MB/s) is much slower than that of the block-based NAND memory architectures (which, by utilizing Fowler-Nordheim tunneling, can utilize lower programming currents that permit many bits to be processed in parallel)\cite{Cappelletti:1999}.   As a consequence, NOR memory offers significantly faster random access with low programming throughput and thus is mainly used for applications such as embedded logic that require fast access to data that is modified only occasionally.  In contrast, NAND memory is a high-density, block-based architecture with slower random access which is mainly used for mass storage applications.

NOR Flash memory cells occupy about 10F$^2$, with an access time upon read of a few tens of nanoseconds or more.  However, the access time upon write for NOR Flash is typically around 10 microseconds, and the write/erase endurance (for both NOR as well as NAND) is only 100k cycles.  These characteristics are well within the capabilities of current PCMs.    NOR Flash with its floating gate technology has difficulties scaling below 45nm, mainly due to difficulties in scaling the thickness of the tunnel oxide.  It is thus no surprise that NOR Flash is the popular target for first replacement by most PCM developers.

NAND Flash, on the other hand, is a much harder target despite PCM's superiority in both endurance and read performance.  Cost is the biggest challenge.  A NAND Flash memory cell occupies only 4\Fsquared of chip area, and as discussed earlier, NAND will be able to maintain this through at least 22nm using trap storage technology\cite{Lee:2003i} and possibly 3-dimensional integration\cite{Fukuzumi:2007}.  Furthermore, MLC NAND has been shipping 2 bits per physical memory cell for years, and is promising to increase this to 4 bits per cell\cite{Shibata:2009}.

\FigureE

NAND Flash is mainly used in consumer electronic devices, where cost is the paramount concern, and in the emerging Solid State Drive (SSD) market to replace magnetic Hard Disk Drives (HDD), where both cost and reliability are important.  The prerequisites for PCM to replace NAND Flash are 4\Fsquared memory cell size, at least 2-bit MLC capability, and 3-dimensional integration to further increase the effective number of bits per unit area of underlying silicon.  A 4\Fsquared cell dictates a memory element that can be vertically stacked over the select device, as shown in Figure~\ref{fig:deviceApplications:FourFsquared}.  Multi-level storage seems to be within reach of PCM given its inherently wide resistance range, and both 2- and 4-bits per cell has already been demonstrated in small-scale demonstrations\cite{Nirschl:2007, Bedeschi:2008}.   Even though write operations are slow for NAND Flash, it tends to achieve an impressive write data-rate because its low write power allows for programming of many bits in parallel.  Thus to deliver equal or better write bandwidth, PCM developers will need to work on reducing the write power so that the data bus can be as wide as possible.

\subsubsection{PCM as Storage-Class Memory}

In addition to the established segments of the memory hierarchy we have described (SRAM, DRAM, and Flash), the gap in access times between 1ms and 100ns shown in Figure~\ref{fig:deviceApplications:HumanPerspectiveAccessTimes} opens up the possibility of Storage-Class Memory (SCM)\cite{Freitas:2008, FAST:2009}.  SCM would blur the traditional boundaries between storage and memory by combining the benefits of a solid-state memory, such as high performance and robustness, with the archival capabilities and low cost of conventional hard-disk magnetic storage. Such a technology would require a solid-state nonvolatile memory that could be manufactured at an extremely high effective areal density, using some combination of sublithographic patterning techniques, multiple bits per cell, and multiple layers of devices (Section~\ref{subsec:ultraHighDensity}).  The target density probably needs to exceed current MLC NAND Flash densities by a factor of 2-8$\times$, in order to bring the cost of SCM down close to the cost of reliable enterprise HDD.

The opportunity for SCM itself actually breaks into two segments: a slower variant, referred to as S--class SCM\cite{FAST:2009}, which would act much like a Flash--based SSD except with better native endurance and write performance.   Here access times of 1-3$\mu$sec would be acceptable, but low cost via high density would be of paramount importance.  The other variant, referred to as M-class SCM\cite{FAST:2009}, requires access times of 300ns or less, with both cost and power as considerations.  This threshold of 300ns is considered to be the point at which an M-class SCM would be fast enough to be synchronous with memory operations, so that it could be connected to the usual memory controller\cite{FAST:2009}. In contrast, S-class SCM, SSD, and HDD would all be accessed through an I/O controller for asynchronous access.  M-class SCM would likely not be as fast as main memory DRAM.  But by being non-volatile, lower in power-per-unit-capacity (via high density), and lower in cost-per-capacity, the presence of M-class SCM could potentially allow the total amount of DRAM required to maintain ultra-high bandwidth to be greatly reduced, thus reducing overall system cost and power.

\section{Physics of PCM}\label{sec:PhysicsOfPCM}
\subsection{Phase change materials and scalability}\label{subsec:materialsScalability}

As discussed in Section~\ref{sec:MotivationForPCM}, the NVM industry faces the prospect of a costly and risky switch from a known and established technology (Flash) into something much less well known (either PCM or something else).  And understandably, the industry wants to make such leaps rare.

The problem here is not that one might fail to create a successful first product.  That would be unpleasant but not devastating, because this would happen during the early development stage, where the level of investment is small and multiple alternative approaches are still being pursued.  Instead, the nightmare scaling scenario is one in which the new technology works perfectly well for the first generation, yet is doomed to failure immediately afterwards.  If only one or two device generations succeed, then the NVM industry, having just invested heavily into this new technology, will be forced to make yet another switch and start the learning process all over again.

Thus scaling studies are designed to look far down the device roadmap, to try to uncover the showstoppers that might bedevil a potential NVM technology at sizes much smaller than what can be built today.  In the case of PCM technology, two aspects of scalability need to be considered: the scaling properties of the phase change materials, and the scaling properties of PCM devices. In this Section, we survey recent literature covering both of these considerations.  In general, experiments have shown that PCM is a very promising technology with respect to scalability.

It is well known that the properties of nanoscale materials can deviate from those of the bulk material, and can furthermore be a strong function of size. For example, it is typical for nanoparticles to have a lower melting temperature than bulk material of the same chemical composition, because the ratio of surface-atoms to volume-atoms is greatly increased. A recurring theme in such studies is the larger role that surfaces and interfaces play as dimensions are reduced.

Phase change material parameters that are significant for PCM applications---and the device performance properties that are influenced by these parameters---are summarized in Table~\ref{table:materScaling:MaterialParameters}. For optical applications the change of optical constants as a function of film thickness is also important, but for this paper we restrict our considerations to material parameters relevant to electronic memory applications.  As can be seen from Table~\ref{table:materScaling:MaterialParameters}, there is a large set of materials parameters which influence the PCM device, either affecting one of the two writing operations (SET to low resistance; RESET to high resistance) or the read operation.

\TableA

A particularly important phase change material parameter is the crystallization temperature, $T_x$.  This is not necessarily the temperature at which crystallization is most likely, but instead is the lowest temperature at which the crystallization process becomes ``fast.''  It is typically measured by raising the temperature slowly while monitoring the crystallinity (either looking for X-ray diffraction from the crystalline lattice or the associated large drop in resistivity).  Thus the crystallization temperature is a good measure of how hot a PCM cell in the RESET state could be made before the data stored by an amorphous plug would be lost rapidly due to unwanted crystallization.  While the crystallization temperature by itself does not reveal how ``slowly'' such data would be lost for slightly lower or much lower temperatures, it sets a definitive and easily measured upper bound on the retention vs. temperature curve for a new phase change material.

The crystallization temperature of phase change materials tends to vary considerably as a function of material composition\cite{Raoux:2007a,Raoux:2009,Okabe:1990}. For example, some materials, such as pure Sb, crystallize {\em below} room temperature.  Yet adding only a few at. \% of Ge to Sb, creating the phase change material Ge$_x$-Sb$_{1-x}$, increases the crystallization temperature significantly above room temperature.  In fact, $T_x$ can reach almost 500\degreesC~ for {GeSb} alloys that are high in Ge content\cite{Raoux:2009,Okabe:1990}. Studies of the crystallization temperature as a function of film thickness show an exponential increase as film thickness is reduced (for phase change materials sandwiched between insulating materials such as SiO$_2$ or ZnS-SiO$_2$)\cite{Wei:2007,Raoux:2008}. However, for phase change materials sandwiched between metals, metal-induced crystallization can occur and the crystallization temperature can be reduced for thinner films\cite{Raoux:2009a}. It is known that for phase change materials the crystallization is typically heterogeneous, starting at defects which can be located in the bulk, but which tend to be more prevalent at surfaces and interfaces.  As film thickness is reduced, the volume-fraction of phase change material that is at or near an interface increases, leading to changes in the externally observable crystallization temperature.

Phase change nanowires are typically fabricated by the vapor-liquid-solid technique, and are crystalline as synthesized\cite{Lee:2008h}. To measure crystallization behavior as a function of wire size, PCM devices were fabricated from single-crystalline, as-grown Ge$_2$Sb$_2$Te$_5$ nanowires using Pt contact pads\cite{Lee:2008h}.  The central section of the nanowire devices was re-amorphized by electrical current pulses and the activation energy was determined by measuring the recrystallization temperature as a function of heating rate. Here, the activation energy was found to fall from 2.34 eV for 190 nm diameter devices to 1.9 eV for 20 nm diameter devices, indicating a deterioration of data retention as the Ge$_2$Sb$_2$Te$_5$ nanowire diameter is reduced. However, Yu and co-workers did not observe a dependence of the crystallization temperature on the device diameter for PCM devices fabricated by contacting GeTe and Sb$_2$Te$_3$ nanowires using Cr/Au contacts\cite{Yu:2008}.

Figure~\ref{fig:materScaling:Nanoparticles} shows phase change nanoparticles fabricated by a variety of techniques including electron beam lithography, solution-based chemistry, self-assembly-based lithography combined with sputter deposition, and self-assembly-based lithography combined with spin-on deposition of the phase change material. When the crystallization temperature of amorphous-as-fabricated  nanoparticles was studied, it was found that larger phase change nanoparticles have a very similar crystallization temperature compared to bulk material\cite{Raoux:2007a,Zhang:2007h}, whereas the smallest nanoparticles in the 10 nm range can show either decreased\cite{Milliron:2007} or increased\cite{Zhang:2008e} crystallization temperature.

\FigureF

In terms of size effects, ultra-thin films still can show crystallization down to thicknesses of only 1.3 nm\cite{Raoux:2008}, and nanoparticles as small as 2-5 nm synthesized by solution-based chemistry have been found to be crystalline\cite{Caldwell:2007}. This is very promising for the scalability of PCM technology to future device generations.

Beyond crystallization temperature, the melting temperature is a parameter which can vary with composition and, at small dimensions, with size.  In fact, reduction in the melting temperature of phase change materials has been observed for very thin films\cite{Raoux:2008d}, nanowires\cite{Sun:2006e} and nanoparticles\cite{Sun:2007b}. This is advantageous for device performance, because a lower melting point implies a reduction in the power (and current) required to RESET such a PCM cell. The electrical resistivity for thin films increases slightly for both phases when film thickness is reduced\cite{Wei:2007}. This is also beneficial for scaling because higher resistivities lead to higher voltage drop across the material and can thus reduce switching currents.

The threshold voltage is a phenomenological parameter of PCM devices that describes the applied voltage (typically around 1V) required to induce an electrical breakdown effect.  Such a sudden increase in electrical conductivity allows the PCM device to rapidly and efficiently attain a significantly lower dynamic resistance (typically 3--10$\times$ lower than the room-temperature SET resistance), allowing efficient heating with moderate applied voltages.  Thus the presence of this electrical switching effect is an important component of PCM technology.

However, a more accurate description of the underlying physical process calls for a threshold electric field, rather than a threshold voltage, that must be surpassed for the amorphous material to become highly conductive.  Studies of phase change bridge devices (described in Section~\ref{subsec:prototypeDevices}) have shown that the threshold voltage scales linearly as a function of the length of the bridge along the applied voltage direction, confirming the role of an underlying material-dependent threshold field\cite{Lankhorst:2005, Krebs:2008}.

No deviation from this linear behavior was observed for bridge devices as short as 20nm. The value of the threshold field varied considerably, from 8 V/$\mu$m for Ge(15 at.\%)-Sb devices to 94 V/$\mu$m for thin Sb devices. For nanowire devices with even smaller amorphous areas, however, Yu and co-workers observed a deviation from this linear behavior\cite{Yu:2008}. Once the amorphous volume spanned less than approximately 10nm along the nanowire, the threshold voltage saturated, at 0.8 V and 0.6 V for GeTe and Sb$_2$Te$_3$ devices, respectively. This scaling behavior was explained with the impact ionization model previously developed to explain the threshold switching phenomenon\cite{Ielmini:2008b}.

Such a saturation in the effective threshold voltage is actually desirable, because for practical device performance a threshold voltage around 1 V is optimum.  This places the switching point well above the typical reading voltage of about 50-100 mV, yet not so far that a transistor or diode access device would fail to easily deliver the required switching pulses with moderate supply voltages. (Note that exceeding the threshold voltage to produce breakdown is not the same as delivering sufficient power to heat the cell to achieve the RESET condition). If the threshold voltage were to continue as a linear function of device size for sub-10-nm devices, then reading the cells without accidentally switching them out of the RESET state could become problematic.

The thermal conductivity of phase change materials is important because it strongly influences the thermal response of a PCM device to an electrical current pulse. However, so far the materials that have been studied (Ge$_2$Sb$_2$Te$_5$, nitrogen-doped Ge$_2$Sb$_2$Te$_5$, Sb$_2$Te and Ag- and In-doped Sb$_2$Te) show only a slight variation in the values for the thermal conductivities between 0.14 and 0.17 W/m$\cdot$K for the as-deposited amorphous phase, and values between 0.25 and 2.47 W/m$\cdot$K for the crystalline phase\cite{Risk:2009}.   Reifenberg and co-workers\cite{Reifenberg:2007} studied the thermal conductivity of Ge$_2$Sb$_2$Te$_5$ with thicknesses between 60 and 350 nm using nanosecond laser heating and thermal reflectance measurements.  They found about a factor of two decrease in the thermal conductivity as film thickness is reduced --- from 0.29, 0.42, 1.76 W/m$\cdot$K in the amorphous, fcc, and hexagonal phases, respectively for 350 nm thick films, to 0.17, 0.28, 0.83 W/m$\cdot$K for 60 nm thick films.  As with earlier results, such a trend leads to advantageous scaling behavior for PCM applications, by helping reduce the energy required for the power-intensive RESET operation.

In addition to these changes in effective material properties as device sizes scale down, there are also simple yet powerful geometric effects which are associated with scaling.  As we will discuss extensively in Section~\ref{subsec:CellStructures}, scaling decreases the size of the limiting cross-sectional aperture within each PCM cell, thus driving down the RESET current.  However, at constant material resistivity, geometric considerations cause both the SET and dynamic resistances to increase.  As a result, the effective applied voltage across the device during the RESET operation remains unchanged by scaling, at least to first order.  These effects can be expected to eventually have adverse effects, as the decreasing read current (from the higher SET resistance) makes it difficult to accurately read the cell state rapidly, and as the non-scaling voltages exceed the breakdown limits of nearby scaled-down access transistors.

To summarize scaling properties of phase change materials, it has been observed that the crystallization temperature is in most cases increased as dimensions are reduced (beneficial to retention), and melting temperatures are reduced as dimensions are reduced (beneficial to RESET power scaling).  Similarly, resistivities in both phases tend to increase (beneficial for RESET power), threshold voltages are first reduced as dimensions are reduced but then level out around 0.6-0.8 V for dimensions smaller than 10 nm (beneficial for voltage scaling), and thermal conductivity seems to decrease as film thickness is reduced (beneficial for RESET current scaling).  As will be seen in more detail in the next section,  the raw crystallization speed can either decrease (detrimental to write performance) or increase (beneficial), and seems to depend strongly on the materials and their environment. Crystallization has been observed to reliably occur for films as thin as 1.3 nm, and crystalline nanoparticles as small as 2-3 nm in diameter have been synthesized. Overall, phase change materials show very favorable scaling behavior --- from the materials perspective, this technology can be expected to be viable for several future technology nodes.

%
\subsection{Speed of PCM}\label{subsec:SpeedOfPCM}

Of all the materials parameters mentioned so far, crystallization speed is probably the most critically important for PCM, because it sets an upper bound on the potential data rate.  And as discussed in Section~\ref{subsec:potentialApplications}, data rate and endurance are the two device characteristics that dictate what possible application spaces could potentially be considered for PCM.  (Of course, cost and reliability are then critically important to {\em succeed} in that space, but without the required speed and endurance for that market segment, such considerations would be moot anyway.)

As mentioned in Section~\ref{subsec:WhatIsPCM}, the early discovery of electronic-induced phase change behavior by Ovshinsky\cite{Ovshinsky:1968} did not immediately develop into the current PCM field.  Early phase change materials simply crystallized too slowly to be technologically competitive, with switching times in the microsecond to millisecond time regimes\cite{Feinleib:1971,Ovshinsky:1973}. Phase change technology began to gain traction in the late eighties with new phase change materials capable of recrystallization in the nanosecond time regime\cite{Yamada:1987,Chen:1986a}.  These discoveries both led to the widespread use of phase change materials in optical re-writable technology (DVDs, CDs and now Blu-Ray), and fostered renewed interest in PCM.

In phase change devices, there are three steps that could determine the overall operating speed: read, RESET (to high resistance), and SET (to low resistance). The read operation depends on the speed with which two (or more) resistance states can be reliably distinguished, and thus is dominated by the circuit considerations (capacitance of the bit-line being charged up, leakage from unselected devices). Although the resistance contrast and absolute resistance of the PCM cell do play a role, the read operation can generally be performed in 1-10ns\cite{Kang:2006}. The SET and RESET steps, however, involve the physical ``brute force'' transformations between distinct structural states.

The energetically less-favorable amorphous phase --- which gives a PCM cell in the RESET state its high resistance --- is attained by melting and then rapidly cooling the material.  As the temperature falls below the glass transition temperature and molecular motion of the under--cooled liquid is halted, a ``kinetically trapped'' phase results. This process can be separated into three steps: 1) current-induced heating above the melting temperature, 2) kinetics of melting\cite{Weidenhof:2000}, and 3) kinetics of solidifying the molten material\cite{Suh:2006}.   Steps 1 and 2 involve the rapid injection of energy to first heat and then melt the material; Step 3 involves the rapid cooling to temperatures {\em below} those favorable to recrystallization (see below).  Thus the most practical method to engineer Step 3 is through design of the PCM cell's thermal environment\cite{Kang:2007b}.

These steps are each quite fast:  electrical pulses as short as 400 picoseconds have been used to switch Ge$_2$Sb$_2$Te$_5$ into the amorphous state\cite{Wang:2008b}.  While the kinetics of phase change devices are strongly material dependant, the generation of the amorphous phase in any practical phase change material is necessarily faster than the speed of crystallization --- because otherwise, the amorphous phase would simply never be observed\cite{Suh:2006}. Note that after a RESET operation, the amorphous phase can continue to evolve extremely slowly at low temperature, undergoing both continued relaxation of the amorphous phase as well as electronic redistribution of the trapped charge that participates in the electrical breakdown phenomenon\cite{Pirovano:2004a,Ielmini:2007a,Ielmini:2007b, Ielmini:2008c,Redaelli:2008c}.
This drift can be an issue for MLC in PCM devices, as discussed in Section~\ref{subsec:MLC}.

While the crystalline form is thermodynamically favorable, its kinetics are much slower than the formation of the high resistance state\cite{Welnic:2006}, by typically at least one order of magnitude. Thus the step that dictates the achievable data rate for PCM technology is the crystallization process associated with the SET operation.

Formation of the crystalline phase involves as many as four steps: 1) threshold switching\cite{Adler:1978}, 2) current-induced heating to elevated temperatures (but below the melting point), 3) crystal nucleation\cite{Kalb:2004}, and 4) crystal growth\cite{vanPieterson:2005}.  The latter two steps are the slowest, and realistically combine to determine the speed of the device. Not all of the steps will be encountered, however.  Step 1 is relevant only if the device started in the high-resistance RESET state, so that a large portion of the applied voltage dropped across material in the amorphous phase.

If all of the contiguous PCM material being heated is in the amorphous phase, then step 3 {\em must} take place before step 4 can begin.  An example is the first crystallization of materials (or devices) containing amorphous-as-deposited material, where no crystalline-amorphous interfaces are present. This nucleation step can be extremely slow in so--called {\em growth--dominated} materials, where nucleation is a highly unlikely event compared to the fast speed of crystal growth.  In fact, frequently the crystallization of microns of surrounding material in such materials can be traced to the creation of a single nanoscopic critical nucleus\cite{Ziegler:2006}.  In contrast, {\em nucleation--dominated} materials tend to have a lower barrier to nucleation, so that a large region of crystalline material stems from the growth of numerous supercritical nuclei\cite{Coombs:1995}.

In a typical PCM cell, only a portion of the phase change material is quenched into the amorphous state, meaning that for both types of materials, step 4 above mainly depends on the crystal growth speed at high temperature.  The main difference between the two classes of phase change materials is that the recrystallization of amorphous  nucleation-dominated material will occur both within the interior (nucleation) as well as from the edge (growth), while for a growth-dominated material only the propagation of the crystalline--amorphous boundaries matters.  This can either be an advantage or disadvantage, depending on whether this added nucleation is desirable (having seeded nuclei that can help speed up the SET operation), or undesirable (by causing data to be lost faster at low to intermediate temperatures).

Examples of growth--dominated materials, where the rate of crystal nucleation $<$ growth rate, include Ge-doped SbTe, GeSb, GeSnSb, and Ge$_3$Sb$_6$Te$_5$.  In contrast, Ge$_2$Sb$_2$Te$_2$, Ge$_2$Sb$_2$Te$_5$, and Ge$_4$Sb$_1$Te$_5$ are considered nucleation-dominated materials, with a rate of crystal nucleation $>$ growth rate. It should be noted that nucleation and growth kinetics have unique responses to temperature, so under certain conditions a material typically considered growth-dominated may appear nucleation-dominated, e.g. AIST\cite{Raoux:2008c}.

The crystallization time of phase change materials, even those intended for use in electrical devices, can be measured relatively easily using optical techniques.  This is because most phase change materials of interest to PCM also have the same large optical contrast between the two phases that originally motivated the rewritable optical storage application.  Somewhat like an optical storage device with a stationary disk, a static laser tester uses a low-power continuous-wave laser to constantly monitor the reflectivity while a high-power pulsed laser induces the desired phase changes.  The pulsed laser heats the material above its crystallization temperature for SET, or above the melting point for RESET.  The advantage of optical testing is that large-area, thin-film samples of new phase change materials can be quickly prepared, without the need for any patterning or other steps required for fabrication of full PCM devices. Then a wide range of powers (e.g., temperatures) and times can be tested out rapidly.

As expected from the above discussion, it has been observed that the recrystallization time of a melt-quenched area in a crystalline matrix is typically orders of magnitude faster than the first-crystallization time of as-deposited amorphous films\cite{Raoux:2008c}.  Because device operation hinges on the repeated cycling between the two phases,
the relevant parameter to use for assessing the viability of a new material for PCM is the recrystallization time.  However, since it is quite difficult to prepare isolated regions of melt-quenched material without introducing new interface effects, amorphous-as-deposited films remain the best way to study the physics of the nucleation process, to avoid the difficulty of deconvolving the entangled roles of nucleation and growth once a crystalline-amorphous boundary is present.
For growth-dominated materials, it is the presence of these crystalline-amorphous boundaries which make the recrystallization speed so much faster than the initial nucleation from the amorphous-as-deposited state.

Figure~\ref{fig:speed:Reflectivity} shows the change in reflectivity of a Ge-Sb thin film with 15 at. \% Ge, as a function of laser power and duration measured by a static laser tester\cite{Raoux:2007a}. The film was first crystallized by annealing it in a furnace for 5 min at 300\degreesC, which is above this material's crystallization temperature\cite{Raoux:2007a}.  A two pulse experiment was then performed. The first pulse, of fixed time and power (100 ns for 50 mW), created a small region of melt-quenched amorphous material at a previously unused spot in the crystalline film.  Then a second laser spot of variable power and duration, applied a few seconds later at the same location, was used to recrystallize the amorphous spot. Figure~\ref{fig:speed:Reflectivity}(b) plots the normalized change in reflectivity caused by this second pulse. Since the crystalline phase has a higher reflectivity than the amorphous, the increase in reflectivity observed for all pulses longer than 5-10 ns, independent of applied power, indicates extremely fast recrystallization.  Because this material is crystallizing the amorphous spot predominantly by growth from the surrounding crystalline-amorphous border, this time scales with the size of the melt-quenched amorphous spot\cite{Raoux:2008c}.  In contrast, Figure~\ref{fig:speed:Reflectivity}(c) shows the much slower initial crystallization from the amorphous-as-deposited phase of Ge-Sb measured in a single-pulse experiment\cite{Krebs:2009}.

\FigureG

Several factors contribute to nucleation and growth kinetics: temperature\cite{Kalb:2005a}, composition\cite{Coombs:1995a,Wuttig:2002,Lankhorst:2005}, material interfaces\cite{Ohshima:1996}, device geometry\cite{Lankhorst:2005,Wu:2008}, device size\cite{Wang:2008b,Breitwisch:2007}, material thickness\cite{Chen:2006t,Raoux:2008}, polarity\cite{TioCastro:2007}, and device history\cite{Raoux:2008c,Coombs:1995}. Of these, the two most important factors governing switching speeds are temperature and local composition.   In fact, most of the macroscopically-observable nucleation effects associated with geometry, size, thickness, device history, polarity and even material interfaces can be understood in terms of the effects of varying local composition on the delicate balance between surface and volume energies that drive crystallization.

For any given composition, the crystallization properties of a phase change material tend to be a strong function of temperature.  As shown in Figure~\ref{fig:speed:GrowthSpeed}, the crystal growth speed can vary by more than 15 orders of magnitude between room temperature (which is off-scale in Figure~\ref{fig:speed:GrowthSpeed}) and the melting point. The symbols in Figure~\ref{fig:speed:GrowthSpeed} correspond to growth speeds of less than 10 nanometers per second, measured by exhaustive atomic force microscopy (AFM) at temperatures below 180\degreesC\cite{Kalb:2004}.  (This data was taken for AIST, a growth-dominated material similar to GeSb, although both {Ge$_2$Sb$_2$Te$_5$} and {Ge$_4$Sb$_1$Te$_5$} were measured to have very similar low-temperature growth velocities\cite{Kalb:2004}).  The solid curve in Figure~\ref{fig:speed:GrowthSpeed} represents a simulation model built to match both this low-temperature experimental data as well as extensive recrystallization data for GeSb measured on a static laser tester.  This  experimental data included both reflectivity measurements similar to Figure~\ref{fig:speed:Reflectivity} as well as AFM measurements of the size of melt-quenched spots before and after laser pulses\cite{Burr:2008b}.  The sharp increase in crystal growth speed above the glass transition temperature at 205\degreesC~ is associated with a sharp drop in viscosity at these temperatures, characteristic of ``fragile glass-forming'' materials\cite{Angell:2000, Salinga:2008}.

This wide range of crystal growth speeds between moderate and high temperatures is one of the most important features of phase change materials.  It allows the amorphous phase to remain unchanged for several years at temperatures near room temperature, while at programming temperatures crystallization can proceed in $<$100 ns.  The kinetic response of phase change materials to temperature has been described by nucleation theory in great detail\cite{Senkader:2004, Salinga:2008, Kalb:2009}.  Perhaps the most influential parameters from nucleation theory involve the relation between the interfacial energy (energetic cost of adding material to a crystalline-amorphous interface) and the free energy of crystallization (thermodynamic driving force for crystallization).  This interplay controls the size of the critical or smallest stable nucleus, which in turn influences the nucleation rate at which such nuclei can be incubated at lower temperatures (typically reaching a maximum rate near the glass transition temperature), and the growth rate at which large nuclei expand into the surrounding undercooled liquid (typically peaking at higher temperatures closer to the melting point).  The presence of these sub-critical nuclei can be detected by Fluctuation Transmission Electron Microscopy (FTEM)\cite{Lee:2009i}, which can explore medium-range spatial correlations beyond the nearest atomic neighbor.  Using FTEM, it was observed that an increase in the population of sub-critical nuclei led to a decrease in the incubation time before crystallization\cite{Lee:2009i}, as predicted by classical nucleation theory\cite{Salinga:2008}.

\FigureH

As discussed earlier, scaling is beneficial for PCM device speed.  For example,  it was observed that the SET time (crystallization time) and RESET times (melt-quenching) were reduced for Ge$_2$Sb$_2$Te$_5$ material  when device dimensions were reduced from 90/1.5 ns for SET/RESET operation for 470 nm diameter ``pore'' cell devices to 2.5/0.4 ns for 19 nm diameter devices\cite{Wang:2008b}.  Such results help move towards one important goal of phase change materials research: the quest for materials that can reliably switch at speeds comparable to RAM (approximately 10-50 ns) without sacrificing retention, endurance, or any other critical performance specification.  Due to the large number of experimental variables that can contribute to switching kinetics (thermal environment, deposition conditions, changes or damage induced during processing, etc.), extrapolating from simple thin-film recrystallization experiments to PCM devices remains difficult.  Crystallization from electrical pulses has been reported to range from 2.5 ns\cite{Wang:2008} to 1 microsecond\cite{Yoon:2006a} for similar materials Ge$_2$Sb$_2$Te$_2$ and Ge$_2$Sb$_2$Te$_5$, respectively.

There are certainly phase change materials which crystallize at much higher speed than the widely used Ge$_2$Sb$_2$Te$_5$ alloy.  For example, GeSb and Sb$_2$Te$_3$ are two high speed materials which have been demonstrated to crystallize in tenths of nanoseconds, comparable to the performance of consumer SRAMs.  Phase-change memory devices fabricated from Ge-Te were shown to switch in resistance by nearly 3 orders of magnitude with SET pulses of one nanosecond\cite{Bruns:2009}.
 Femtosecond laser pulses have been demonstrated to induce disorder-to-order transition in amorphous GeSb films\cite{Chen:2006t,Sokolowski-Tinten:1998,Callan:2001}, indicating very high speed potential.

However, pulse duration is often confused with material switching speed.  This is an oversimplification: crystallization and growth are thermally activated processes\cite{Wuttig:2007,Kalb:2005} and removal of stimulation is followed by a cooling period where additional crystal growth can contribute to observed phenomenon.  Nevertheless, from a large aggregate of reports we can expect first generation devices to attain switching speeds of 20-200ns.  These phase change materials have the significant advantage of being highly nonvolatile at temperatures near room temperature, while retaining fast switching speeds at high temperature.  Extensive materials and device research continues to decrease crystallization times below these values\cite{Lee:2007u,Lee:2007o}, offering hope of a bright future for PCM technology in application niches that call for rapid switching.

\subsection{Modeling of PCM physics and devices}

Because of the large number of factors influencing the performance of PCM devices, a number of groups have begun to perform predictive numerical simulations.  Particularly for the consideration of reducing the RESET current that can limit density by requiring a overly large access device, even straightforward electrothermal modeling of the temperature produced by a particular injected current can be highly revealing.
Such electrothermal studies typically need to simultaneously solve the heat diffusion equation,
\begin{equation} \label{eq:heatDiffusion}
d C_p \frac{d T}{d t} = \nabla \cdot \left( \kappa \, \nabla T \right) \,+\, \left| \rho J^2 \right|,
\end{equation}
and Laplace's equation,
\begin{equation}
\nabla \cdot \left( \sigma \, \nabla V \right) = 0.
\end{equation}
In these equations, temperature $T$ and voltage $V$ are each computed as a function of time $t$.  Even inside each material of density $d$, parameters such as specific heat ($C_p$), thermal conductivity ($\kappa$), and electrical resistivity and conductivity ($\rho$ and $\sigma$) are frequently functions of both position and temperature.  The current density $J$ and the temperature dependence of the electrical conductivity serve to intimately cross-couple these two equations.

A number of studies have used analytical equations\cite{Senkader:2004,Ielmini:2007b,Kim:2006b,Wright:2007,Sonoda:2008,Rajendran:2009}, finite-element techniques\cite{Lacaita:2004a,Kim:2005o, Wicker:1999, Gille:2006, Yin:2006f,Russo:2008,Russo:2008a,Zhang:2008a,Liu:2009}, and finite-difference techniques\cite{Chen:2006t, Kim:2007c} to analyze either PCM cells or phase change material. Pirovano et al. studied the RESET current and the thermal proximity effect of scaled PCM by both simulation and experiment\cite{Pirovano:2003}.  Although analytical techniques are attractively simple, and work well for explaining the incubation of new crystalline nuclei\cite{Senkader:2004} or threshold switching\cite{Ielmini:2007b}, it is difficult to include the effects of inhomogeneous temperature distributions and temperature-dependent resistivity, which critically affect the RESET current through their effect on the dynamic resistance of the cell.  Finite element techniques can include these effects, and work well for cylindrically symmetric cell designs such as the conventional ``mushroom'' cell, since such structures can be reduced to a single ($r,z$) plane.  However, because of the inherent computational difficulty in inverting matrices as they grow very large, these techniques are difficult to extend to three-dimensional cell designs.  And finally, even though nucleation is unlikely to play an effect during the fast RESET pulse, recrystallization at the end of a RESET pulse does play an important role in the value of the RESET current, especially for the fastest-crystallizing phase change materials that hold the most attraction for applications.  The best case scenario would be to have a finite-difference simulation tool capable of handling large and arbitrary 3-D structures, which could potentially be matched against fast electrical SET and RESET experiments, slow thin-film crystallization experiments, and optical pulse experiments performed with the same material.

From our experience with such a simulation tool\cite{Chen:2006t}, the RESET condition is not dictated by the maximum temperature at the cell-center, but by what happens at the edge of the cell.  Typically, a voltage pulse just below the RESET condition leaves a small portion of the limiting cross-sectional aperture remaining in the crystalline state, usually at the extreme edges of the cell\cite{Chen:2006t}. In general, besides the obvious choice of reducing the diameter of this limiting aperture, the best way to reduce the RESET current is to improve the efficiency with which injected electrical power heats the cell.  In the best case scenario, this power would heat just the portion of the cell needed to block all of the cross-sectional aperture and produce a high resistance state.  However, in any practical case, the surrounding material is also heated to some degree.  Optimization can be performed by ensuring that the heated volume is minimized and by reducing the heat-loss through the thermally-conductive electrodes as much as possible.  Another popular way to decrease RESET current is to increase the overall resistance of the cell by increasing the series resistance of the contact electrode\cite{Czubatyj:2006}, although it is not clear how much of this benefit may be due to associated changes in thermal resistance.

As with any simulation, care must be taken to establish the boundary conditions correctly, because the computer memory available for simulation is inevitably finite. For instance, Dirichlet boundary conditions, which call for the edge of the cell to be held at room temperature, are frequently used\cite{Russo:2008} and are easy to program.  However, in a simulation where the hot central region of the PCM cell is not very far from this boundary, then the effective heat transfer over this boundary can become unphysically large, skewing the results.  In contrast, Neumann boundary conditions, in which the spatial derivative of temperature (or equivalently, outgoing heat flow) is held constant at the boundaries, allow a truncated simulation to act as if it is embedded within a large expanse of surrounding material.

Another important consideration is the optimization metric.  It is conventional in the PCM community to discuss the importance of RESET current.  However, it is the dissipated power, not current, which leads to the heating of the cell.  The focus on current comes from the assumption that a co-located access transistor will have a current-voltage characteristic which saturates. If the applied current needs to be higher than this saturation value, there is no way for that particular transistor to supply it. This in turn implies that density will need to be sacrificed in order to provide this current.  In actuality, the transistor and PCM device will interact in a complex manner, with the transistor supplying power to the PCM device as if it were a load resistor.  The added complexity is that the dynamic resistance of the PCM device itself will in turn be a strong function of this supplied power\cite{Rajendran:2009}. Thus too tight a focus on RESET current can end up optimizing into a shallow minimum in RESET current which in fact is quite disadvantageous in terms of RESET power\cite{Russo:2008}.

\subsection{Scalability of prototype PCM devices}\label{subsec:prototypeDevices}

Since so many interlocking parameters influence the performance of PCM devices, especially as they become ultra-small, an important part of scalability studies is the fabrication and testing of prototype PCM devices.  Because only a modest number of such prototype devices are typically fabricated and tested in research environments, these types of experiments cannot hope to predict the actual device reliability statistics (yield, endurance, and resistance distributions) that can be expected from full arrays. However, prototype devices are an extremely important test for the scalability of PCM --- if you cannot get {\em any} ultra-small devices to operate correctly, then this is a bad sign for the future of the technology. Here we will report on the properties of one such prototype device: the bridge cell.

The phase change bridge cell is a relative simple testing vehicle for studying novel phase change materials\cite{Chen:2006t}, extending the line-device concept that had been introduced earlier\cite{Lankhorst:2005} to ultra-small dimensions. In these studies, two 80nm-thick TiN electrodes separated by a planarized dielectric layer were typically used as the contacts, with a thin phase change bridge fabricated to connect the two electrodes.  Bridge devices have been fabricated from various materials including undoped and doped Ge-Sb with 15 at. \% Ge, Ge$_2$Sb$_2$Te$_5$, Ag- and In-doped Sb$_2$Te, Ge-Te with 15 at. \% Ge, and thin Sb phase change materials\cite{Krebs:2008, Chen:2006t, Krebs:2009a}.

After fabrication of these TiN bottom electrodes using KrF lithography and chemical-mechanical polishing, a thin layer of phase change material (down to 3nm thick) was deposited by sputter deposition and capped with a thin SiO$_2$ layer to prevent oxidation.  Electron-beam lithography was used to define the phase change bridge itself.  Bridge widths (set by the e-beam lithography) varied between 20nm and 200nm, and the length (determined by the spacing between the underlying electrodes) ranged from 20nm to 500nm. Figure~\ref{fig:devScaling:PCBTEM}(a) shows a scanning electron microscope image of the phase change bridge and the TiN electrodes, while Figure~\ref{fig:devScaling:PCBTEM}(b) shows a cross-sectional transmission electron microscope image of a Ge-Sb bridge that is only 3nm thick.  Negative photoresist was used to define the bridge so that the resist did not need to be removed after the fabrication process. Ar ion-milling was applied to transfer the exposed photoresist pattern into the phase change material, and a 50 nm thick layer of SiO$_2$ was subsequently deposited for protection.

These devices could be cycled through more than 30,000 SET-RESET cycles and the stored data was shown to survive temperature excursions up to 175\degreesC\cite{Chen:2006t}.  These ultra-small devices --- down to devices with cross-sectional apertures as small as 60nm$^2$ --- correspond to effective switching areas that will not be encountered by mainstream device technology until the 22nm node, which Flash is expected to reach in 2015\cite{ITRS:2008}.  Thus these bridge-device demonstrations show that PCM will remain not only functional but robust through at least the 22nm technology node.

\FigureI

Several device parameters were measured as a function of device geometry. Current-voltage (I-V) curves revealed typical PCM behavior with threshold switching.  By modifying the fabrication procedure so that the phase change material never experienced any temperatures over 120\degreesC, bridges that remained in the amorphous-as-deposited phase could be produced\cite{Krebs:2009, Krebs:2009a}.  This allowed the precise measurement of threshold switching as a function of device length in a well-known geometry.  Each material was found to have a unique threshold field, measured by determining the threshold voltage as a function of device length. These fields were 8, 19, 39, 56 and 94 V/$\mu$m for Ge-Sb with 15 at. \% Ge, Ag- and In-doped Sb$_2$Te, Ge-Te with 15 at. \% Ge, Ge$_2$Sb$_2$Te$_5$, and thin Sb phase change materials, respectively\cite{Krebs:2009, Krebs:2009a}. No leveling-off of the threshold voltage with length was observed, although the shortest bridge devices were 20nm, as opposed to the nanowire devices where amorphous plugs shorter than 10nm were studied (Section~\ref{subsec:materialsScalability}). Unfortunately, line-edge-roughness in the TiN electrodes (see Figure~\ref{fig:devScaling:PCBTEM}(a)) led to shorts between the long TiN electrodes for separations smaller than 20nm.  Figure~\ref{fig:materScaling:PCBRESETcurrentScaling} demonstrates the scaling behavior of bridge devices in terms of the RESET current. Both measured RESET current (dots) and the predictions of numerical simulations (lines)\cite{Chen:2006t} decrease linearly with the cross-sectional area of the bridge device.  This indicates a favorable scaling behavior because the required RESET current determines the size of the access device which in turn determines the effective density of the PCM array.

It was possible to repeatedly cycle bridge devices fabricated from fast-switching Ge-Sb material with SET and RESET pulses of 10ns\cite{Krebs:2009, Krebs:2009a}, confirming the observations of fast crystallization for this material seen in optical testing (Figure~\ref{fig:speed:Reflectivity}). Very short switching times and reduced switching times with reduced device dimensions have also been observed for ultra-scaled pore devices by Wang and co-workers\cite{Wang:2008b}.

\FigureJ

\section{Design and fabrication of PCM}\label{sec:DesignFabricationOfPCM}

In this section, we discuss issues relevant to the design of PCM cells, such as cell structures and access circuitry, as well as those related to fabrication, such as the effects of variability and the deleterious effects of semiconductor processing on PCM materials and devices.

\subsection{Cell structures}\label{subsec:CellStructures}

Over the next few technology generations, the most serious consideration for PCM is the large current needed to switch PCM cells. We can roughly estimate this required current using ``back-of-the-envelope'' calculations.  The operation of the PCM cell relies on Joule heating, so the cell structure and operating conditions are dictated by the electro-thermal diffusion equation shown earlier (Equation~\ref{eq:heatDiffusion}).

We can assume that the critical volume undergoing phase change within the cell in a closely packed memory array is laid out at a pitch of $2F$. This immediately imposes the restriction that, to first order, the applied electric pulse width should be such that the thermal diffusion length should not exceed the half-pitch distance, $F$, in order to avoid cross-talk during programming,
\begin{equation}\label{eq:thermalDiffusionIn1D}
F \;>\: \sqrt{2 D \tau},
\end{equation}
where $D$ is the diffusion constant defined as $(\kappa/d C_p)$ and $\tau$ is the time duration of the applied electric pulse.  This approximation is only good to first order, however, because thermal diffusion through a 1--D geometry is {\em not} identical to a 1--D slice of diffusion through a 3--D geometry.  We address this issue again in Section~\ref{subsec:crosstalk} where we discuss cell-to-cell thermal crosstalk.

However, this computation allows us to set a lower bound for the required current density to achieve melting.  We can use the expected pulse duration $\tau$ from Equation~\ref{eq:thermalDiffusionIn1D} to satisfy the minimum condition that the supplied energy should be large enough to raise the temperature of the critical volume above the melting point,
\begin{equation}
J^2 \rho \;>\; d C_p \frac{\Delta T}{\tau},
\end{equation}
where $\Delta T$ is the difference between the melting point of the phase change material and ambient temperature.
Note that the energy spent to heat neighboring material, as well as any inhomogeneous heating of the center of the critical volume beyond the melting point, is not included.

Figure~\ref{fig:cellStructures:ExpectedCurrentVsPitch} shows a plot of pulse duration $\tau$ and, more importantly, the lower bound on current density required to heat the minimum volume. Here typical material parameters for phase change materials have been used.  This analysis thus suggests that we will need to supply a current density of \emph{at least} 10$^6$A/cm$^2$ (10$^4$ $\mu$A/$\mu$m$^2$) to be able to melt the critical volume within the cell for RESET.

\FigureK

In contrast, the measurements and simulations shown in Figure~\ref{fig:materScaling:PCBRESETcurrentScaling} would seem to indicate that the current density required is actually much larger, possibly as high as 300$\mu$A for a 300nm$^2$ aperture, or 10$^8$A/cm$^2$ (10$^6$ $\mu$A/$\mu$m$^2$).  However, the large expanse of metallic electrodes in close proximity to these tiny prototype devices actually makes this number more pessimistic than is warranted.  Other demonstrations, such as the 160$\mu$A RESET current shown for a 7.5nm $\times$ 65nm dash-type cells\cite{Im:2008}, seem to suggest a number like $J_{PCM}$ $\sim$ 3$\times$10$^7$A/cm$^2$ = 3$\times$10$^5 \mu$A/$\mu$m$^2$.  Because the estimated numbers plotted in Figure~11 represent a lower bound, we will instead use this higher empirical value in the remainder of this discussion.

In a full memory array, an access device such as a diode or transistor must be included at each memory cell to ensure that the read and write currents on each bitline are interacting with one and only one memory device at a time.  The amount of current that this access device can supply must comfortably exceed the required RESET current of the worst-case PCM element. Unfortunately, the CMOS FETs often considered for use as access devices in a PCM memory array have limited current drive capability; most optimized devices can provide only about $I_{acc}$ $\sim$ 800-1500 $\mu$A/$\mu$m, where this current capability scales linearly with the effective gate width of the drive transistor. One way to solve this problem is to simply make the access device larger so that it can drive a larger current.  However, since this immediately sacrifices memory density which subsequently drives up the cost per MByte, such a move would be economic suicide for a prospective memory technology.

Given a transistor of width $F$, the available transistor drive-current, $I_{acc} F$, must exceed the required current for RESET, $J_{PCM} \eta F^2$, where $\eta$ is an area factor between 0 and 1. Although returning to the SET state does involve exceeding the threshold voltage, the amount of power (and current) in the SET pulse is typically 40-80\% that of the RESET pulse.  Thus it is almost always the RESET pulse that must be considered when determining if the access device will supply sufficient current, while the SET pulse is typically the factor that dictates the write speed of PCM technology. Putting in the numbers above produces
\begin{eqnarray}\label{eq:pcmRESETscaling}
I_{acc} F &>& J_{PCM} \eta F^2 \nonumber\\
\left( 1.5 \times 10^3 \mu\mathrm{A}/\mu\mathrm{m} \right) F &>&
\left( 3 \times 10^5 \mu\mathrm{A}/\mu\mathrm{m}^2 \right) \eta F^2 \nonumber\\
5\mathrm{nm} &>&  \eta F,
\end{eqnarray}
which implies that transistor current scaling could only hope to catch up with the RESET current of lithographically--defined PCM devices at ultra-small technology nodes.  This is especially sobering given that PCM RESET current, as shown in Figure~\ref{fig:materScaling:PCBRESETcurrentScaling},
has been empirically observed to scale with CD at a pace
somewhere between $F^{1.5}$ and $F^{1.0}$, rather than as $F^2$.

This analysis implies that minimal-width FETs cannot supply the necessary current {\em if the dimensions of the phase change volume are determined lithographically}. However, use of a factor $\eta \sim 0.1$, corresponding to a sub-lithographic CD for the phase change element of roughly $F/3$, allows Equation~\ref{eq:pcmRESETscaling} to be satisfied for $F \sim$ 45nm, precisely where industry has been targeting first PCM products\cite{Numonyx_reverseEngineering}.  At this node, the required RESET current of 61$\mu$A (not that far from the demonstrated RESET current in Ref~\cite{Im:2008}) could be supplied by the CMOS transistor capable of 67$\mu$A.

Thus there are two parallel routes to ensuring sufficient RESET current for PCM devices:
\begin{itemize}
\item Use access devices that have higher current drive capability, including either known devices such as BJTs\cite{Pellizzer:2006} or diodes\cite{Oh:2006b} or novel devices such as surrounding-gate transistor\cite{Schmidt:2006} or FinFETs\cite{Huang:1999}.
\item Locally increase the current density within the phase change element and decrease the switching volume by creating sub-lithographic features in the current path through the PCM element.
\end{itemize}
Both of these approaches have been pursued aggressively to demonstrate the basic operation of PCM technology. In this section, we will focus on the latter path: optimization of the phase change element itself by creation of a sub-lithographic aperture.

A typical PCM cell is designed so that the only current path through the device passes through a very small aperture.  As this aperture shrinks in size, the volume of phase change material that must be melted (and quenched into the amorphous state) to completely block it is reduced.  In turn, this decreases the power (and thus the current) requirements.  If this current is low enough, then a minimum-size access device can provide enough power to switch the cell from the SET state to the RESET state.

In order to fabricate a PCM cell that will work even with these small currents, an innovative integration scheme is needed which creates a highly sub-lithographic yet controllable feature size.  Subtle variations in cell design may have a large impact on critical device characteristics, including endurance, retention, SET and RESET resistance distributions, and SET speed.  These considerations will be the subject of Section~\ref{subsec:Variability}. The cell design must be scalable as well as highly manufacturable, since scaling implies not only a shrink in physical dimensions of the memory cell, but also an increase in the number of memory cells per chip.  Lastly, to maximize the number of bits per cell, a cell structure which allows multi-bit functionality is highly desirable\cite{Happ:2006, Nirschl:2007}.

Depending on how this sub-lithographic aperture is implemented, PCM cell structures tend to fall into one of two general categories:  those which control the cross-section by the size of one of the electrical contacts to the phase change material (contact-minimized, Figure~\ref{fig:cellStructures:twoTypes}(a))\cite{Pirovano:2003,Pellizzer:2004,Ahn:2005a,Lai:2001,Jeong:2006,Song:2006,Ryoo:2007}, and those which minimize the size of the phase change material itself at some point within the cell (volume-minimized, also known as confined, Figure~\ref{fig:cellStructures:twoTypes}(b))\cite{Pirovano:2003, Tyson:2000,Kim:2005o,Cho:2005,Lankhorst:2005,Chen:2006t,Happ:2006,Oh:2006b,Breitwisch:2007,
Pirovano:2004b}. The typical volume-minimized cell structures tend to be a bit more thermally efficient, offering the potential for lower RESET current requirements compared to the contact-minimized structures\cite{Pirovano:2003}.

\FigureL

\subsubsection{Contact-minimized cell}

The most common contact-minimized cell structure is the mushroom cell, where a narrow cylindrical metal electrode contacts a thin film of phase change material. Figure~\ref{fig:cellStructures:TEM_mushroom} shows TEM images of mushroom cells in the SET state (a) and in the RESET state (b). In the RESET state, an amorphous dome of the phase change material --- resembling the cap of a mushroom, thus the name --- plugs the critical current path of the memory cell, resulting in an overall high resistance state for the cell. The bottom electrode contact (BEC), typically made of TiN, is the smallest and thus most critical dimension (CD) in this cell.  It is common to see this described as the ``heater,'' although the cell works most efficiently when the heat is mostly generated in the phase change material at the top of this BEC.

The sub-lithographic BEC can either be formed by a spacer process\cite{Horii:2003}, resist trimming\cite{Nirschl:2007}, or by the key-hole transfer process\cite{Breitwisch:2007}, followed by Chemical Mechanical Polishing (CMP) for planarization.  The processing of phase change materials is discussed in more detail in Section~\ref{subsec:Processing}. Prototype devices can use e-beam patterning to define the heater but this is too slow to be used commercially.  The thin film of the phase change material (or a stack of different phase change materials\cite{Lai:2006} with varying alloy concentrations) can then be deposited over the planarized feature using standard techniques such as Physical Vapor Deposition (PVD) or Chemical Vapor Deposition (CVD).  The top electrode contact (TEC) is also deposited, usually without breaking vacuum. The simplicity of the phase change materials portion of the process --- and the ability to define the CD before any novel materials are introduced --- represent two of the most attractive features of the mushroom cell.

The deposited films are then patterned into islands using conventional lithography to form individual cells, and isolated and encapsulated using thermally-insulating dielectric materials such as Si$_3$N$_4$\cite{Song:2006}. A variety of materials engineering techniques have been introduced to optimize the cell performance, especially the minimization of RESET current. These include increasing the resistivity of either the electrode material\cite{Lee:2005q} or the phase change material\cite{Czubatyj:2006}, and decreasing the thermal diffusive losses through both the top and bottom electrode region\cite{Rao:2008a, Zhang:2008a}.

Although cells with horizontal heater electrodes have been demonstrated\cite{Ha:2003}, the vast majority of contact-minimized cells closely resemble the mushroom cell.  One popular variant is the ring-electrode mushroom cell, where the heater electrode consists of a thin ring of metal surrounding a center dielectric core\cite{Jeong:2005}.  The incentive here is to reduce both RESET current and variability by decreasing the effective area, since compared to
a normal heater the metal annulus has a smaller area which also scales only linearly with CD.

\FigureM

\subsubsection{Volume-minimized cell}

Significant research efforts have been spent exploring a variety of volume-minimized cell structures, owing to their superior scaling characteristics. However, achieving such a structure can be a challenge, requiring the development of processing technologies that can successfully confine the phase change material within a sub-lithographic feature. The most obvious structure in this category is the pillar cell (Figure~\ref{fig:cellStructures:FETaccessDevice}), where a narrow cylinder of phase change material sits between two electrodes\cite{Happ:2006}.  This cell is fabricated in a similar fashion to the mushroom cell, with a thin film stack of the phase change material and top electrode material  deposited atop a bottom electrode, and then patterned.  However, in the pillar cell, the BEC is large, and it is the phase change material that must be successfully and reliably patterned into sub-lithographic islands.  This patterning can be performed in various ways, including lithography followed by resist trimming\cite{Happ:2006}. In addition to the challenges in controlling the size of the patterned islands, this cell structure also suffers from the drawback that the Reactive Ion Etch (RIE) of the phase change material can form a thin layer of altered alloy composition at the surface, strongly affecting the performance and yield of the cell\cite{Joseph:2008}.  This is discussed in further detail in Section~\ref{subsec:Processing}.

A modified version of the pillar cell structure is the pore cell\cite{Breitwisch:2007}, where a sub-lithographic hole formed in an insulating material atop the BEC is filled with the phase change material(Figure~\ref{fig:cellStructures:TEM_subLithoPore}). Conformal filling of nanoscale holes with high aspect ratio is difficult using conventional PVD processes; hence development of CVD or Atomic Layer Deposition (ALD) technology for phase change materials will be necessary to enable continued scaling of pore cell devices\cite{Lee:2007j}. As with nearly all PCM cell designs, the dimensions and aspect ratio of the phase change material region critically influences the RESET current.

\FigureN

\FigureO

\subsubsection{Hybrid PCM cells}

The most advanced scaling demonstration of PCM technology to date was realized using the volume-confined bridge cell\cite{Chen:2006t}, which consists of a narrow line of ultrathin phase change material bridging two underlying electrodes (Figure~\ref{fig:devScaling:PCBTEM}). The cross-sectional area of this device is determined by film thickness in one direction and by electron-beam lithography in the other, allowing the realization of functional cells with cross-sectional area of about 60 nm$^2$ and RESET current requirement of about 80$\mu$A.

The $\mu$-trench cell\cite{Pellizzer:2004} and the dash-confined cell\cite{Im:2008} are examples of PCM cells that combine the contact-minimized and volume-minimized approaches.  The $\mu$-trench cell is an extension of the bridge concept, with a PCM element formed at the intersection of an underlying sidewall-deposited CVD TiN bottom electrode and a trench of phase change material formed at right angles across this electrode (Figure~\ref{fig:cellStructures:MicroTrench}).  The dash-confined cell\cite{Im:2008} is an extension of the pore cell idea, except that the bottom electrode contact is formed by a spacer process.  A CVD process is then employed to fill in the rectangular sub-lithographic holes formed by the recess etch into the metal BEC (Figure~\ref{fig:cellStructures:DashConfinedCell}). In all three of these cases, one critical dimension associated with the limiting cross-sectional aperture is controlled by a thin film deposition process.  As a result, only one dimension inherits the variability of the lithography process. Conversely, of course, only one dimension will enjoy the associated scaling benefits as $F$ shrinks from one technology node to the next.

\FigureP

\FigureQ

\subsection{Access circuitry}

In order to fully leverage the scalability of PCM and thus achieve the very high densities needed for Storage Class Memory (SCM), the most ideal implementation of PCM would be a cross-point array architecture, where each memory element in the array is directly connected to two orthogonal lines (Figure~\ref{fig:deviceApplications:FourFsquared}).  In fact, a novel architecture has been proposed for such a direct cross-point memory, in which PCM devices are switched not between SET and RESET, but between the RESET state and an over-RESET state (strongly RESET state)\cite{Chen:2003k}.  Even though these resistances may differ, the difference in read current would be too low to support rapid read.  But because the threshold voltage varies between these two states, the device state can be sensed by detecting whether a ``read'' voltage intermediate between these levels produces an electrical switching event.  While this scheme does cleverly avoid an access device, it has several serious issues.  The margins between the read and the two threshold voltages are uncomfortably tight, and it would be difficult to detect the breakdown without potentially heating up the cell, which means that reads must be treated as destructive.  Even so, typical RESET resistances are still low enough that the leakage through ``half-selected'' devices (those that share either the same bit-line or the same word-line as the ``selected'' device) would be quite high, thus limiting the maximum array size (and the effective memory density) that could be built.  Worst of all, the strong negative correlation between switching energy and endurance (which will be discussed in detail in Section~\ref{subsec:Endurance}) means that improving the sense margin by increasing the resistance (and thus the threshold voltage) of the over-RESET state will sharply reduce endurance.

Thus the integration of PCM into an array architecture seems to require the use of an access device: either a diode\cite{Oh:2006b, Zhang:2007d}, a field-effect transistor\cite{Hwang:2003c, Ahn:2004a}, or a bipolar junction transistor\cite{Pellizzer:2006, Bedeschi:2008}). The main role of this device is to minimize the leakage current that would otherwise arise from the non-selected cells in the array.  As has been mentioned, the most important unknown for the success of PCM technology is whether this memory access device is able to provide sufficient current to RESET the PCM cell. While a diode can provide a current-to-cell size advantage over a planar transistor down to the 16nm node\cite{Lung:2007}, the diode scheme is more vulnerable to write disturbs due to bipolar turn-on of nearest-neighbor cells\cite{Oh:2006b}.  A 5.8\Fsquared PCM diode cell has been demonstrated using a 90nm technology in which the diode was able to supply 1.8mA at 1.8V\cite{Oh:2006b}.  In comparison, a 90nm 10\Fsquared tri-gate FET could only supply approximately half as much current\cite{Oh:2006b}.

Although the raw footprint of the access device is of primary concern for memory density, other considerations can also come into play.  Peripheral circuits, such as charge pumps to increase the voltage supply level or special read and write circuitry for MLC operation\cite{Bedeschi:2009}, reduce the portion of the chip that can be dedicated to memory devices (the ``area efficiency'' of the chip).  In addition, the effective area per cell can grow because of additional vias or wiring that may be required within the memory array.  For instance, given the high currents required for PCM programming, the voltage drop along metal bit- and word-lines can become significant, further reducing the power that can be delivered to the actual PCM cell.  Thus splitting an array in half to reduce the maximum line length is attractive because it reduces the worst-case line loss, but detrimental because chip real-estate is now being used for redundant wiring rather than memory devices.  Particularly problematic is the wiring that must sit under the PCM layer (such as common source lines, as well as the word-lines to transistor gates), since these lines must typically use high-resistance tungsten or polysilicon.  Because of the significant voltage losses in such lines, a via must often be introduced every few (e.g., 4 or 8) cells in order to strap this line to an overlying low-resistance copper line.  This extra via immediately increases the effective footprint per memory cell.  Although copper cannot be introduced near the transistors, lest the CMOS devices become degraded, one solution would be to move the PCM devices higher up away from the silicon so that a layer of copper interconnect can be fabricated under the PCM devices.

\subsection{Variability}\label{subsec:Variability}

Different types of variability can affect the operation, performance, endurance, and reliability of PCM devices, ranging from inter-cell variability introduced during processing, inter-cell variability as resistances change over time after programming, and cycle-to-cycle variation of the SET and RESET resistances of any given cell (intra-cell variability). While the read voltage, the RESET pulse, and the SET pulse can be optimized for the \emph{average} memory cell, variations between cells must be minimized so that these same choices can successfully operate all the cells in the memory array.  The same is true for the performance of the access device.

Any fabrication- or process-induced variability in the physical structure of the phase change element can result in devices which react differently to the same stimuli.  Thus any large collection of phase change elements, which because of variability are in similar yet non-identical states, may then propagate in time on different resistance trajectories.  In addition to group behavior caused by inter-cell variability, there is also intra-cell variability produced by motion and rearrangement of the atoms within the active region of the phase change device during the programming of the phase change device. Understanding the variability of PCM devices is particularly important in the context of multi-bit storage capability, due to the reduced margin between resistance levels that must be correctly sensed in order to successfully retrieve data.  And finally, variability can be expected to play an increasingly important role in the scaling of PCM technology into future technology nodes.

\subsubsection{Structural variability}

Each step in the process of fabricating a wafer of PCM memory devices is typically associated with one or more physical attributes (e.g., thickness, CD, sidewall-angle, etc.), each with a nominal target value.  To illustrate the variability challenge inherent in wafer-level fabrication process, features on the order of nanometers (10$^{-9}$ meters) must be accurately controlled across the surface of a wafer which spans nearly a third of a meter (0.3 meters).  Exact control over such a large range of dimensions is impossible.  Consequently, each process is associated with an acceptable range around some target value.

Furthermore, this fabrication process must construct a fully integrated set of devices comprising a CMOS technology (CMOS field effect transistors, diodes, resistors, capacitors, wiring levels, and the vias connecting them) in addition to the PCM devices (which usually reside on top of the CMOS devices, at the bottom of the wiring levels).  Thus a wafer will undergo hundreds of processing steps (including photolithography, atomic implant, Reactive Ion Etch, material deposition, chemical mechanical planarization, wet etches, etc.), each with associated variability, before the final processed wafer is ready for the dicing and packaging of the chips.  Variability in processes which affect either the structure of the phase change element or the access device (transistor, diode, etc.) can contribute to the overall phase change device variability. These structural variations then translate into variations in electrical (device operation) properties of the device.

Most relevant to the operation of PCM are the structural physical properties which affect the temperature profile (how much heat is generated and where), the critical limiting cross-sectional aperture (which must be fully blocked to get high resistance contrast), and the volume where the phase change material undergoes repeated melting and crystallization.  Crucial features of the cell are the aperture size and shape, the thickness and uniformity of the phase change material (in both stoichiometry and doping), the resistivity and interface resistance of electrodes, the thermal conductivities of surrounding materials, and the stresses on the active switching volume introduced by surrounding material.

As discussed in the previous section, several cell structures have been proposed and developed in order to minimize the required power (current) for RESET. Each of these structures employs a similar strategy for minimizing the RESET power:  at one and only one point within the cell, the electrical current is forced to pass through a small aperture. This aperture increases the current density, maximizing the thermal power which is generated, and reducing the volume of high resistivity material needed to significantly alter the external device resistance. Together with the electrical and thermal properties of the phase change material, electrodes, and surrounding materials, the size and shape of this aperture determines the temperature profile obtained within the phase change element for a given SET or RESET programming current pulse.

In order to estimate the impact of small variations of the aperture size on the cell operation, it is instructive to examine the functional dependence of the aperture size on the required RESET programming current.   We show the published dependence of the RESET current on the aperture size for the Bridge cell (Figure~\ref{fig:materScaling:PCBRESETcurrentScaling}), the Pillar and Mushroom cells (Figure~\ref{fig:variability:RESETcurrentScaling}(a)), and the Pore cell (Figure~\ref{fig:variability:RESETcurrentScaling}(b)).  In each case, the required current is a steep function of critical dimension.  Variability introduces a distribution of PCM cells of different sizes and thus different RESET currents. In order to be sure to be able to RESET all of the devices, it is thus the largest diameter cell that dictates the required current-driving capability of the access device.  Variability directly reduces density by mandating a larger-area access device.  At the same time, as we will see in Section~\ref{subsec:Endurance}, switching cells with more power than is necessary has a strong and negative influence on device endurance.  Thus this same variability may also reduce endurance in the smaller-area devices, which are being driven much harder than they really need to be.

\FigureR

\subsubsection{Sources of structural variability}

There are several ways a sub-lithographically sized aperture can be defined. Not surprisingly, the degree to which the aperture size can be controlled varies for each of these methods\cite{Breitwisch:2009}. A direct way to reduce the size of a lithographically-defined hole is to implement a collar process, as shown in Figure~\ref{fig:variability:CollarProcesses}. Similarly, a subtractive method can be used (as in the Pillar cell scheme\cite{Happ:2006}), where an island of lithographically-patterned photoresist is trimmed in size using RIE and then used as a mask to transfer the sub-lithographic pattern down into the underlying layers. Unfortunately, in both of these techniques, any variability in the diameter of the original hole (or photoresist pillar) introduced by either lithography or etch (or resist development) transfers directly to the final CD.  Thus the fractional variability ($\Delta$CD~$/$~CD) can become uncomfortably large.

\FigureS

The $\mu$-trench cell\cite{Pellizzer:2004}, the ring bottom-electrode Mushroom cell\cite{Jeong:2005}, the dash-confined cell\cite{Im:2008} and the Bridge cell\cite{Chen:2006t}  define one of the dimensions of the cross-sectional area of the aperture through film deposition.  The thickness of such deposited film can be tightly controlled, especially for CVD and even more so for ALD techniques. This thickness can easily be much thinner than the lithographic dimension $F$ (at least for current and near-future technology nodes).  This method of defining one dimension of the aperture by film thickness is inherently decoupled from any lithographic variability.  However, for these schemes, lithography is still needed to define the ``other'' dimension of the aperture.

A keyhole-transfer process has been developed for PCM devices which decouples both dimensions of the final aperture from lithography.  Here, a keyhole is defined within a lithographically-defined hole by film deposition.  Typically, a keyhole is undesired and indicates a failure to fill the hole.  However, the advantage is that the dimension of the keyhole itself can be tightly controlled by accurate control over the hole depth (through film deposition) and over the etched undercut into the sidewalls of the hole, despite poor control over the actual lithographically-defined diameter of the hole itself.  Figure~\ref{fig:variability:LithoPoreIdea} describes this process and shows how the keyhole process can produce identical sub-lithographic holes (30nm) despite significant variation in the much larger lithographically-defined holes (243nm and 302nm).

\FigureT

The keyhole-transfer process has been experimentally demonstrated to decouple the final aperture size from the initial lithographically-defined hole size\cite{Breitwisch:2007}. Over a span of initial lithographically-defined hole sizes,  the distribution of RESET current was found to be consistently narrower for pore cells fabricated through the keyhole-transfer method than for those fabricated with a collar process. Similar results have been demonstrated for mushroom cells\cite{Rajendran:2009}.  Here, mushroom cells with heaters defined with the keyhole-transfer process were compared to otherwise identical mushroom cells with heaters defined by trimming of photoresist islands.  This comparison was performed by examining the dynamic resistance $R_{dyn}$ during programming, which tends to exhibit a dependence on programming current $I$ as
\begin{equation}
R_{dyn} = \frac{A}{I} + B.
\end{equation}
Here the term $A$ depends only on material characteristics, while $B$ incorporates both material and structure-dependent factors\cite{Rajendran:2009}.  Thus the lower variability in $B$ empirically observed for mushroom cells
with heaters defined by the keyhole--transfer method (as compared to those defined by
trimmed-photoresist) is indicative of the tighter CD control offered by the keyhole--transfer method.

\subsubsection{Impact of structural variability}

Structural variability gives rise to variability in electrical response, which leads to broader resistance distributions after single-shot programming.  Figure~\ref{fig:variability:Rdistributions} shows a series of SET resistance distributions, including one (for 100ns SET pulses) which is strikingly broad.  Although the majority of cells can reach a low resistance of approximately 2k$\Omega$ with a SET pulse of 100ns duration, for this collection of cells there is a subset of devices whose resistances after such a single SET pulse extend all the way out to the fully RESET resistance of several hundred k$\Omega$.  For a given programming current amplitude, devices with different diameters will present different dynamic resistances during programming, thus dissipating different amounts of power despite the same drive voltages.  This variable power will lead to a different maximum temperature, and even a different temperature distribution within the cell because variations in aperture or heater size affect the thermal resistances within the cell.  For RESET pulses, the size of the amorphous plug required to significantly affect the room-temperature low-field resistance of the cell changes drastically with changes in the cross-sectional aperture of the cell.  In terms of SET pulses shown in Figure~\ref{fig:variability:Rdistributions}, rapid SET requires that the optimal temperatures for crystal growth be present at the crystalline--amorphous boundary (growth-dominated material) or within the cell interior (nucleation-dominated material). Since the crystallization speed is a strong function of temperature (Figure~\ref{fig:speed:GrowthSpeed}), variability in aperture size can result in a variety of incompletely-SET cells if the pulse is too short.  As Figure~\ref{fig:variability:Rdistributions} shows, increasing the duration of SET pulses tends to overcome this effect.  Another, even more effective route is to ramp down the SET pulse slowly, allowing each cell to pass through the temperature for maximum crystal growth\cite{Oh:2005}.

\FigureU

In addition to broadened resistance distributions, device variability also affects how these resistances evolve over time.  It is well known that the resistance of cells in the RESET state tends to increase slowly over time, an effect which has been attributed to either mechanical relaxation forced by the reduced density associated with the amorphous state\cite{Karpov:2007a, Braga:2009},  to the formation of electronic traps associated with lone-pair states which increases resistance by repositioning the Fermi level\cite{Redaelli:2008c}, the annihilation of defects by trap-filling which reduces transport and thus increases resistance\cite{Ielmini:2007a, Ielmini:2008a}, or to some combination of these effects.  Since this drift increases the already high resistance of the RESET state, it is not an issue for binary PCM devices.  However, drift can be particularly problematic in the context of multiple bits per cell, as discussed in Section~\ref{subsec:MLC}.

Drift would be bad enough if all cells changed in resistance along the same trajectory.  However, variability introduces different drift coefficients, so that cells with similar resistances immediately after programming tend to separate in resistance over time. In addition, some variability in crystallization speed has been observed over large arrays of PCM devices, so that some cells tend to SET more easily (both at elevated temperature and after low-temperature anneals) than the average cell\cite{Mantegazza:2008}. Figure~\ref{fig:variability:RESETtailModulation} illustrates the signature of this effect, with RESET tail bits emerging when a RESET programming pulse with an insufficiently short quench time is used for programming. Although such anomalous devices can be RESET to high resistance by using pulses with rapid quench, such tail-bit devices will still be associated with increased long-term retention loss and resistance distributions that broaden more rapidly over time. In general, combining the effects of drift and recrystallization, variable drift effects tend to broaden the resistance distribution of a large collection of cells over time:  some cells increase in resistance due to structural relaxation, some cells decrease due to partial recrystallization.  While some may first increase and then only later decrease, the random walk nature of these effects typically leads to broader distributions over time.

\FigureV

\subsubsection{Intra-device variability}

The act of programming a PCM cell involves the rearrangement and movement of atoms within the cell.  When a given cell, starting from the polycrystalline SET state, is melt-quenched and then crystallized back to the SET state, the distribution of crystal grains within the cell is not identical.  The nucleation of these crystalline grains within the amorphous plug leads to a decrease of RESET resistance, which accelerates at elevated temperature\cite{Gleixner:2007a, Shih:2008}.  It has been shown that a small fraction of cells in the RESET state will tend to show these effects more rapidly than the average cell, but that the {\em location} of these cells is random from cycle to cycle.  A cell may participate in this tail of the resistance distribution on one cycle, yet show much improved resistance to the anneal upon the next, consistent with random nucleation of crystalline grains within the amorphous plug\cite{Shih:2008}.  Thus we can expect that this random formation of crystalline grains is present throughout the portion of the cell that reaches elevated temperatures.

Upon the application of a subsequent programming pulse, the temperature distribution of the cell may not be exactly identical to the previous cycle, leading to variations in resistance.  This intra-device programming variability can be readily observed in any single-cell endurance plot, as illustrated in Figure~\ref{fig:variability:CyclingPerformance}.

\FigureW

\subsubsection{Variability and MLC}

Variability also forces the use of iterative write schemes for programming of multi-level resistance states, as illustrated in Figure~\ref{fig:variability:BeforeMLCdistributions}\cite{Nirschl:2007}.  Here, a collection of cells is programmed with single current pulses, with the trailing edges controlled to produce intermediate resistance values.  The highest and lowest resistance levels have the smallest variation in resistance --- for these states the RESET and SET resistances tend to saturate, as overly large voltages produce little additional increase in RESET resistance and overly long pulses produce little additional decrease in SET resistance.  The intermediate resistance levels, which represent hybrid states where a smaller portion of the cell is amorphous than in the full RESET state\cite{Itri:2004}, are much more sensitive to variability.

\FigureX

Even so, since each given cell does respond in a somewhat reproducible and hence predictable manner, the desired resistance value can be produced by an iterative write scheme.  As illustrated in Figure~\ref{fig:variability:AfterMLCdistributions}, the same cells used in Figure~\ref{fig:variability:BeforeMLCdistributions} can be programmed into relatively tight resistance distributions using iterative write attempts, where the length of the trailing edge of the pulse is carefully controlled based on the just-accumulated prior experience with that cell.  Such iterative write schemes and other considerations of Multi-Level Cell programming are discussed in more detail in Section~\ref{subsec:MLC}.

\FigureY

\subsubsection{PCM scaling and variability}

As PCM devices scale into future technology nodes, variability can be expected to become an even more important consideration.   Difficulties in scaling lithography and processing techniques may lead to even looser specifications on relative CD uniformity in future technology nodes, since the acceptable range of control will be driven by considerations for conventional CMOS devices and not by PCM technology.  At future technology nodes, PCM cell designs which depend on control over film thickness may no longer be as attractive, if CD is shrinking while both the minimum thickness $t$ and the achievable control of this variable, $\Delta t$, remain relatively static.  As the number of atoms participating in the active portion of the PCM cell decreases, any variations in local doping or stoichiometry can be expected to affect individual cells, reducing yield and further broadening distributions.  Eventually, in the same way that Poissonian effects have arrived for CMOS devices (e.g., as the number of dopants in the channel becomes countable)\cite{Liao:2009}, such few-atom effects can be expected to be observable for PCM.  That said, all logic and memory devices seeking to operate in this regime will have some variant of these effects --- the winners will be those with not-yet-known physics which keeps those effects sufficiently unlikely, or those amenable to not-yet-invented engineering techniques which can finesse these issues to an acceptable degree.

\subsection{Processing}\label{subsec:Processing}

Most studies related to the integration and processing of PCM technology focus on the popular variant, Ge$_2$Sb$_2$Te$_5$, here abbreviated simply as GST, along with various choices of dopant.  This reflects the fact that GST possesses many favorable characteristics and has thus received the bulk of attention of technologists attempting to take PCM technology from research to development.  Although these studies reveal numerous characteristics which might be considered less than perfect, the mere existence of such a large base of knowledge also represents a substantial hurdle for any alternative phase-change meterial hoping to challenge GST.

\subsubsection{Deposition}

The vast majority of phase change materials reported in the literature have been deposited by sputtering, i.e. Physical Vapor Deposition (PVD).  Sputter deposition from multiple targets makes it very simple to try different compositions.
However, sputtered films typically do not have good step coverage and cannot completely fill high aspect-ratio vias without keyhole formation\cite{Breitwisch:2007}. A number of interesting PCM cell structures have been discussed (Section~\ref{subsec:CellStructures}) that call for confining the phase change volume inside a contact hole (or ``pore'') formed in a dielectric, primarily to reduce the RESET current. It is therefore essential to examine alternatives to sputter deposition for GST and other phase change materials that can successfully fill higher aspect-ratio vias.

Cho et al\cite{Cho:2005} developed a novel GST sputter deposition process with in-situ deposition/etch/deposition, in order to fill GST into high aspect-ratio ($>$ 2:1) pores of approximately 50 nm diameter. A number of groups\cite{Kim:2006h, Lee:2007j, Kim:2008, Im:2008} have investigated chemical vapor deposition (CVD) of GST and related phase change materials. Kim et al\cite{Kim:2006h} deposited hexagonal phase GST by MOCVD at temperatures in the range 330-370\degreesC~using precursors that were bubbled at different temperatures, demonstrating complete filling into 120nm diameter trenches with aspect-ratio larger than 1.6.  However, it should be noted that in contrast to sputter deposition, CVD processes are far less flexible with respect to changes in material composition.

Lee and coworkers\cite{Lee:2007j} used CVD to deposit GST from metal-organic precursors and hydrogen at 350\degreesC~into high aspect-ratio (3:1) 50nm contact holes. X-ray diffraction (XRD) measurements showed that the GST had a hexagonal phase and was thermally stable up to 400\degreesC. CMP was used to planarize the GST, confining it to the contact holes (and removing it from everywhere else). Devices fabricated with this technique showed a significant reduction in RESET current as well as good endurance. Im et al\cite{Im:2008} used CVD GST to fill high aspect-ratio (4:1) dash-type contacts of 7.5nm width. As expected from device modeling, these cells showed very low (160uA) RESET currents. In addition, these devices also had fast SET speeds and very good endurance.

Lee et al\cite{Lee:2007a} developed atomic layer deposition (ALD) of GST with the assistance of an H$_2$ plasma. They synthesized new precursors for the GST deposition and obtained good step coverage (90\%) in 7:1 aspect-ratio holes. Finally, Milliron et al\cite{Milliron:2007} developed a solution-based deposition method for GeSbSe films using hydrazine and Se to form soluble precursors, so that film composition and properties were tunable through appropriate combination of these novel precursors in solution. XRD and laser pulse annealing were used to study the phase change and crystallization speed, and complete filling was demonstrated in 2:1 aspect-ratio vias patterned in thermal silicon oxide.

\subsubsection{Etching}

Etching of phase change materials has been explored using both wet as well as dry etching schemes. Such steps are important for electrical isolation in most cell designs, and are absolutely critical for cell designs such as the pillar cell that depend on subtractive processing to produce a confined volume of phase change material. Some important parameters that are studied in developing etch processes are etch rate, selectivity (i.e. how fast the desired PCM etches compared to the masking layer and other surrounding films), and anisotropy (etched sidewall angle; steeper profiles are usually desirable since they enable higher resolution patterning). It is also important to understand etch-induced material modification and other sidewall damage effects since these could impact device operation, especially if the damaged portion is close to or part of the active switching volume of the PCM.

A number of groups have reported wet etching of phase change materials using alkaline\cite{Shintani:2004, Anzai:2004, Lin:2007d} as well as acidic\cite{Cheng:2005b, Cheng:2005c} solutions. Depending on the etchant and specific phase change material etched, in some cases the crystalline material was found to have etched faster than the amorphous phase while the opposite was true in other cases. Cheng et al\cite{Cheng:2005b} used X-ray Photoelectron Spectroscopy (XPS) and Inductively Coupled Plasma (ICP) to study the wet etch mechanism of GST. They explained their results, obtained with a 20\% aqueous solution of nitric acid, as a chemical etching process that involves bond-breakage, oxidation of the various constituents, and dissolution of those oxides. They inferred that the Sb component was hardest to etch. The authors used this wet etching process to fabricate large PCM cells (400$\mu$m $\times$ 400$\mu$m) and performed I-V measurements showing successful switching of devices from the high to low-resistance states.

There have been a number of studies of plasma etching of phase change materials such as GST and its doped variants. A variety of gas chemistries have been reported in the literature, including Cl$_2$/Ar\cite{Yoon:2005c, Park:2006a, Min:2007, Yang:2007a, Min:2008, Yoon:2006d}, CHF$_3$/Ar\cite{Yoon:2005c}, CF$_4$/Ar\cite{Yoon:2006d, Feng:2008a}, HBr/Ar\cite{Lee:2007g}, CHF$_3$/O$_2$\cite{Feng:2007a, Xu:2008b}, and CHF$_3$/Cl$_2$/Ar\cite{Joseph:2008}. Some of these studies have examined the effects of varying process parameters such as reactant gas concentration fraction, chamber pressure, coil power, and dc bias on the etch rate and anisotropy (i.e. etched sidewall profile).

Yoon et al\cite{Yoon:2005c} indicated that the GST removal mechanism in high density helicon plasma etching using Cl$_2$/Ar chemistry was due to ion-assisted chemical etching. They reported GST etching selective to SiO$_2$ and did not notice any significant change in GST composition after etching. In the same work, the authors also studied the CHF$_3$/Ar chemistry and observed that the GST removal in this case was by physical etching due to ion bombardment. They noticed a small decrease in the Ge/Te ratio after this etching process. Min and coworkers\cite{Min:2008} also performed a systematic study of GST etching in Cl$_2$/Ar using an inductively coupled plasma (ICP). Using a model that analyzed etch kinetics, they explained the GST removal mechanism as a combination of spontaneous and ion-assisted chemical etching. They suggested that TeCl$_4$, not being as volatile as the Ge and Sb chlorides, accumulated on the surface and was removed by ion-stimulated desorption. In their etching studies of GST films using HBr/Ar in an ICP tool, Lee et al\cite{Lee:2007g} pointed to the role of hydrogen in passivating the etched surface as well as the sidewalls, leading to more anisotropic etching and lower etch rates at higher HBr concentrations. Their XPS analysis showed that as the etch progressed, the surface became Te-deficient and Ge- and Sb-rich implying that Te reacted most readily with the bromine and was preferentially etched.

Attempting to form nanoscale features into GST with SiO$_2$ hard mask and e-beam lithography (HSQ resist), Yoon et al\cite{Yoon:2006d} reported that using the Cl$_2$/Ar chemistry to etch the GST resulted in undercut and collapse of small GST features due to the isotropic component of the etching. By switching to a TiN hard mask process and using a CF$_4$/Ar chemistry, they were able to successfully pattern sub-100nm features in the GST.  Joseph and colleagues\cite{Joseph:2008} developed etch processes to pattern small features in N-doped GST films deposited on and capped with TiN. After using trimming processes to shrink the critical dimension (CD) of 248nm photolithography-patterned resist down to 100nm, ICP RIE in a Ar/Cl$_2$/CHF$_3$ gas mixture was used to first etch the TiN and then the N-GST. The authors pointed to the importance of carefully choosing the amount of Cl$_2$ so as to minimize undercut and obtain anisotropic profiles.

In the same work \cite{Joseph:2008}, the authors also reported the presence of a uniform 10nm thick damaged layer on all plasma-exposed surfaces of the N-GST, suggesting that the material modification was chemically driven and not enhanced by ion bombardment. TEM-EELS and EDS analysis were used to show that the damaged layer had depletion of Sb and/or Te along with potential oxidation. Depth profile XPS analysis on blanket films subjected to etch (and etch  followed by a strip process in an oxygen plasma) revealed selective loss of N and metallic Sb, and increased amount of Ge oxide. Finally, they also showed that the damaged layer could be removed selectively to the undamaged N-GST.

\subsubsection{Chemical-mechanical polishing}

Chemical mechanical polishing (CMP) is now a ubiquitous process in CMOS technology, especially in the Back-End-Of-the-Line (BEOL) flow. It enables the formation of inlaid, or `damascene,' structures that are used in copper interconnect fabrication\cite{Andricacos:1998}. Such structures are created by first forming a hole (or trench) in a dielectric, then filling the hole, usually with a metal, followed by CMP to planarize the metal with the surrounding dielectric. Thus CMP allows patterning of the metal without needing an explicit metal etching step. There have been a number of publications on the CMP of phase change materials such as GST.

Liu and coworkers\cite{Liu:2006e} made arrays of damascene PCM cells using CMP process. SEM and EDS were used to verify that the GST was properly filled in the contact holes and that the material composition of the GST was not changed by the CMP process. DC current sweeps on fabricated cells showed successful switching from the high to low resistance state.

Zhong et al\cite{Zhong:2008a} used CMP to fabricate damascene-type PCM cells. An alkaline slurry was used to polish GST deposited into 300nm wide vias. AFM measurements confirmed that a very smooth (0.8nm rms roughness) surface was obtained as a result of the CMP. A thin TiN layer was placed between the top surface of the GST and the top electrode. The programming endurance of such devices was an order of magnitude better than devices built without CMP. In addition, SET and RESET state resistance fluctuations along the cycling sequence were greatly reduced in the CMP-processed devices. The authors attributed this to the smooth surface and good quality TiN/GST interface leading to lower and more uniform contact resistance. They also suggested that better confinement of Joule heating and enhanced heat flux in the damascene structure was responsible for more uniform temperature distribution across the GST and thus more homogeneous material composition distribution as the cycling progressed.

The effect of adding oxidants such as H$_2$O$_2$ to the GST CMP slurry was studied by Zhong et al\cite{Zhong:2008b}. This study showed that while CMP of GST in a pure acidic silica slurry resulted in a rough surface with microscratches, the addition of 2 wt \% H$_2$O$_2$ allowed them to obtain a smooth surface that was free of damage. XPS analysis on an unpolished GST sample dipped into the oxidant for 10 min showed that the surface was oxidized. Though the component elements were oxidized by different amounts, i.e. Ge $>$ Sb $>$ Te, the CMP process was believed to be able to remove all of these oxides. The authors suggest that the GST CMP mechanism is similar to that of metals, i.e. oxidation followed by removal of this oxide due to friction with the abrasives in the slurry. The same group also compared `RIE' cleaning in an Ar plasma post-CMP of GST with conventional ultrasonic cleaning\cite{Zhong:2008}. An alkaline silica slurry was used to polish GST into sub-micron contact holes. AFM measurements showed that low surface roughness (0.64nm rms) was obtained after the CMP and RIE cleaning. DC I-V sweep measurements on fabricated devices also showed a significant reduction in the threshold switching voltage as a result of the RIE cleaning method.

Lee et al\cite{Lee:2007j} used CMP of GST in their demonstration of a scalable confined PCM cell concept. CVD GST deposited into high aspect-ratio 50\,nm diameter contact holes was planarized by CMP, thus fully confining the phase change material in the contact. SEM images showed that the GST CMP was successful with no scratches, excessive dishing, or void formation in the GST. Electrical tests on these devices showed low RESET currents and good cycling endurance.

\subsubsection{Process-induced damage}

A number of cell failures observed in fully integrated PCM chips have been attributed to process-related damage. Both the phase change material itself as well as its interfaces with the top and bottom electrodes are susceptible to degradation as a result of steps in the integration flow.

Lee et al\cite{Lee:2004a} introduced suitable interface cleaning processes in order to obtain good contact resistance in the TEC/GST/BEC current path and thus reduce the write current. They also noticed that edge damage in small GST cells could lead to increased initial cell resistance. Ahn et al\cite{Ahn:2004a} used nitrogen-doped GST so as to increase the dynamic resistance and lower the writing current. However, they found that higher nitrogen doping increased the cell's contact resistance and broadened its distribution across cells. They suggested that this was due to instability between the BEC and GST caused by the nitrogen doping and exacerbated by interface defects and the subsequent thermal processing. In addition, they observed that smaller cells exhibited a wider SET resistance distribution. This was attributed to the effects of contaminants and GST etch-related damage. By appropriate interface treatment, optimization of the GST etch process, minimization of process damage, and thermal budget reduction, they were able to obtain sharp SET and RESET resistance distributions.

As part of `product-level reliability verification' for 64Mb PCM chips, Kim et al\cite{Kim:2005ab} reported that the activation energy for loss of data retention (i.e. due to crystallization of amorphous volume) in fully processed PCM cells (2.1 eV) was lower than that measured on as-deposited films of the phase change material (2.46 eV). They suggested that the lower retention time in the fully processed cells could be attributed to higher nucleation probability resulting from processing damage on the GST or defects at the BEC/GST interface. In addition to developing an optimized etch chemistry for patterning the GST/TE stack, Oh et al\cite{Oh:2006b} used a line-type GST layout (as opposed to an island-type GST layout) thereby maximizing the size of the patterned GST region and thus reducing the effect of RIE damage on the GST switching volume.

There have been multiple reports on the development of confined PCM cell structures\cite{Cho:2005, Lee:2007j}. While the most striking advantage of such cells is the reduced RESET current, the authors also point out that these cells are also more resilient to sidewall edge damage during BEOL processes since the volume of material undergoing phase change is farther away from the damaged edge regions of the cell.

In particular, in their `on-axis confined cell' structure, Cho et al\cite{Cho:2005} confirmed that there was no degradation, such as an increase in SET resistance, as the cell size was reduced. Song et al\cite{Song:2006} pointed to the problem of increased SET resistance in devices due to oxygen penetration during BEOL processing steps that degraded the BE/GST interface through oxidation. In order to address this issue, they developed encapsulation processes for the GST cell. Operating under the constraints that the oxygen-blocking encapsulation layers should be scalable and should not chemically react with GST films or degrade their electrical properties during deposition, they investigated a variety of encapsulation schemes. They found that a `double-capping' technique worked best --- resulting in reduced SET resistance while still being able to achieve low RESET currents. Oh et al\cite{Oh:2006b} also used encapsulation in their 90nm diode-accessed high-density PCM demonstration.

While studying SET and RESET resistance distributions in 4Mb-level PCM cell arrays, Mantegazza et al\cite{Mantegazza:2006} found two kinds of anomalous cells that contributed to low resistance tails in the RESET distributions. One of those anomalous cell types showed saturation of RESET resistance at lower-than-desired values even if higher RESET currents were applied. These cells also had a distinctly different (higher) slope in the plot of threshold voltage (Vt) vs. cell resistance. The authors found that these cells could be modeled as having a conducting path in parallel with the amorphous plug. They pointed to contamination or impurities in the active PCM volume as likely causes. Solving this problem required a modified GST integration scheme that involved reducing the number of contamination sources, improving cell encapsulation, and cleaning the heater/PCM interface.

Modified PCM cell structures sometimes lead to distinct problems with GST integration and related failure modes. A case in point is the ring-type bottom electrode cell, introduced by Ahn and coworkers\cite{Ahn:2005a} in order to reduce the GST/bottom electrode contact area as well as its dependence on the patterned contact diameter. Subsequent researchers\cite{Song:2006, Ryoo:2007} pointed out that a failure mechanism in these ring bottom electrode cells involved the formation of a recess in the core dielectric (that is surrounded by the ring electrode) due to the wet cleaning solution used in the CMP process. This led to increased effective GST/BE contact area and resulted in reduced RESET resistance. The authors solved this problem by optimizing the CMP process and using a more resilient core dielectric.

In summary, addressing yield loss in fully integrated PCM chips requires that adequate care be taken to minimize process-related damage to the phase change material and its various interfaces. While the specific details of optimized integration flows are not easily gleaned from the published literature, some of  techniques that are usually mentioned include contamination reduction, GST/BEC interface optimization, minimizing the phase change material sidewall etch damage and/or keeping such damage as far away as practical from the active switching volume, and the use of encapsulating layers as protection from oxidation during BEOL processing.

\section{PCM reliability}\label{sec:PCMinOperation}

\subsection{Retention}\label{subsec:Retention}

In order for a nonvolatile memory candidate to be considered a viable next-generation memory option, it must demonstrate long term retention of stored data.  The typical criterion is 10 years (or 100,000 hours) at 85\degreesC, with fewer than 1 PPB (part per billion) retention failures at the array level over this entire lifetime, independent of the previous cycling history of the memory array\cite{Gleixner:2007,Gleixner:2007a}. Since the crystalline SET state is a stable low-resistance state, it is the stability of the quenched high-resistance RESET phase which dominates retention issues.  Not surprisingly. the stability of this phase has been investigated widely. As discussed earlier, the amorphous phase suffers from two independent resistance-altering processes: resistance drift and spontaneous crystallization. The drift process is a steady increase in the resistivity of the amorphous phase, related to structural rearrangement of the amorphous chalcogenide and the dynamics of intrinsic traps. Since this process increases the ON/OFF ratio, it does not cause any data-loss for binary PCM devices.

On the other hand, thermally activated crystallization of the amorphous material eventually leads to significant reduction in the resistance of the active layer, causing eventual retention failures for both binary and MLC storage. Data retention measurements typically involve monitoring the resistance of the cell that has been put in the RESET state, as shown in Figure~\ref{fig:retention:simpleLoss}. When the resistance of the cell falls below a threshold resistance (between the SET and RESET resistance values), the cell is said to have suffered a retention failure. Similar to accelerated tests used on other nonvolatile memory candidates, PCM retention measurements are done at higher temperatures to speed up this crystallization process and the subsequent resistance change. Using measurements done at a number of high temperatures (typically near 160\degreesC) and a reasonable activation model, the retention properties of PCM at 85\degreesC~can be predicted fairly well. Of course, the validity of the activation model and of the extrapolation from higher temperature measurements is also strongly dependent on the nucleation and grain-growth properties of the specific phase change material being used in the memory cell.

\FigureZ

Early work\cite{Lai:2003} used the activation energy of blanket GST (3.5 eV) to estimate that a 10 year lifetime at 120\degreesC~should be possible for PCM cells. Using data on mushroom PCM cells, Pirovano et. al.\cite{Pirovano:2004b} showed a retention activation energy of 2.6eV, adequate for more than 300 years of data retention at 85\degreesC.

Later Redaelli et. al.\cite{Redaelli:2005, Russo:2006, Redaelli:2006} proposed a percolation model for retention on 180~nm  $\mu$-trench PCM cells and analyzed this model using a temperature dependent percolation effect. This is the most widely accepted model for PCM retention. By showing that repeated retention measurements on the same device resulted in widely varying retention times, they demonstrated the stochastic component to GST crystallization discussed earlier. They proposed that this variation has its origin in the random spatial configuration of the as-nucleated grains in the amorphous region. As nucleation proceeds with time, the cell resistance decreases significantly when a percolation path finally appears through the amorphous layer. Since the crystalline state is so much more conductive, the occurrence of even a partial path through the amorphous plug can strongly reduce the overall device resistance.

Due to these percolation effects, the measured retention times of an ensemble of cells tend to obey Weibull statistics, where the cumulative distribution of retention times represent a line of constant slope $\beta$ on a Weibull plot (which plots the logarithm of the fraction of failed cells against the logarithm of elapsed time). It was also experimentally observed that $\beta$ increases (e.g., distributions become tighter) at lower measurement temperatures. This temperature dependency was explained using a Monte Carlo model which randomly generated crystalline nuclei and let them grow in the thin amorphous region (note that heterogeneous nucleation was neglected). By comparing this model to experimental data, they also showed that the effective grain size $r$ at most temperatures is higher than the as-nucleated grain radius, with the difference being attributed to grain growth. By assuming a temperature-activated Arrhenius model for failure times, retention exceeding 10 years at 105\degreesC~was predicted using this model. Below 120\degreesC, the model postulated that grain growth is negligible, so that the size $r$ of each grain is no larger than nucleation-limited FCC GST-crystal monomer radius.  Thus nucleation becomes solely responsible for retention failures, and this causes the Weibull slope $\beta$ to increase significantly. This also implies that the distributions of retention times could become very tight in the temperature regime of interest for retention (around 85\degreesC) which led them to conclude that PPB retention failures should still exceed 10 years at 103\degreesC. Modeling and simulations done on different PCM cell structures also showed that the retention properties are identical  when the effective amorphous layer thickness is the same.

The previous papers also assumed that failure times have an Arrhenius temperature dependency and that defect mechanisms played no role in tail-bit (PPB) failures. Russo et. al.\cite{Russo:2007a} showed that a pure Arrhenius extrapolation\cite{Redaelli:2005, Redaelli:2006} would be pessimistic in its estimation of retention and concluded that GST-based PCM cells should show data retention exceeding 10 years at 118\degreesC~ (instead of 10 years at 105\degreesC~shown earlier). This is largely because the energy barrier for nucleation is larger at higher temperatures (as the driving force for crystallization reduces), causing the increase of nucleation rate with temperature to be less than Arrhenian.

Gleixner et. al\cite{Gleixner:2007a} studied large-sized PCM arrays fabricated with 180\,nm and 90\,nm technology in order to study new defect failure modes and retention loss at the PPB level. They noticed that even at the lowest times used to measure retention (at elevated temperatures), a small fraction of the bits ($<$ PPM) had already failed, indicating that the time to failure for the first cell is not accurately predicted by the failure curve. It was also shown that these tail bits show similar initial resistances to the nominal bits and similar activation energy of the time-to-failure (2.4 eV). Furthermore these tail bits show a Weibull distribution consistent with a weak-link failure mechanism, in which rare combinations of closely-set nuclei can lead to rapid retention loss.  Although the same fraction of bits fail (to within 10\%) during each test, it was different bits that occupy this tail (and thus fail) each time.  This indicates that manufacturing defects were not responsible for these tail-bit failures. It was therefore postulated that the most likely cause of this failure comes from a scenario where the nucleation sites are arranged such that when thermal energy is applied, very little growth is required before a quick resistance decrease can be observed. Data also showed that cycling has no impact on retention --- if anything a slight improvement was observed. Gleixner et. al\cite{Gleixner:2007a} also showed that optimization of the process and the write scheme could be used to significantly suppress these tail bits, although the exact nature of this optimization was not discussed in detail.

Work done by Shih et. al\cite{Shih:2008}, albeit with slightly worse data retention, concluded that a large fraction of the bits fail due to grain growth from the amorphous / crystalline boundary. While this retention loss mechanism has not been observed before (presumably due to a reduction in grain growth velocity at lower temperatures), these differences could be attributed to differences in nucleation and growth properties of the materials used in the different experiments. Clearly, reducing the crystal growth component at retention temperatures would be an important step in improving the retention times.

Finally, Redaelli et. al\cite{Redaelli:2008} showed that the value of the threshold resistance that is chosen to define the retention failure criterion directly impacts the calculation of the activation energy ($E_A$) extracted for failure times.  This might explain the wide variability in previously reported $E_A$ values, ranging from 1.9 eV to 3.5 eV. They concluded that reliable extraction of the activation energy of crystallization can be obtained by using a critical threshold resistance that is close to the crystalline resistance, or by performing all resistance measurements at room temperature.

Finally, it should be mentioned that while much of the work described here deals with GST, the exact composition and doping of the GST that has been explored by different groups is very likely different.  This itself could alter the value of the measured activation energy. A number of groups have also investigated alternate PCM materials and doping as a way to increase the activation energy for crystallization and to increase retention margins. Matsuzaki et. al\cite{Matsuzaki:2005} showed that oxygen doping of GST films results in a much higher activation energy due to a smaller grain size --- which leads to improved retention.  Kim et. al\cite{Kim:2005ab} also showed that N-doped GST single-cell PCM devices show retention lifetimes exceeding 10 years at 85\degreesC. They also concluded that the activation energy is lower for devices compared to blanket films because of process-induced damage and defects at the interface between the bottom electrode and the GST --- effects that have to be minimized in order to maximize retention. Morikawa et. al\cite{Morikawa:2007} have also explored In-Ge-Te as a phase change material and have shown that its retention properties are significantly better than GST, ranging from 10 years at 122\degreesC~to 156\degreesC~depending on the Indium fraction. Other materials explored for improved PCM retention include doped GeSb\cite{Chen:2006t}, Si-doped Sb$_2$Te$_3$\cite{Lin:2007b} and Si$_x$Sb$_{1-x}$\cite{Zhang:2007i},\cite{Zhang:2008b}.  However, only very basic materials studies have been carried out for most of these new materials, and detailed array data such as the study of tail bits is lacking. Further materials development will be key for improved retention as the size of the amorphous plug shrinks with scaling, as well as for improved tail-bit retention, MLC capability and proximity-disturb performance.

\subsection{Crosstalk}\label{subsec:crosstalk}

From the discussion in Section~\ref{subsec:CellStructures}, it is clear that during the RESET operation the peak temperature within a PCM device exceeds the melting point of the phase change material.  In fact, since
the extent of the amorphous plug quenched from the molten state must be larger than the critical dimension of the limiting aperture,
the material at the edge of this aperture is just over the melting point, and the peak temperature in the center of the cell is well over the melting temperature.
As an example, Ge$_2$Sb$_2$Te$_5$ melts at $\sim$630\degreesC~\cite{Yamada:1991}, and other phase-change materials have similar melting temperatures\cite{RaouxWuttigBook:2009}.  However, from Section~\ref{subsec:Retention}, any significant exposure of this amorphous plug (once it has been formed) to temperatures exceeding 150\degreesC~ or so will lead to the recrystallization of enough crystalline nuclei within the plug to induce data loss.

It is thus not surprising that, upon learning these two facts, many engineers introduced to PCM immediately ask about the thermal crosstalk between cells.  In particular, the worst-case scenario is the effect on an amorphous plug encoding the RESET state when one or more of the immediately neighboring cells is programmed through multiple RESET-SET cycles.  What is truly surprising is that, for most researchers working in the PCM field, thermal crosstalk or ``proximity disturb'' is considered to be a second--order rather than first--order problem. In fact, the small amount of empirical data available on proximity is focussed primarily on the {\em absence} of any crosstalk effects\cite{Pirovano:2004b,Kang:2008}.

Pirovano et. al. showed with careful thermal simulations that, at least out to the 65nm node, the thermal crosstalk between cells should remain low enough that 10-year lifetime will not be significantly reduced by write disturbs from neighboring cells\cite{Pirovano:2003}.  Figure~\ref{fig:retention:PirovanoSimulations} shows their simulated temperature profiles for the 180nm and 65nm nodes, assuming ``micro-trench''-type PCM cells.  The expected temperature rise at the position of the neighboring cell remains well below 100\degreesC.  Russo et. al. did a similar study extending down to the 16nm node\cite{Russo:2008a}.  They found that isotropically-scaled devices, where all cell dimensions scale with the technology node, can be expected to experience no thermal crosstalk problems.  However, because their cell designs had been optimized to emphasize low RESET current without any consideration to efficiency, RESET power or proximity effects\cite{Russo:2008}, the hottest point in their cells is deep within the highly resistive heater electrode (both for the mushroom cell, which they refer to as a ``lance'' cell, and to a lesser extent for the micro-trench cell).  As a result, the temperature at the neighboring cell grows rapidly if only the lateral dimensions are scaled (non-isotropic scaling).  Despite these highly power-inefficient design points, though, Russo et. al. found it straightforward to avoid thermal crosstalk simply by choosing a mixed scaling approach that decreased both the lateral and thickness dimensions\cite{Russo:2008a}.

\FigureAA

One observation made by Russo et. al. is that as the spacing between devices drops,
the time dimension becomes much less effective as an avenue for avoiding proximity issues\cite{Russo:2008a}.  As the size of the heated volume decreases, the thermal time-constant $\tau_{th}$ decreases, but the characteristic thermal diffusion length $L_{th}$ drops more slowly (scaling as $\sqrt{\tau_{th}}$).  Thus as the technology node scales down, the neighboring device eventually moves inside the characteristic thermal diffusion length.  This effect is noticeable even in the data of Pirovano et. al. -- in Figure~\ref{fig:retention:PirovanoSimulations}(a), one can avoid experiencing the maximum steady-state temperature change by simply using a very short RESET pulse, while in Figure~\ref{fig:retention:PirovanoSimulations}(b), the difference between the temperature rise due to a short RESET pulse and the steady-state temperature rise has become much smaller.

To a certain extent, knowledge of the temperature to which neighboring cells will be heated during RESET can be combined with
retention measurements which reveal a failure time (curves of $t_X$ vs. 1/$k_B T$\cite{Russo:2007a, Redaelli:2005, Redaelli:2006}) in order to estimate the effect of proximity disturb. The additive effects of retention and proximity both need to be considered to account for data integrity loss in high density PCM based crosspoint memory arrays.  In particular, the presence of even modest proximity effects can push the temperature for which acceptable retention is required significantly higher than the actual maximum average operating temperature of the memory chip.

However, it is important to keep in mind that the temperature distribution during a retention-failure measurement is completely homogeneous across the PCM cell.  In contrast, in proximity disturb situations, the temperatures are strong functions both of position within the neighboring PCM cell and of time, due to the fast thermal transients during RESET pulses and any slow ramp-downs used at the end of long SET pulses.  This is particularly relevant when considering that early-to-fail retention problems are attributed to the unpleasantly rapid generation of a percolation path that connects a statistically--rare chain of crystalline nuclei.  For proximity disturb, this unlucky combination of closely-set nuclei could only lead to trouble if it were also located at the extreme edge of the amorphous plug, where the maximum temperature excursion generated by the frequently-cycled neighbor could then rapidly drive retention failure.  This additional criterion will further suppress the likelihood of such an initial failure, but it also complicates the incorporation of such early-to-fail retention data together with modeled temperature distributions (within the ``neighboring'' PCM cell) for the accurate prediction of proximity disturb.

\subsection{Endurance}\label{subsec:Endurance}

Cycling endurance has long been one of the strengths of PCM, especially in comparison to established Flash technologies, where Stress-Induced Leakage Current (SILC) frequently limits device endurance to 10$^4$-10$^5$ program-erase cycles.
The demonstration, as early as 2001, of 10$^{12}$ SET-RESET cycles in PCM devices without any significant degradation of resistance contrast, as shown in Figure~\ref{fig:variability:CyclingPerformance}\cite{Lai:2001}, was almost certainly a significant factor in the surge of interest in PCM technology that followed.  Of course, while it is telling that single devices could be operated reliably for so many cycles, the more critical question is what happens to the worst-case device in a large array.  Subsequent large-scale PCM integration experiments have tended to show endurance numbers in the range of 10$^8$-10$^{10}$ cycles\cite{Kim:2005ab,Gleixner:2007} --- still easily exceeding the endurance of Flash, but coming somewhat short of what would be necessary for DRAM replacement without wear-leveling (Equation~\ref{eq:wearLeveling}).

\FigureBB

Two different failure modes have been observed to occur after cycling, termed ``stuck-RESET'' and ``stuck-SET'' failure\cite{Kim:2005ab,Gleixner:2007}, as illustrated by Figure~\ref{fig:endurance:cyclingFailures}.  In a stuck-RESET failure, the device resistance suddenly and irretrievably spikes, entering a ``blown-fuse'' state that is much more resistive than the RESET state. This sometimes occurs after some degradation in resistance contrast (as in Figure~\ref{fig:endurance:cyclingFailures}), but can also suddenly occur, with no prior indication that failure is imminent. These failures are typically attributed to void formation or delamination that catastrophically severs the electrical path through the device, typically at a material interface such as the heater-to-GST contact in a mushroom cell.

In contrast, in a stuck-SET failure, a gradual degradation of resistance contrast is typically observed, as if the cell were being slowly but inexorably altered by the SET-RESET cycling. The cell seems to change its characteristics so much that the original RESET pulse is, at some later point in time, somehow less effective at creating an amorphous plug in the device than before.  Eventually, the RESET step of the cycling fails to induce any change in the device resistance, and the device becomes ``stuck'' in the SET state.

\FigureCC

Typically, for a cell in this stuck-SET condition, a larger amplitude RESET pulse proves sufficient to RESET the device and cycling can resume, albeit with a larger RESET power. However, this continued operation inevitably hastens the onset of a stuck-RESET failure.  Figure~\ref{fig:endurance:vsEnergy} shows the strong correlation between pulse energy and device failure: every order-of-magnitude increase in pulse energy implies three orders-of-magnitude lower endurance\cite{Lai:2003}.  Unfortunately, it is not clear from
Figure~\ref{fig:endurance:vsEnergy} whether it is pulse amplitude or pulse duration that is critical, nor does the plot differentiate between stuck-SET and stuck-RESET failure.

Fortunately, Goux et. al. have carefully studied endurance failure due to stuck-SET in phase change bridge cells\cite{Goux:2009a}.  By measuring the resistance-vs-current curves at various points during cycling, they clearly demonstrated that stuck-SET failure is due to a change in the RESET condition (i.e., the pulse amplitude required for RESET) that is induced by cycling. They also observed that while pulse amplitude had a fairly minor impact on device endurance, pulse duration had a strong effect. By using pulses ranging from 10ns to 10$\mu$s in length, they were able to show that the time-spent-melting their PCM material (in their case, Ge-doped SbTe) was the critical factor.  Their endurance data suggests that endurance scales inversely with $t_m^{3/2}$, where $t_m$ is the time-spent-melting during each RESET pulse.  The way to reach a very large number of SET-RESET cycles is thus to minimize the time-spent-melting during each RESET pulse.  This data also fits with observations that repeated cycling with only SET pulses shows greatly extended endurance ($>$ 10$^{12}$ cycles) over RESET-SET cycling (10$^{10}$)\cite{Kim:2005ab}.  Together these results imply that the gradual cell degradation associated with stuck-SET is strongly correlated to the melting inherent in each RESET operation.

A number of groups have been using techniques such as Energy Dispersion Spectroscopy (EDS), Secondary Ion Mass Spectrometry (SIMS), and Energy Dispersive X-ray (EDX) spectrometry to perform elemental analysis
on failed cells\cite{Park:2007, Rajendran:2008a, Ryu:2006, Krusin-Elbaum:2007, Yoon:2007, Yoon:2007b, Lee:2008c, Nam:2008, Kang:2009a, Kim:2009a, Nam:2009, Yang:2009c, Yang:2009d}.
Measurements on mushroom cells built from Ge$_2$Sb$_2$Te$_5$ material\cite{Park:2007, Rajendran:2008a} tend to show agglomeration of Antimony (Sb) at the bottom electrode at the expense of Tellurium (Te).  (The tendency of Germanium (Ge) is
not clear --- Reference~\cite{Rajendran:2008a} shows it clearly depleting from the repeatedly molten mushroom cap over the heater, while Reference~\cite{Park:2007} indicates no motion).  Sarkar et. al. used these results to explain why their Ge$_2$Sb$_2$Te$_5$ material actually improves slightly, in characteristics such as resistance contrast and RESET current, over the first 10,000 SET-RESET cycles\cite{Sarkar:2007}.  They surmise that as the phase-change material evolves in composition through cycling, the volume melted by each RESET pulse increases (increasing RESET resistance) while the inherent lower resistivity of the more Sb-rich GST material provides a lower SET resistance.

Eventually, however, it would appear that such compositional changes driven by cycling will steadily decrease the dynamic resistance of the active region, shifting the required RESET current to larger values.  This leads to stuck-SET failure if the RESET pulse is not adaptively increased, or to stuck-RESET failure if the RESET pulse energy is increased to compensate.

\subsection{Polarity issues}\label{subsec:Polarity}

A number of recent measurements of phase segregation, in various types of
phase change bridge devices, have forced a re-interpretation of the failure analysis results just described within the context of two previously unknown bias-polarity-dependent effects.

\FigureDD

Tio Castro et. al. showed convincing top-down TEMs of bridge devices in the RESET state (Figure~\ref{fig:polarity:TioCastro}) that demonstrate that the amorphous plugs in their devices were shifted by the polarity of the applied bias, by as much as 100nm\cite{TioCastro:2007}.  They attributed this effect to the thermo-electric Thomson effect, in which the overlap of temperature gradients with electrical current can lead to additional heat generation or absorption.  Because the hot spot in the center of a phase-change device is surrounded by
temperature gradients of opposite signs but the current flow is uni-directional, the Thomson effect acts to shift the centroid of the hot spot depending on the polarity of the applied voltage.  Tio Castro et. al. estimated from their observations that the Thomson coefficient in their material might be in the range of -100$\mu$V/K\cite{TioCastro:2007}.

In those same experiments, it was observed that for intentionally-asymmetric ``dog-bone'' bridge devices (somewhat like a high-aspect-ratio pore device on its side), there was a ``bad'' polarity of operation (large-area electrode negative and small-area electrode positive) for which subsequent SET operations were unable to return the device to low resistance. In symmetric bridge devices, other researchers have reported that the most reliable SET operations can only be produced by {\em alternating} the polarity between SET and RESET, with little dependence on the absolute sign of the bias polarity\cite{Lin:2008a}.  Tellingly, these results were only observed for bridges fabricated from GST, and not for ultra-thin bridges fabricated from doped--GeSb\cite{Chen:2006t}.

Similarly, other researchers have reported bias-dependent operation of pore devices where only the ``good'' choice of polarity (positive on large-area electrode) can be used to produce low SET resistances, while the ``bad'' polarity is associated with ``hard-to-SET'' operation\cite{Lee:2008f}.  In these experiments, only a narrow voltage window could be used for the SET operation, which also required longer pulses and which never produced the same low SET resistances as the ``good'' polarity\cite{Lee:2008f}.

Fortunately, this ``good'' polarity corresponds exactly to the typical operation of integrated PCM devices where the positive voltage is applied to top of the PCM device built over the underlying transistor\cite{Happ:2006}.  In fact, operation of such integrated devices in the ``bad'' polarity is difficult to study, since in that configuration the
gate-to-source voltage changes dynamically during each pulse  as the PCM device resistance changes.  Thus it is not surprising that such effects have not been widely reported for PCM devices integrated together with access transistors.

The Thomson effect would seem to be inadequate by itself to explain all of these bias-polarity effects.  However, a few groups have been performing bias-dependent failure analysis experiments on various types of phase change bridge devices.  Early versions of these experiments were affected by lingering uncertainties related
to the role of metallic electrodes at the failure point\cite{Nam:2008}, and to the difficulty in
understanding material desegregation during long-term cycling by studying the aftermath of a
single-pulse ``blown-fuse'' failure\cite{Nam:2008, Kang:2009a, Nam:2009}.  However, the most recent
experiments have investigated the cycling failure of tapered bridge structures located far from any metal electrodes\cite{Kim:2009a}, and the controlled fast melting of large symmetric bridge devices\cite{Yang:2009c, Yang:2009d}.

The results of these experiments, together with the earlier failure analysis data on mushroom devices\cite{Park:2007, Rajendran:2008a}, sketch out a convincingly consistent story for Ge$_2$Sb$_2$Te$_5$ devices: Te moves towards the positive electrode (anode), while Sb moves toward the negative electrode (cathode)\cite{Park:2007, Rajendran:2008a, Nam:2008, Kang:2009a, Nam:2009, Kim:2009a, Yang:2009c, Yang:2009d}.  This motion is attributed to the higher electronegativity (5.49eV) of Te compared to Ge and Sb (4.6eV and 4.85eV)\cite{Yang:2009d}.
Most of the data seem to indicate that Ge moves together with the Sb towards the cathode, but as mentioned earlier, there are some data which indicate otherwise.  In connecting these interpretations of bridge and mushroom devices, we assume that the mushroom cycling was performed in the ``good'' polarity direction, with positive voltage on the large-area top electrode.

\FigureEE

One particularly interesting study was performed by Yang et. al. on very large symmetric devices (20$\mu$m long by 2$\mu$m wide bridges of 300nm-thick Ge$_2$Sb$_2$Te$_5$), where both high-amplitude millisecond-long pulses and day-long exposure to low-amplitude 10Mhz pulsed DC were used to explore elemental segregation through Wavelength Dispersive Spectroscopy (WDS)\cite{Yang:2009d}. They were able to show that the material segregation is very rapid in the molten state, observing nearly complete desegregation along a 10$\mu$m length of bridge after a 1.5ms pulse, as shown in Figure~\ref{fig:polarity:rapidDesegregation}.  This works out to an
effective diffusion coefficient of 1-2 $\times 10^{-5}$ cm$^2$/s\cite{Yang:2009d}, which roughly corresponds to a field- or current-driven migration of one nanometer every nanosecond.  In contrast, their long-term measurements of bias-induced elemental desegregation through the crystalline state seemed to suggest diffusion coefficients nine orders of magnitude lower\cite{Yang:2009d}, implying that the drift of elements through that same one nanometer would take one second.

While it is not yet clear how these observed effects (the Thomson effect and polarity-dependent elemental segregation via electromigration) can be combined into a unified theory that quantitatively explains PCM polarity and cycling endurance, it is already quite clear that such bias-polarity effects and cycling endurance are intimately related. It has already been independently observed that switching from one cycling polarity to the other can be used to continue cycling of bridges after a stuck-SET failure\cite{Goux:2009a}.  In fact, as shown in Figure~\ref{fig:polarity:rescuePulses}, even as few as 10 aggressive pulses applied in the opposite polarity direction can allow cycling to not only resume after a stuck-SET failure, but to continue for another 10$^5$ cycles with the original RESET pulse conditions\cite{Lee:2008k, Lee:2009a}.

\FigureFF

Continued understanding of what differentiates the ``good'' and ``bad'' polarity should allow researchers to continue to improve PCM cycling endurance, through a combination of creative use of the ``bad'' polarity\cite{Goux:2009a, Lee:2008k, Lee:2009a}, improved cell design, and new materials that show greater resistance to elemental segregation.  As with Section~\ref{subsec:Processing}, most of the detailed data available is limited to GST because of its ubiquity.  However, it has been observed that the phase change material GeSb phase-segregates quite readily\cite{Cabral:2008}, implying that simply reducing the number of atomic species involved is not necessarily the best approach.

\section{The future of PCM}\label{sec:FutureOfPCM}
\subsection{Multi-level cells (MLC)}\label{subsec:MLC}

Judging from the recent history of the aggressive nonvolatile memory market currently dominated by Flash, it is clear that all available directions for improving effective density (e.g., the average number of information bits that can be stored per unit area) will be exploited. One direction, both promising and challenging, is that of the so-called multilevel cell (MLC) technology, which exploits the intrinsic capability of a memory cell to store analog data in order to encode more than one bit of digital data per cell. The feasibility of MLC for PCM has already been shown\cite{Nirschl:2007, Bedeschi:2008, Kang:2008,Bedeschi:2009}, including the demonstration of programming into both 4 and 16 sharply distinct analog levels, corresponding to two and four bits per cell, respectively. Such intermediate resistance levels are obtained by properly modulating the electrical signals used to program the PCM element. In Figure~\ref{fig:MLC:pulseFamilies}, two examples of such signals are shown: (a) rectangular current pulses of different height $h$ and width $w$\cite{Bedeschi:2008,Bedeschi:2009} and (b) variable slope pulses, with different duration $d$ of the trailing edge of a trapezoidal pulse\cite{Nirschl:2007}. By controlling these parameters carefully, one can control the analog resistance of the PCM element and thus enable MLC operation.

\FigureGG

The degree of success of such an MLC writing scheme can be characterized by the resistance distributions over a large ensemble of PCM devices. Figure~\ref{fig:MLC:histograms} illustrates this concept: each of the four levels labeled ``00,'' ``01,'' ``10,'' and ``11'' is associated with a resistance distribution. In a perfect world, these distributions would be delta functions, simplifying the classification process into a straightforward thresholding operation. If the distributions overlap, however, then there is a non-zero probability of level mis-detection at the receiver, resulting in the retrieval of erroneous data. The use of the logarithm of the resistance is expedient to obtain more uniform shapes of the distributions across all levels.  However, it should be pointed out that since these levels are classified using read current, the optimal configuration may not necessarily call for spacings between levels that are uniform in either resistance or log(resistance).

\FigureHH

There are several factors that can limit the number of effective levels that can be reliably stored in a PCM cell. Among them are
\begin{itemize}
\item The intrinsic randomness associated with each write attempt, or \emph{write noise};
\item Resistance drift, which we'll refer to as \emph{short term drift};
\item Array variability, which includes any variability during the lifetime of the PCM array;
\item Crystallization of the amorphous phase, which we'll refer to as \emph{long term drift}.
\end{itemize}

Some of these factors, such as short and long term drift, represent a fundamental limitation to the \emph{storage capacity}---the maximum number of bits that can be stored in the average PCM cell. As such, these factors cannot be overcome but only mitigated, as we will discuss later.

In contrast, factors such as write noise and, to a certain extent, array variability do not directly limit the storage capacity, but instead make it harder to achieve a given storage capacity. In order to deal with these kinds of limiting factors, resistance distribution tightening techniques have been developed based on \emph{write-and-verify} procedures. These iterative techniques consist of applying programming pulses and verifying that a specified precision criterion is met, along the lines of what is currently done in Flash memories\cite{Grossi:2003}. These methods have been used to demonstrate 16 level PCM\cite{Nirschl:2007}, 4 level PCM\cite{Nirschl:2007, Bedeschi:2008, Kang:2008,Bedeschi:2009}, and to tighten the distribution of SET state resistances for binary PCM\cite{Mantegazza:2008}.

The effect of a write-and-verify technique is to reshape the conditional probability density function, as shown in Figure~\ref{fig:MLC:writeAndVerify}.  This can be obtained by successively refining\cite{Philipp:2008} the write procedure until the verify step finds the resistance value within the desired range around the nominal resistance target. When properly used, a write-and-verify algorithm produces tighter resistance distributions (compare Figure~\ref{fig:variability:BeforeMLCdistributions} to Figure~\ref{fig:variability:AfterMLCdistributions}), and therefore allows the packing of more MLC levels into the same resistance range.  This increase in the number of levels obtained with write-and-verify reflects an actual increase in the information-theoretic storage capacity. There exists an intricate tradeoff between storage capacity and the average number of write-and-verify iterations. In particular, References [228,229] show that, for a simple cell model affected by write noise, the achievable storage capacity tends to increase logarithmically with the number of write-and-verify iterations\cite{Lastras-Montano:2008, Mittelholzer:2009}. This logarithmic increase is expected to hold even for more realistic cell models at a
sufficiently large number of write iterations.

\FigureII

Among the factors representing a fundamental limitation to the storage capacity, resistance drift plays an important role. Short term drift manifests itself as a slow but steady increase of the resistivity of the amorphous material (see Reference~\cite{Pirovano:2004a} and references therein). The resistance drift has been shown to follow a power law,
\begin{equation}
R(t) \;=\; R_0 \, \left( \frac{t}{t_0} \right)^{\nu}
\end{equation}
where $R(t)$ denotes the resistance at time $t$, $R_0$ denotes the initial resistance at time $t_0$, and $\nu$ is a drift coefficient\cite{Ielmini:2008c}. Typical values for $\nu$ for thin amorphous GST layers are on the order of 0.05-0.1\cite{Pirovano:2004a}. This phenomenon has been explained in terms of structural relaxation in the amorphous material influencing a Poole or Poole-Frenkel conduction\cite{Ielmini:2007b, Ielmini:2008c}, and in terms of kinetics of electrically active defects in the amorphous GST material\cite{Redaelli:2008c}. The drift process is fairly predictable in the case of thin films of amorphous GST, which would suggest that a few cells of known state in a block that evolved in time together might serve to identify the needed shifts in threshold bias.  However, short term drift appears to be a random process, which can be expected to vary from cell to cell\cite{Mantegazza:2007,Ielmini:2007a, Redaelli:2008c, Kang:2008}.
By introducing yet another source of unpredictability, this short-term drift reduces the effective storage capacity of PCM. The phenomenon can be perceived as a broadening of the resistance distributions over time.
Figure~\ref{fig:MLC:KangData} compares the cumulative distributions for four resistance levels measured immediately after programming to the distributions after programmed cells have been drifting, both at room temperature and then at elevated temperature\cite{Kang:2008}.

\FigureJJ

A number of techniques have been proposed for coping with drift. These include changing the write target resistances to take into account the expected broadening of the resistance distributions due to drift\cite{Kang:2008}, and compensation techniques at read time, where pulses are used upon readout to return the device to its initial as-written resistance (presumably without accidentally reprogramming the cell)\cite{Kostylev:2005}. We remark that these topics are the subject of current active research in the PCM research community.

\subsection{Role of coding}

Although little or no published literature yet exists on coding techniques designed expressly for PCM, the success that these techniques have had in established memory and storage technologies would imply that coding will be extensively used in PCM. Moreover, careful adaptation of existing codes  as well as development of new coding technologies suited to the physical characteristics of PCM could prove essential to unlocking much of its inherent potential.

Although there are many different types of coding, the most prevalent technique is Error Correcting Coding (ECC),  which allows for the detection and correction of bit or symbol errors. Other coding techniques, called modulation codes,  are designed to ensure that the patterns stored in memory are adapted to particular characteristics of the physical medium. Here we discuss how both of these approaches might be incorporated into PCM, with emphasis on ECC.

ECC technology is now a standard feature in most storage and high-end computing systems, finding applications in caches and buffers built using SRAM, main memory which generally uses DRAM, and hard disks, solid state drives and  tape.  The sophistication of the coding technology employed at each layer is often inversely related to the proximity of a memory technology to the computing element.  Much of this depends on the speed with which such coding can be implemented --- close to the processor, a few extra nanoseconds spent on decoding may represent a significant (and unacceptable) delay, while far from the processor those same nanoseconds represent a tiny rounding error.  Processor caches tend to utilize regular or extended binary Hamming codes\cite{Hamming}, which are arguably among the simplest codes that can be found in a pervasive manner.  In contrast,  main memory uses symbol-based codes such as Reed-Solomon codes\cite{ReedSolomon}, while disks use extremely sophisticated constructions involving error correcting codes, modulation codes, and advanced signal-detection and signal-processing technology.  Since PCM holds promise both as a storage and as a memory device, it is reasonable to expect that it will draw coding and signal processing ideas from all of these technologies.

To begin with a relatively simple example, consider a single-bit PCM cell. Here, a zero or a one correspond to a cell being SET or RESET, with the dramatic difference in resistance that accompanies these two states.  Detection of a zero or one can be accomplished by a simple ``hard decision'' based on a resistance threshold in between these two resistance states based on read current.  In PCM, one of the complications is that the measured resistance may change over time (short term drift, see Section~\ref{subsec:MLC}) or with the instantaneous temperature of the cell, requiring a threshold that can be shifted with time and temperature.  However, since the short-term drift is associated with the amorphous phase, if the SET state resistance is sufficiently dominated by the crystalline resistivity, then its resistance can be considered relatively independent of elapsed time.  (Note that temperature-dependent thresholds can be produced by circuits designed around inherent temperature-dependent changes in the silicon underlying the PCM devices.) Thus for binary storage, simple thresholding may prove adequate. Another solution is to have ``pilot cells'' that are known to be in the RESET and/or SET states.  These pilot cells are programmed and read at the same time as the data-bearing cells in the same block, and thus can convey the information necessary to construct a suitable threshold for discriminating between a zero and a one.  This is where device-to-device variability in this short-term drift proves to be the real issue.

Errors in data that has been retrieved can have their origin in either a failure of the cell to switch to the desired state, an erroneous reading caused by noise, quantization or resistance fluctuations, or in the state of a cell switching over time. As discussed in Section~\ref{subsec:Endurance}, the most likely such state-change event is the gradual transition from RESET to SET caused by crystallization of the phase change material.  Although it is difficult to place a definite bound on the number of errors that these problem sources will cause, it is likely a safe statement to say that standard coding and decoding techniques such as Reed-Solomon codes based on the BCH (Bose, Ray-Chaudhuri, and Hocquenghem) code family\cite{BCH1,BCH2,BCH3} will be adequate to address these problems for a good number of applications.

Having stated this, it is entirely possible that more sophisticated error control coding techniques may find their way even in single bit PCM. These techniques would allow a greater number of errors to be corrected for a given number of redundant (``check'') bits. The benefits of this include extending of the lifetime of PCM.  In fact, as currently happens in Flash memory, the error rate of PCM devices can be expected to increase gradually during the lifetime of the memory, due to wearout mechanisms whose physics are currently not completely understood. A higher error correction capability would imply, in  this case, a longer lifetime for the memory.   Other benefits include the possibility of more relaxed engineering requirements on a cell's expected physical behavior, which could greatly accelerate the introduction of PCM in the marketplace. Examples of such relevant coding techniques include  enhancing traditional algebraic coding techniques with soft decoding capabilities, as well as using powerful coding mechanisms such as Low Density Parity Check (LDPC) codes with iterative decoding methods\cite{LDPC}. The additional complexity demanded by these newer techniques may be well within reach given the significant progress that has been achieved both from the algorithmic and logic device technology fronts.

The development of Multibit PCM is a significant engineering challenge, in many ways similar to the challenges faced by multibit NAND Flash manufacturers.
However, unlike block--based Flash memories, single PCM bits can be erased and re-programmed.
As discussed previously, due to variability in the response of different cells to the same input signal, as well as the smaller yet still significant variability in the response of the same cell to repeated applications of the same input signal, it appears impractical, at least presently, to attain a desired resistance level in a PCM cell by the application of a single write pulse.  Instead, write-and-verify techniques will be necessary to sharpen the distribution of the outcome of each write procedure. A write-and-verify procedure is associated with a probability of failure, because even after exhausting the allowed resources (in terms of time, iterations, energy, etc), the resistance of a cell may still not be within the desired range around the target value. This probability of failure generally decreases as more resources are devoted to the  write procedure, but in general will not be negligible.

Error control coding can be pivotal in the management of these iterative write failures, provided that a good upper estimate can be established for how often they are expected to happen.   One possibility for handling these errors is to employ non-binary codes, that is, codes that detect and correct errors in symbols with more than 2 states. For example, a 4-bit/cell PCM may employ BCH codes defined over 4-bit symbols. If reliable information on the shape of the output distribution of an iterative write procedure is available, soft decoding procedures can be employed to improve the likelihood of successful decoding of data.

The change of programmed resistance due to short and long term drift are further exacerbated in the case of multibit PCM, and constitute the fundamental limitation in storage capacity. The effects of both of these phenomena are limited whenever the PCM devices have access to a reliable  power source that allows them to do regular refreshes (known as ``scrubbings'') over time.  This is often the case for PCM devices employed in a memory context.  In the case of a device intended for storage applications, no such guarantee of refresh power can be assumed. Thus the problem of recovering the stored information is markedly harder.  Fortunately, the relatively relaxed bandwidth and latency requirements demanded by storage applications allow for the possibility of more complex processing at a decoder. Such more complex processing can in general include signal processing to recover levels that have drifted (for short and long term drift) as well as advanced error control coding techniques that might incorporate soft information  from the drift recovery layer to enhance decoding success.

From the perspective of a read operation, after adjustments for drift have taken place, a multibit PCM cell appears to be an analog write medium with some noise around a written level, with restrictions on the minimum and maximum value we might write on the medium. Many techniques can be borrowed from communication system theory that are relevant to this setting. For example, a technique that might prove relevant is the notion of \emph{trellis coded modulation} (TCM).  At a very high level, TCM is a method for obtaining highly reliable multibit cells that exploits the idea of writing coded levels in the memory at a precision higher than that ultimately intended. The specific manner in which this coding is designed is one of the cornerstone successes of communication theory\cite{Trellis}.

A complementary idea is the concept of adding cells with redundant content, rather than writing more precise signals, which is a paradigm that is much more accepted in the memory community as it matches established methods for designing reliable memories. The correct balance between these two kinds of redundancies in a memory system design will ultimately depend on the design point for the memory, including expected density, power, bandwidth, latency, etc.
One example of the options available here is ``endurance'' coding, in which effective PCM device endurance can be greatly extended by using full RESET pulses which melt the PCM material only sparingly\cite{Lastras-Montano:2008}, albeit at the cost of reduced storage capacity. Once the design guidelines are established and available technologies for doing iterative programming and analog to digital conversion are put forth, it is possible to objectively identify the best method for designing the various redundancies that will be necessary for the attainment of an extremely reliable and high density PCM system.

\subsection{Routes to ultra-high density}\label{subsec:ultraHighDensity}

As discussed earlier (Section~\ref{subsec:potentialApplications}), the cost of a semiconductor technology depends strongly on its device density.  Even though PCM technology already appears to have a good chance of matching or exceeding Flash technology in terms of performance and endurance, neither of these will matter if the cost of PCM does not (eventually) match or improve upon Flash.  For instance, one possible scenario might find PCM perpetually more expensive than Flash, either because of cost issues related to large cell size or reliability issues that dampen achievable yield.  Along this path, the future of PCM is dim --- few customers will be willing to pay more for the better performance and/or endurance of what is essentially a new, unproven, and low-volume technology.  But in the alternative scenario, where PCM can pass Flash in terms of cost, then not only would PCM be able to compete in all the markets that Flash now occupies, but it would be immediately more suitable for Solid-State Disk devices and other not-yet-existing Storage Class Memory applications that may develop.

Thus the eventual cost of PCM technology is absolutely key.  While some of this depends on high-yield processing of robust PCM memory devices in high-volume manufacturing, a significant component of the cost equation depends on implementation of ultra-high density.  In particular, it is already clear that even 1 bit per 4F$^2$ will not catch up with NAND flash, since MLC Flash is already 2$\times$ better than this and moving towards 4$\times$ higher densities\cite{ITRS:2008}.

Thus other techniques must be invoked in order to achieve the ultra-high memory
densities that PCM will need, both in order to succeed as a successor to Flash and to enable new Storage Class Memory applications. We have already extensively discussed one of these, that is, multiple bits per cell using MLC techniques in Section~\ref{subsec:MLC}.  Two other approaches that have been discussed are the implementation of a
sublithographic crossbar memory to go beyond the lithographic dimension, $F$\cite{Zhong:2003, Gopalakrishnan:2005}, and
3-D integration of multiple layers of memory, currently implemented commercially for write-once
solid-state memory\cite{Johnson:2003b}.

A sublithographic crossbar memory requires a scheme for connecting the ultra-small memory devices laid out at tight pitch to the ``larger'' wiring created at the tightest pitch offered by lithography.  One scheme that was proposed used a Micro-to-Nano Addressing Block, in which current injected into a lithographically-defined via was steered into one of several sublithographic wires using either precise control over depletion regions\cite{Gopalakrishnan:2005} or binary gating by overlying control gates\cite{Hart:2007}.  These schemes work because lithography is typically capable of overlay errors that are 5-10$\times$ smaller than the minimum size feature.  Thus overlying control gates can be placed to cover two but not three sub-lithographic wires, even though the control gate cannot possibly be made as narrow as the sub-lithographic wire.

The weakness of such a sublithographic crossbar scheme is that it requires the creation and careful placement of sublithographic wire arrays of non-trivial complexity.  Next-generation techniques such as imprint lithography may soon be capable of delivering small arrays at roughly the same pitch as cutting-edge lithography\cite{Hart:2007}, but unfortunately pitches that are 4$\times$ more dense than cutting-edge lithography is what would be required.  Intriguingly, large portions of such sublithographic wire arrays would resemble simple grating patterns, suggesting the use of techniques such as interferometric lithography.  Unfortunately, the addressing schemes require that wires at the edges of such arrays terminate precisely yet non-uniformly along the edge, thus complicating the task greatly for an interferometric exposure scheme\cite{Gopalakrishnan:2005, Hart:2007}.

A more flexible approach is to build layers of PCM memory devices, stacking the memory in 3--D above the silicon wafer. This is not the same as 3--D packaging, where devices originally fabricated on separate silicon wafers are connected together using vias that punch through the upper silicon layers to connect to the underlying circuitry.  Instead, the entire memory is built above a single layer of silicon just as the multiple wiring levels of a conventional semiconductor product are built in the ``back end'' of a CMOS process.

This approach has several constraints, including the need to tolerate a significant BEOL temperature budget (an example might be $\sim$400\degreesC~for $>$1 hour) and the need to implement an access device for the PCM devices which can be produced in the metal-and-dielectric layers above the original silicon wafer.  Given the difficulty of growing single-crystal silicon without a seed layer, this implies that the access device must be implemented with either a polysilicon or non-silicon device.  (Note of course that the ability to easily grow multiple layers of high-quality silicon would likely enable a straightforward path to multi-layer Flash memory.)

One example of such a stacked memory is the write-once anti-fuse memory technology developed by Matrix semiconductor (now part of SanDisk), which uses a high-performance polysilicon diode\cite{Johnson:2003b}.  However, it is difficult to obtain the high currents and current densities needed for PCM from such diodes.  This is the case even after accounting for the consideration that the lithographically-defined polysilicon diode can be 10$\times$ larger in area than the sub-lithographic PCM device without increasing the 4$F^2$ footprint.

Thus the advent of 3D-in-the-BEOL PCM technology depends on either dramatic improvements in the current-carrying capability of polysilicon diodes, or on the development of a high-performance non-silicon access device.  Such a device would need to be BEOL-compatible yet not require any temperatures higher than $\sim$400\degreesC, would need to readily pass the high currents (50-150$\mu$A) needed for PCM, yet must provide ultra-low leakage for all non-selected devices.  For instance, in a ``half-select'' scheme implemented across a 1000 $\times$ 1000 device array, at the same instant that a selected device is receiving its RESET current, there are $\sim$2000 devices that share either the same word-line or bit-line with the selected device.  Although these devices are each ``seeing'' half the voltage across the selected device, the total leakage through all these devices must remain much lower than the RESET current value, implying that the required ON-OFF ratio should be significantly in excess of 2000.  However, if such an access device could be developed, since PCM itself has been proven to be BEOL-compatible, the path to 4-8$\times$ increases in effective areal density would be available.  In combination with 2-4 bits of MLC, this would provide an extremely attractive density-- (and thus cost--) differential over even 4-bit MLC Flash.

\section{Conclusions}\label{sec:Conclusions}

PCM has made great strides over the past decade.  Ten or so years ago, PCM was merely a long-dead technology that had, before expiring, helped point the way to fast-crystallizing materials and the mass-market success of read-write optical storage. At that point, any worries about the future of Flash technology were safely covered by the promise of Ferroelectric and Magnetic RAM.  Since that time, both FRAM and MRAM have proved to be less scalable than had been hoped\cite{Burr:2008a}, although the original MRAM concept has since been mostly replaced by the more promising Spin-Transfer Torque (STT-RAM)\cite{Hosomi:2005} and Racetrack memory\cite{Parkin:2004} concepts.  In addition, as is often the case, the large and talented body of engineers working on Flash technology managed to hold off its ``impending'' demise and successfully scale their technology to smaller and smaller technology nodes.

However, these continuing worries about ultra-scaled Flash devices have not gone away --- and the rapid increases in the size of the NAND Flash market, as driven by consumer-oriented devices such as cell phones and MP3 players, now means that significant financial implications are associated with such worries.  Thus PCM was given another opportunity, which it seized by quickly demonstrating better endurance than Flash and near-DRAM switching speeds using those new materials\cite{Lai:2001}, and later CMOS-compatible integration\cite{Hwang:2003c}, scalability to future technology nodes\cite{Chen:2006t}, and the capability for robust Multi-Level Cell (MLC) operation\cite{Nirschl:2007}. All this despite needing to perform high-temperature melting or recrystallization on every writing step.

That said, there remain significant hurdles standing between PCM and its success in the NVM market. These high temperatures force the associated transistor or diode used as an access device to supply a significant amount of current, and lead to PCM cell designs built around aggressively sub-lithographic features.  In turn, the need to define such tiny features with high yield yet low variability, when coupled with the sensitivity of phase change materials to process-related damage, leads to fabrication processes that are difficult to perfect.
MLC performance is frustrated by long-term drift of resistance in the amorphous phase, while PCM retention is bedeviled by early failure of the amorphous plugs in a few ``unlucky'' RESET cells.   Cycling endurance is affected by slow yet steady separation of the constituent atoms, which may be dependent on bias-polarity, leading to void formation (stuck-RESET) or to significant shifts of the cell's operating characteristics (stuck-SET).

It remains to be seen if PCM researchers and developers will be able to successfully navigate these hurdles, allowing the strengths of PCM technology (its high endurance and performance relative to Flash) to shine through in marketable products.  Given these strengths, one can surmise that PCM either will succeed in the long run or will fail completely, but will not be condemned to serving a few niche markets.  Instead, if PCM fails, it will be on a cost basis --- because either tricky processes proved too difficult to implement, delivered unacceptable yields even after many months of effort, or designs were constrained to large cell-sizes and thus uninteresting density points.  In the cutthroat memory and storage landscape, few customers can be expected to be interested in paying significantly more for the better endurance and performance characteristics of PCM.  However, if researchers can finesse the issues of resistance drift and deliver high-current-capable non-silicon access devices, and if
developers can take these advances and implement robust, high-yielding processes that combine MLC and multiple layers, then the resulting ultra-high memory densities will put PCM in a highly advantageous position.  It would be well-positioned to compete directly with Flash, while simultaneously creating new applications ranging from ``Storage-type'' Storage Class Memory (high performance PCM-based SSDs for HDD replacement) to ``Memory-type'' Storage Class Memory (synchronously accessed fast PCM that could bring down the cost and power of DRAM-based systems).

\section{Acknowledgements}\label{sec:Acknowledgements}

There are many people who have helped the authors prepare for this paper.  Some authors participated in the
PCM device learning performed as part of the IBM/Qimonda/Macronix PCRAM Joint Project, involving,
in addition to the present authors, R.M. Shelby, C. T. Rettner, S.-H. Chen, H.-L. Lung, T. Nirschl, T.D. Happ, E. Joseph, A. Schrott, C.F. Chen, J.B. Philipp, R. Cheek, M.-H. Lee, W.P. Risk, G.M. McClelland, Y. Zhu, B. Yee, M. Lamorey, S. Zaidi, C.H. Ho, P. Flaitz, J. Bruley, R. Dasaka, S. Rossnagel, M. Yang, and R. Bergmann.  We also gratefully acknowledge our collaborators D. Milliron, M. Salinga, D. Krebs, B.-S. Lee, R. Mendelsberg, and J. Jordan-Sweet, as well as processing support
from the Microelectronic Research Line at Yorktown Heights; expert analytical and other support from the Almaden
Research Center (R. King, K. Nguyen, M. Jurich, D. Pearson, A. Friz, A. Kellock, V. Deline, T. Topuria, P. Rice, D. Miller, C.M. Jefferson, J. Cha, Y. Zhang, M. Caldwell,
P. Green, and K. Appavoo); physical failure analysis (M. Hudson, L. Garrison, M. Erickson); and valuable discussions
with M. Wuttig, C. Narayan, W. Wilcke, R. Freitas, W. Gallagher, R. Liu, G. Mueller, and T.C. Chen.

\bibliography{PCM_review,PCM_extra}

\ifthenelse{\forSubmission = 1}{
\renewcommand{\atend}{1}

\pagebreak
\TableA
\clearpage

\leftline{\bf Figure Captions:}
\begin{longtable}{p{0.75in}p{5.5in}}
Figure~1 & \FigureCaptionA \\
Figure~2 & \FigureCaptionB \\
Figure~3 & \FigureCaptionC \\
Figure~4 & \FigureCaptionD \\
Figure~5 & \FigureCaptionE \\
Figure~6 & \FigureCaptionF \\
Figure~7 & \FigureCaptionG \\
Figure~8 & \FigureCaptionH \\
Figure~9 & \FigureCaptionI \\
Figure~10 & \FigureCaptionJ \\
Figure~11 & \FigureCaptionK \\
Figure~12 & \FigureCaptionL \\
Figure~13 & \FigureCaptionM \\
Figure~14 & \FigureCaptionN \\
Figure~15 & \FigureCaptionO \\
Figure~16 & \FigureCaptionP \\
Figure~17 & \FigureCaptionQ \\
Figure~18 & \FigureCaptionR \\
Figure~19 & \FigureCaptionS \\
Figure~20 & \FigureCaptionT \\
Figure~21 & \FigureCaptionU \\
Figure~22 & \FigureCaptionV \\
Figure~23 & \FigureCaptionW \\
Figure~24 & \FigureCaptionX \\
Figure~25 & \FigureCaptionY \\
Figure~26 & \FigureCaptionZ \\
Figure~27 & \FigureCaptionAA \\
Figure~28 & \FigureCaptionBB \\
Figure~29 & \FigureCaptionCC \\
Figure~30 & \FigureCaptionDD \\
Figure~31 & \FigureCaptionEE \\
Figure~32 & \FigureCaptionFF \\
Figure~33 & \FigureCaptionGG \\
Figure~34 & \FigureCaptionHH \\
Figure~35 & \FigureCaptionII \\
Figure~36 & \FigureCaptionJJ \\
\end{longtable}

\clearpage \FigureA
\clearpage \FigureB
\clearpage \FigureC
\clearpage \FigureD
\clearpage \FigureE
\clearpage \FigureF
\clearpage \FigureG
\clearpage \FigureH
\clearpage \FigureI
 \clearpage \FigureJ
 \clearpage \FigureK
 \clearpage \FigureL
 \clearpage \FigureM
 \clearpage \FigureN
 \clearpage \FigureO
 \clearpage \FigureP
 \clearpage \FigureQ
 \clearpage \FigureR
 \clearpage \FigureS
 \clearpage \FigureT
 \clearpage \FigureU
 \clearpage \FigureV
 \clearpage \FigureW
 \clearpage \FigureX
 \clearpage \FigureY
 \clearpage \FigureZ
 \clearpage \FigureAA
 \clearpage \FigureBB
 \clearpage \FigureCC
 \clearpage \FigureDD
 \clearpage \FigureEE
 \clearpage \FigureFF
 \clearpage \FigureGG
 \clearpage \FigureHH
 \clearpage \FigureII
 \clearpage \FigureJJ
}{
}

\end{document}